\documentclass[longauth]{aa}  

\usepackage{graphicx}
\usepackage{txfonts}
\usepackage{amsmath}	
\usepackage{subcaption}
\usepackage{hyperref}
\hypersetup{
    colorlinks=true,
    linkcolor=blue,
    filecolor=blue,      
    urlcolor=blue,
    citecolor=blue,
    }
\usepackage{xcolor}

\mathchardef\mhyphen="2D

\newcommand{\di}{\mathrm{d}}

\newcommand{\bfB}{\mathbf{B}}

\newcommand{\bfv}{\mathbf{v}}
\newcommand{\bfu}{\mathbf{u}}

\newcommand{\pc}{\,{\rm pc}}
\newcommand{\kpc}{\,{\rm kpc}}
\newcommand{\Myr}{\,{\rm Myr}}

\newcommand{\cm}{\,{\rm cm}}
\newcommand{\kms}{\,{\rm km\, s^{-1}}}

\newcommand{\cs}{c_{\rm s}}

\newcommand{\pa}{\partial}

\newcommand{\Msun}{\, \rm M_\odot}
\newcommand{\Msunyr}{\, \rm M_\odot\, yr^{-1}}

\newcommand{\Omegap}{\Omega_{\rm p}}
\newcommand{\G}{\mathrm{G}}

\newcommand{\hatex}{\hat{\mathrm{\mathbf{e}}}_x}
\newcommand{\hatey}{\hat{\mathrm{\mathbf{e}}}_y}
\newcommand{\hatez}{\hat{\mathbf{\mathbf{e}}}_z}
\newcommand{\hateR}{\hat{\mathrm{\mathbf{e}}}_R}
\newcommand{\hatephi}{\hat{\mathrm{\mathbf{e}}}_\phi}

\DeclareMathOperator{\sech}{sech}

\begin{document} 

\title{Magnetic field morphology and evolution in the Central Molecular Zone and its effect on gas dynamics}

\newcommand{\aifa}{Argelander-Institut f\"{u}r Astronomie, Universit\"{a}t Bonn, Auf dem H\"{u}gel 71, 53121, Bonn, Germany}
\newcommand{\ita}{Universit\"{a}t Heidelberg, Zentrum f\"{u}r Astronomie, Institut f\"{u}r Theoretische Astrophysik, Albert-Ueberle-Str. 2, 69120 Heidelberg, Germany}
\newcommand{\iwr}{Universit\"{a}t Heidelberg, Interdisziplin\"{a}res Zentrum f\"{u}r Wissenschaftliches Rechnen, Im Neuenheimer Feld 205, D-69120 Heidelberg, Germany}
\newcommand{\eso}{ESO Southern Observatory, Karl-Schwarzschild-Stra{\ss}e 2, 85748 Garching, Germany}
\newcommand{\mcmaster}{Department of Physics and Astronomy, McMaster University, 1280 Main Street West, Hamilton, ON L8S 4M1, Canada}
\newcommand{\cita}{Canadaian Institute for Theoretical Astrophysics (CITA), University of Toronto, 60 St George Street, Toronto, ON M5S 3H8, Canada}
\newcommand{\lyon}{Univ Lyon, Univ Lyon1, ENS de Lyon, CNRS, Centre de Recherche Astrophysique de Lyon UMR5574, F-69230 Saint-Genis-Laval France}
\newcommand{\OSU}{Department of Astronomy, The Ohio State University, 140 West 18th Avenue, Columbus, Ohio 43210, USA}
\newcommand{\CCAPP}{Center for Cosmology and Astroparticle Physics, 191 West Woodruff Avenue, Columbus, OH 43210, USA}
\newcommand{\ljmu}{Astrophysics Research Institute, Liverpool John Moores University, 146 Brownlow Hill, Liverpool L3 5RF, UK}
\newcommand{\mpia}{Max-Planck-Institut f\"ur Astronomie, K\"onigstuhl 17, D-69117 Heidelberg, Germany}
\newcommand{\ugent}{Sterrenkundig Observatorium, Universiteit Gent, Krijgslaan 281 S9, B-9000 Gent, Belgium}
\newcommand{\ign}{Observatorio Astron{\'o}mico Nacional (IGN), C/Alfonso XII, 3, E-28014 Madrid, Spain}
\newcommand{\oxford}{Sub-department of Astrophysics, Department of Physics, University of Oxford, Keble Road, Oxford OX1 3RH, UK}
\newcommand{\durham}{Institute for Computational Cosmology, Department of Physics, Durham University, South Road, Durham, DH1 3LE, UK}
\newcommand{\aapf}{NSF Astronomy and Astrophysics Postdoctoral Fellow}
\newcommand{\arizona}{Steward Observatory, University of Arizona, Tucson, AZ 85721, USA}
\newcommand{\anu}{Research School of Astronomy and Astrophysics, Australian National University, Canberra, ACT 2611, Australia}  
\newcommand{\arc}{ARC Centre of Excellence for All Sky Astrophysics in 3 Dimensions (ASTRO 3D), Australia}
\newcommand{\epfl}{Institute of Physics, Laboratory for Galaxy Evolution and Spectral Modelling, EPFL, Observatoire de Sauverny, Chemin Pegasi 51, 1290 Versoix, Switzerland}
\newcommand{\ualberta}{Dept. of Physics, 4-183 CCIS, University of Alberta, Edmonton, Alberta T6G 2E1, Canada}
\newcommand{\manch}{Jodrell Bank Centre for Astrophysics, Department of Physics and Astronomy, University of Manchester, Oxford Road, Manchester M13 9PL, UK}
\newcommand{\UCSD}{Center for Astrophysics and Space Sciences, Department of Physics, University of California San Diego, 9500 Gilman Drive, La Jolla, CA 92093, USA}
\newcommand{\ari}{Astronomisches Rechen-Institut, Zentrum f{\"u}r Astronomie der Universit{\"a}t Heidelberg, M{\"o}nchhofstra{\ss}e 12-14, 69120 Heidelberg,Germany}
\newcommand{\surrey}{Department of Physics, University of Surrey, Guildford GU2 7XH, UK}
\newcommand{\insubria}{Universit{\`a} dell’Insubria, via Valleggio 11, 22100 Como, Italy}
\newcommand{\naoc}{Chinese Academy of Sciences South America Center for Astronomy, National Astronomical Observatories, CAS, Beijing 100101, China}
\newcommand{\ucn}{Instituto de Astronom\'ia, Universidad Cat\'olica del Norte, Av. Angamos 0610, Antofagasta 1270709, Chile}
\newcommand{\uconn}{University of Connecticut, Department of Physics, 196A Auditorium Road, Unit 3046, Storrs, CT 06269 USA}
\newcommand{\ice}{Institut de Ci\`encies de l’Espai (ICE, CSIC), Can Magrans s/n, E-08193, Bellaterra, Barcelona, Spain}
\newcommand{\uwm}{Department of Astronomy, University of Wisconsin-Madison, Madison, WI 53706, USA}
\newcommand{\MIT}{MIT Haystack Observatory, 99 Millstone Road, Westford, MA, 01886}
\newcommand{\racah}{Racah Institute for Physics, The Hebrew University, Jerusalem 91904, Israel}
\newcommand{\pu}{Department of Astrophysical Sciences, Princeton University, Princeton, NJ 08544, USA}
\newcommand{\cut}{Space, Earth and Environment Department, Chalmers University of Technology, SE-412 96 Gothenburg, Sweden}
\newcommand{\ieec}{Institut d’Estudis Espacials de Catalunya (IEEC), E-08034, Barcelona, Spain}
\newcommand{\inaf}{INAF Astronomical Observatory of Trieste, via G.B. Tiepolo 11, I-34143 Trieste, Italy}
\newcommand{\cab}{Centro de Astrobiolog{\'i}a (CAB), CSIC-INTA, Ctra. de Ajalvir Km. 4, Torrej{\'o}n de Ardoz, 28850 Madrid, Spain}
\newcommand{\cit}{Jet Propulsion Laboratory, California Institute of Technology, 4800 Oak Grove Drive, Pasadena, CA, 91109, USA}
\newcommand{\uflor}{Department of Astronomy, University of Florida, PO Box 112055, Gainesville, FL 32611-2055, USA)}
\newcommand{\harvard}{Center for Astrophysics, Harvard \& Smithsonian, MS-42, 60 Garden Street, Cambridge, MA 02138 USA}
\newcommand{\ucal}{Department of Physics and Astronomy, University of California, Los Angeles, Box 951547, Los Angeles, CA 90095-1547 USA}
\newcommand{\mpira}{Max-Planck-Institut f{\"u}r Radioastronomie, Auf dem H{\"u}gel 69, 53121 Bonn, Germany}
\newcommand{\tum}{Technical University of Munich, School of Engineering and Design, Department of Aerospace and Geodesy, Chair of Remote Sensing Technology, Arcisstr. 21, D-80333 Munich, Germany}
\newcommand{\cool}{Cosmic Origins Of Life (COOL) Research DAO, Munich, Germany;}
\newcommand{\oab}{INAF -- Osservatorio Astronomico di Brera, via E. Bianchi 46, Merate, 23807, Italy}
\newcommand{\iaps}{Istituto di Astrofisica e Planetologia Spaziale (IAPS). INAF. Via Fosso del Cavaliere 100, 00133 Roma, Italy}
\newcommand{\unam}{Instituto de Radioastronomía y Astrofísica, UNAM. Apdo. Postal 72-3 (Xangari), Morelia, Michocán 58089, México}
\newcommand{\stand}{SUPA, School of Physics and Astronomy, University of St Andrews, North Haugh, Fife, KY16 9SS, UK }
\newcommand{\saclay}{Université Paris-Saclay, Université Paris Cité, CEA, CNRS, AIM, 91191, Gif-sur-Yvette, France}


\author{R.~G.~Tress     \inst{\ref{epfl}}\thanks{robin.tress@epfl.ch}
        \and
        M.~C.~Sormani        \inst{\ref{insubria}, \ref{surrey}}\thanks{mattiacarlo.sormani@gmail.com}
        \and
        P.~Girichidis        \inst{\ref{ita}}
        \and
        S.~C.~O.~Glover      \inst{\ref{ita}}
        \and 
        R.~S.~Klessen        \inst{\ref{ita},\ref{iwr}}
        \and
        R.~J.~Smith         \inst{\ref{stand},\ref{manch}}
        \and
        E.~Sobacchi          \inst{\ref{oab}}
        \and
        L.~Armillotta        \inst{\ref{pu}}
        \and
        A.~T.~Barnes          \inst{\ref{eso}}
        \and
        C.~Battersby         \inst{\ref{uconn}}
        \and
        K.~R.~J.~Bogue       \inst{\ref{manch}}
        \and 
        N.~Brucy             \inst{\ref{ita}}
        \and
        L.~Colzi             \inst{\ref{cab}}
        \and
        C.~Federrath         \inst{\ref{anu}, \ref{arc}}
        \and
        P.~Garc\'ia          \inst{\ref{naoc}, \ref{ucn}}
        \and
        A.~Ginsburg          \inst{\ref{uflor}}
        \and
        J.~G{\"o}ller        \inst{\ref{ita}}
        \and
        H~P.~Hatchfield      \inst{\ref{cit}, \ref{uconn}}
        \and
        C.~Henkel            \inst{\ref{mpira}}
        \and
        P.~Hennebelle        \inst{\ref{saclay}}
        \and
        J.~D.~Henshaw        \inst{\ref{ljmu}, \ref{mpia}}
        \and
        M.~Hirschmann        \inst{\ref{epfl}, \ref{inaf}}
        \and
        Y.~Hu                \inst{\ref{uwm}}
        \and
        J.~Kauffmann         \inst{\ref{MIT}}
        \and
        J.~M.~D.~Kruijssen   \inst{\ref{tum}, \ref{cool}}
        \and
        A.~Lazarian          \inst{\ref{uwm}}
        \and
        D.~Lipman            \inst{\ref{uconn}}
        \and
        S.~N.~Longmore       \inst{\ref{ljmu}, \ref{cool}}
        \and
        M.~R.~Morris            \inst{\ref{ucal}}
        \and
        F.~Nogueras-Lara     \inst{\ref{eso}}
        \and
        M.~A.~Petkova        \inst{\ref{cut}}
        \and
        T.~G.~S.~Pillai      \inst{\ref{MIT}}
        \and
        V.~M.~Rivilla        \inst{\ref{cab}}
        \and
        \'A.~S\'anchez-Monge \inst{\ref{ice}, \ref{ieec}}
        \and
        J.~D.~Soler          \inst{\ref{iaps}}
        \and
        D.~Whitworth         \inst{\ref{unam}, \ref{ita}}
        \and
        Q.~Zhang             \inst{\ref{harvard}}
         }

\institute{\epfl   \label{epfl}
           \and
           \insubria \label{insubria}
           \and
           \surrey \label{surrey}
           \and 
           \ita    \label{ita}
           \and
           \iwr    \label{iwr}
           \and
           \stand  \label{stand}
           \and
           \manch  \label{manch}
           \and
           \oab  \label{oab}
           \and
           \pu     \label{pu}
           \and
           \eso    \label{eso}
           \and
           \uconn  \label{uconn}
           \and
           \cab    \label{cab}
           \and
           \anu    \label{anu}
           \and
           \arc    \label{arc}
           \and
           \naoc   \label{naoc}
           \and
           \ucn    \label{ucn}
           \and
           \uflor  \label{uflor}
           \and
           \cit    \label{cit}
           \and
           \mpira  \label{mpira}
           \and
           \saclay \label{saclay}
           \and
           \ljmu   \label{ljmu}
           \and
           \mpia   \label{mpia}
           \and
           \inaf   \label{inaf}
           \and
           \uwm    \label{uwm}
           \and
           \MIT    \label{MIT}
           \and
           \tum    \label{tum}
           \and
           \cool   \label{cool}
           \and
           \ucal   \label{ucal}
           \and
           \cut    \label{cut}
           \and
           \ice    \label{ice}
           \and
           \ieec   \label{ieec}
           \and 
           \iaps   \label{iaps}
           \and    
           \unam   \label{unam}
           \and
           \harvard \label{harvard}
           }

\newcommand{\ackHelix}{The authors acknowledge support by the state of Baden-Württemberg through bwHPC and the German Research Foundation (DFG) through grant INST 35/1597-1 FUGG and DFG grant INST 35/1134-1 FUGG. }
\newcommand{\ackMCS}{MCS acknowledges financial support from the European Research Council under the ERC Starting Grant ``GalFlow'' (grant 101116226) and from the Royal Society (URF\textbackslash R1\textbackslash 221118). }
\newcommand{\ackRSK}{RSK, SCOG and PG gratefully acknowledge support from the ERC via the ERC Synergy Grant ``ECOGAL'' (grant 855130),  from the German Excellence Strategy via the Heidelberg Cluster of Excellence (EXC 2181 - 390900948) ``STRUCTURES'', and from the German Ministry for Economic Affairs and Climate Action in project ``MAINN'' (funding ID 50OO2206). }
\newcommand{\ackJDH}{JDH gratefully acknowledges financial support from the Royal Society (University Research Fellowship; URF/R1/221620)}
\newcommand{\ackAMS}{ASM acknowledges support from the RyC2021-032892-I grant funded by MCIN/AEI/10.13039/501100011033 and by the European Union `Next GenerationEU’/PRTR, as well as the program Unidad de Excelencia Mar\'ia de Maeztu CEX2020-001058-M. }
\newcommand{\ackPG}{PG was supported by the Chinese Academy of Sciences (CAS), through a grant to the CAS South America Center for Astronomy (CASSACA) in Santiago, Chile, and by the “Comisión Nacional de Ciencia y Tecnología (CONICYT)" now ANID via Project FONDECYT de Iniciación 11170551. }
\newcommand{\ackVMR}{VMR also acknowledges support from the grant number RYC2020-029387-I funded by MICIU/AEI/10.13039/501100011033 and by "ESF, Investing in your future", and from the Consejo Superior de Investigaciones Cient{\'i}ficas (CSIC) and the Centro de Astrobiolog{\'i}a (CAB) through the project 20225AT015 (Proyectos intramurales especiales del CSIC). }
\newcommand{\ackLC}{LC and VMR acknowledge funding from grants No. PID2019-105552RB-C41 and PID2022-136814NB-I00 by the Spanish Ministry of Science, Innovation and Universities/State Agency of Research MICIU/AEI/10.13039/501100011033 and by ERDF, UE. }
\newcommand{\ackES}{ES acknowledges support from the Marie Sk{\l}odowska-Curie Grant 101061217. }
\newcommand{\ackJMDK}{JMDK gratefully acknowledges funding from the European Research Council (ERC) under the European Union's Horizon 2020 research and innovation programme via the ERC Starting Grant MUSTANG (grant agreement number 714907). COOL Research DAO is a Decentralised Autonomous Organisation supporting research in astrophysics aimed at uncovering our cosmic origins. }
\newcommand{\ackAG}{AG acknowledges support from the NSF under grants AAG 2206511 and CAREER 2142300. }
\newcommand{\ackCB}{CB gratefully  acknowledges  funding  from  National  Science  Foundation  under  Award  Nos. 1816715, 2108938, 2206510, and CAREER 2145689, as well as from the National Aeronautics and Space Administration through the Astrophysics Data Analysis Program under Award No. 21-ADAP21-0179 and through the SOFIA archival research program under Award No.  09$\_$0540. }
\newcommand{\ackJG}{JG is a member of the International Max Planck Research School for Astronomy and Cosmic Physics at the University of Heidelberg (IMPRS-HD) and acknowledges financial support by the European Research Council via the ERC Synergy Grant “ECOGAL” (project ID 855130). }


 
  \abstract{The interstellar medium in the Milky Way's Central Molecular Zone (CMZ) is known to be strongly magnetised, but its large-scale morphology and impact on the gas dynamics are not well understood. We explore the impact and properties of magnetic fields in the CMZ using three-dimensional non-self gravitating magnetohydrodynamical simulations of gas flow in an external Milky Way barred potential. We find that: (1) The magnetic field is conveniently decomposed into a regular time-averaged component and an irregular turbulent component. The regular component aligns well with the velocity vectors of the gas everywhere, including within the bar lanes. (2) The field geometry transitions from parallel to the Galactic plane near $z=0$ to poloidal away from the plane. (3) The magneto-rotational instability (MRI) causes an in-plane inflow of matter from the CMZ gas ring towards the central few parsecs of $0.01\mhyphen0.1\Msunyr$ that is absent in the unmagnetised simulations. However, the magnetic fields have no significant effect on the larger-scale bar-driven inflow that brings the gas from the Galactic disc into the CMZ. (4) A combination of bar inflow and MRI-driven turbulence can sustain a turbulent vertical velocity dispersion of $\sigma_z \simeq 5\kms$ on scales of $20\pc$ in the CMZ ring. The MRI alone sustains a velocity dispersion of $\sigma_z \simeq 3\kms$. Both these numbers are lower than the observed velocity dispersion of gas in the CMZ, suggesting that other processes such as stellar feedback are necessary to explain the observations. (5) Dynamo action driven by differential rotation and the MRI amplifies the magnetic fields in the CMZ ring until they saturate at a value that scales with the average local density as $B \simeq 102 \, (n/10^3 \, {\rm cm^{-3}})^{0.33} \, {\rm \mu G}$.  Finally, we discuss the implications of our results within the observational context in the CMZ.} 
   \maketitle


\section{Introduction} \label{sec:intro}

The Milky Way's Central Molecular Zone (CMZ) is a ring-like $\sim 3\times10^7\Msun$ accumulation of molecular gas within Galactocentric radius $R\simeq 200\pc$ \citep{Morris1996,Henshaw2023}. It is generated by the Galactic bar, which efficiently transports gas from Galactocentric radius $R\approx 3 \kpc$ down to $R \approx 200\pc$ \citep{Sormani2019,Hatchfield2021}. The CMZ is the most extreme star-forming environment in the entire Milky Way and has emerged in the last decade as an important astrophysical laboratory to study the physics of the interstellar medium (ISM) and star formation \citep{Henshaw2023}.

The CMZ is permeated by a strong magnetic field that is likely to play an important role in the dynamics of the ISM and in the process of star formation \citep{Ferriere2009,Morris2015,Butterfield2023}. However, many aspects of the magnetic field configuration are poorly understood. For example, it is unclear whether the CMZ is immersed in a pervasive $|\bfB|\gtrsim 1$~mG field \citep{Morris1989,Morris2006}, or whether the magnetic field drops to $|\bfB| \sim 100 \; \mu$G or less in the more diffuse inter-cloud medium \citep{Tsuboi1985,Yusef-Zadeh1987,Lang1999a,Lang1999b,LaRosa2005,Ferriere2009,Yusef-Zadeh2022} while reaching mG strengths only inside very dense gas \citep{Schwarz1990,Killeen1992,Plante1995,Uchida1995,Marshall1995,Yusef-Zadeh1999,Pillai2015}.

The large-scale geometry of the magnetic field is also unclear. Observations show that the magnetic field in the dense and cold ISM is oriented predominantly parallel to the Galactic plane (in projection on the plane of the sky), while the magnetic field in the more diffuse ISM, including a population of prominent filamentary structures known as non-thermal filaments (NTFs, \citealt{Yusef-Zadeh1984,Tsuboi1986,Heywood2022}), is predominantly perpendicular to the Galactic plane \citep{Novak2003,Chuss2003,Nishiyama2010,Mangilli2019,Guan2021,Hu2022a,Hu2022b,Butterfield2023,Pare2024}. This can be appreciated for example in fig.~6 of \cite{Guan2021}, which shows a change in the observed magnetic field geometry as the fractional contribution of synchrotron radiation (tracing ionised gas) and thermal dust emission (tracing cold neutral gas) varies with frequency, or in fig.\ 1 of \cite{Nishiyama2010}, which shows that $\bfB$ is prevalently parallel to the Galactic plane at latitudes $|b|<0.4^\circ$, where the dense gas dominates their measurements, and becomes perpendicular to the Galactic plane at $|b|>0.4^\circ$, where the diffuse ionised gas dominates the measurements. This might point to the presence of two separate magnetic field systems, one predominantly perpendicular and one predominantly parallel to the plane \citep{Morris2015}. It is presently unclear how the two systems relate to each other and what is their three-dimensional geometry.

Since the CMZ is a star-forming nuclear ring similar to those that are commonly found at the centres of barred galaxies \citep{Mazzuca2008,Comeron2010,Ma2018}, it is reasonable to expect that its magnetic field system shares many similarities with those of other nuclear rings. Measurements have been performed for a handful of galaxies \citep[for a review, we refer to e.g.][]{Beck2015}, the best studied example probably being NGC 1097 \citep{Beck2005,Tabatabaei2018,Lopez-Rodriguez2021,Hu2022c}. These measurements lead to the following general conclusions: (i) The magnetic field lines inferred from radio polarisation maps of synchrotron-emitting gas in the bar region surrounding the nuclear ring are approximately aligned with the gas streamlines. In particular, the magnetic field in the bar lanes that transport the gas towards the nuclear ring is approximately parallel to the lanes in the frame co-rotating with the bar (e.g.\ fig.\ 2 of \citealt{Beck2005}). (ii) The magnetic field in the ring spirals towards the centre with a relatively large pitch-angle both in radio and far-infrared polarisation maps (e.g. fig.\ 1 of \citealt{Lopez-Rodriguez2021}). The pitch-angle and general geometry of the magnetic field inside the ring are tracer dependent \citep{Lopez-Rodriguez2021,Hu2022c}, which is reminiscent of the Milky Way, where as mentioned above the projected orientation of the field near the mid-plane depends on the tracer. Measurements of the magnetic field in dense molecular gas are currently not available for external galaxies \citep{Lopez-Rodriguez2021}. (iii) The equipartition magnetic field in synchrotron-emitting gas in the nuclear ring of NGC 1097 is $\sim$60 $\mu$G, which is of the same order of magnitude as similar estimates for the diffuse gas in the CMZ \citep{LaRosa2005,Morris2006,Yusef-Zadeh2022}.

The effects of the magnetic fields on the dynamics of the ISM in the CMZ are also poorly understood \citep[e.g.][]{Morris2006,Morris2015}. Magneto-hydrodynamic (MHD) instabilities are believed to be one of the primary mechanisms for mass and angular momentum transport in astrophysical accretion discs \citep{Balbus1998}. A back-of-the-envelope calculation suggests that the magneto-rotational instability (MRI; \citealt{Balbus1991}) should transport gas in the CMZ at a rate given by:
\begin{equation} \label{eq:Mdot}
\dot{M} =  3 \alpha {\Msun \, yr^{-1}} \left( \frac{M}{5 \times 10^7 \Msun} \right) \left( \frac{\sigma}{15 \kms } \right) \left( \frac{h}{40 \pc } \right) \left( \frac{R}{100 \pc } \right)^{-2} ,
\end{equation}
where $M$ is the total gas mass of the CMZ, $\sigma$ is the gas velocity dispersion, $h$ is the gas scale-height, $R$ is the Galactocentric radius, $\alpha$ is the \cite{Shakura1973} coefficient which we have assumed to be determined by the MRI, and we have inserted typical CMZ values in the denominators. Assuming $\alpha\simeq 0.01$, the predicted inflow rate is $\dot{M}\simeq0.03 \Msunyr$. This value would be significant, because at this rate the entire circum-nuclear disc \citep[CND;][]{Genzel1985,Mills2017,Hsieh2021}, which is the closest large gas reservoir to SgrA* with a mass of $M_{\rm CND}\sim 5 \times 10^4 \Msun$ \citep{Etxaluze2011,Requena-Torres2012}, would build up on a rather short timescale of $M_{\rm CND}/\dot{M} \sim 1.7\Myr$. However, simple order-of-magnitude estimates such as these are inherently limited. The MRI-driven transport is traditionally studied in the context of Keplerian, weakly magnetised accretion discs \citep[e.g][]{Balbus2003}, and it is much less understood in the context of non-Keplerian, non-axisymmetric potential of the Galactic centre \citep{Kim2012b}, and in the cold ISM regime with small plasma $\beta$ that is relevant there \citep{Kim2000,Kim2003,Piontek2007,Jacquemin-Ide2021}. Therefore, it is currently unclear how effective MRI-driven transport is in the Galactic centre, and how important it is compared to other effects such as redistribution of angular momentum due to stellar feedback \citep{Tress2020}.

There have been a number of theoretical studies of magnetised gas flow in barred galaxies, which can broadly be divided in two groups. The first group uses dynamo theory to follow the evolution of magnetic fields using an approximate set of equations in a prescribed velocity field \citep[e.g.][]{Otmianowska-Mazur2002,Moss2001,Moss2007}. The velocity field can be taken for example from purely hydrodynamical simulations of gas flow in barred galaxies \citep[e.g.][]{Athan92b}. This approach is computationally fast but requires assumptions on how velocity fluctuations act on the magnetic fields on unresolved scales and ignores the back reaction of the magnetic field on the gas. The second approach uses the full set of MHD equations without approximations. For example, \cite{Kim2012b} performed MHD simulations of barred galaxies, finding that magnetic fields can enhance inflows and that the bar potential plays a role in dynamo action. However, their simulations are two-dimensional and therefore unable to capture the potential effects of poloidal fields and other dynamical processes that may be important in three dimensions. \cite{Suzuki2015} and \cite{Kakiuchi2024} performed three-dimensional MHD simulations of gas flow in the innermost few kiloparsec of the Milky Way, but they assumed an axisymmetric potential and neglected the presence of the Galactic bar. \cite{Moon2023} run semi-global MHD simulations of nuclear rings in an axisymmetric potential subject to a prescribed mass inflow rate, and found that magnetic fields can drive radial flows from the ring inwards and that they can suppress star formation in the ring. However, to the best of our knowledge, a global sub-pc 3D MHD simulation of gas flow in a barred potential is still missing.

In this paper we use global 3D MHD simulations of gas flow in a Milky Way barred potential to address the following open questions:
\begin{itemize}
    \item How do magnetic fields affect the gas morphology in the CMZ? (Sect.~\ref{sec:gasmorphology})
    \item What is the geometry of the magnetic field in the CMZ and in the surrounding bar region? (Sect.~\ref{sec:BfieldsCMZbar})
    \item Do magnetic fields drive turbulence? (Sect.~\ref{sec:turbulence})
    \item How are magnetic fields amplified and maintained? (Sect.~\ref{sec:Bevolution})
    \item Do magnetic fields enhance the bar-driven inflow from the Galactic disc to the CMZ? Do magnetic fields drive a nuclear inflow from the CMZ towards the central few parsecs? (Sect.~\ref{sec:inflow}) 
\end{itemize}
The aim is to study magnetic fields and their effect on the gas dynamics. We therefore deliberately choose not to include the gas self-gravity, nor any type of star formation and stellar feedback. Such additional processes would make it very difficult to isolate the contribution of magnetic fields for example in driving turbulence or changing the probability density function of the gas. Studying these processes and their (possibly non-linear) interaction with the magnetic fields is an important avenue for future work.

This paper is structured as follows. In Sect.~\ref{sec:methods} we describe our numerical methods. Sections~\ref{sec:gasmorphology}-\ref{sec:inflow} are dedicated to addressing the open questions listed above. We sum up our results in Sect.~\ref{sec:conclusion}.

\section{Numerical methods} \label{sec:methods}

We run three-dimensional MHD simulations of gas flow in the inner regions of the Milky Way using the moving-mesh code {\sc arepo} \citep{Springel2010,Pakmor2016,Weinberger2020}. The simulated gas disc covers the entire region within Galactocentric radius $R=5\kpc$. We assume ideal MHD which is generally an excellent approximation for the ISM \citep[e.g.][]{Marinacci2018} and use the standard MHD implementation contained in {\sc arepo}, which has been previously employed for galaxy simulations and tested against a number of standard test problems including the development of the magneto-rotational instability \citep[e.g.][]{Pakmor2013,Marinacci2018b}. The equations we solve are:
\begin{align}
    \label{eq:Continuity}
    & \frac{\partial \rho}{\partial t} + \nabla \cdot (\rho \bfv ) = 0 \,, \\
    \label{eq:euler}
    & \frac{\partial ( \rho \bfv) }{\partial t} + \nabla \cdot \left( \rho \bfv \bfv + P \mathbb{I} + \mathbb{T} \right) = - \rho \nabla \Phi \,, \\
    \label{eq:energy}
    & \frac{ \partial \left(\rho e \right) }{\partial t} + \nabla \left[ \left( \rho e  \right) \bfv + (P \mathbb{I} + \mathbb{T} ) \cdot \bfv \right] = \rho \frac{\partial \Phi}{\partial t} - \mathcal{L} \,, \\
     \label{eq:induction}
    & \frac{\partial \bfB}{\partial t} = \nabla \times \left( \bfv \times \bfB \right)\,,
\end{align}
where $\rho$ is the gas density, $\bfv$ is the velocity, $P$ is the thermal pressure, $\mathbb{I}$ is the identity matrix, $\bfB$ is the magnetic field, $\mathbb{T}=B^2/(8\pi)\mathbb{I} - \bfB \bfB/(4\pi)$ is the Maxwell stress tensor under the approximation of non-relativistic ideal MHD, $\Phi$ is the external gravitational potential, $\rho e = \rho e_{\rm th} + \rho \bfv^2/2 + \bfB^2/(8 \pi) + \rho \Phi$ is the total energy per unit volume, which is the sum of the thermal ($\rho e_{\rm th}$), kinetic ($\rho \bfv^2/2$), magnetic ($\bfB^2/(8\pi)$), and gravitational ($\rho\Phi$) contributions, and $\mathcal{L}$ is the net cooling (or heating) rate per unit volume. Magnetic field divergence errors can arise as a result of the discretization of the MHD equations. In Appendix~\ref{sec:divB} we checked that these are always under control and do not dominate the dynamics of the ISM. 

The gas is assumed to flow in an externally imposed Milky Way barred potential. The potential is identical to that used in \cite{Ridley2017} and is described in detail in section 3.2 of that paper (their fig.\ 4 shows the rotation curve). The bar rotates rigidly with a pattern speed $\Omegap=40 \kms \kpc$, consistent with recent determinations for the Milky Way \citep[e.g.][]{Sormani2015,Portail2017,Sanders2019,Li2022,Clarke2022}. This places the (single) inner Lindblad resonance (ILR) calculated in the epicyclic approximation at $R_{\rm ILR} = 1.1 \kpc$ and the corotation resonance at $R_{\rm CR} = 5.9 \kpc$. This potential was chosen to allow direct comparison with previous simulations in \cite{Ridley2017} and \cite{Sormani2018} that used the same potential. We did not include gas self-gravity nor the consequent star formation in this paper as we aim to isolate the effects of the magnetic fields from other competing effects on the dynamics of the ISM.

\begin{figure}
	\includegraphics[width=\columnwidth]{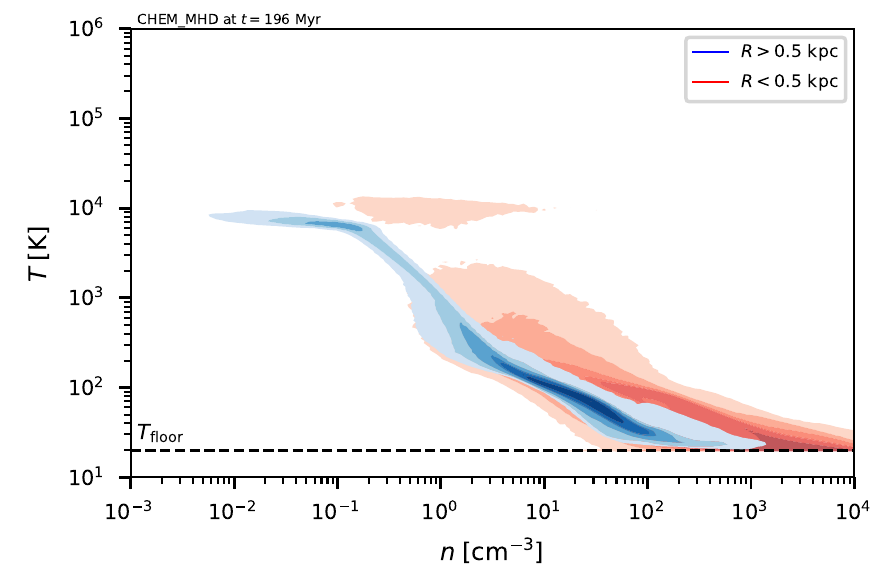}
        \caption{Gas temperature as a function of total gas density in our fiducial CHEM\_MHD model at $t=196$~Myr. Blue and red indicate the regions at $R>0.5$~kpc and $R<0.5$~kpc, respectively. Contours contain [99, 95, 90, 80, 75, 50]\% of the total mass respectively. The gas is a two phase medium with a cold ($T=10^2$~K) and warm ($T=10^4$~K) phase. The hot phase ($T=10^6$~K) is absent since our simulations do not include stellar feedback or other processes that can create it.}
    \label{fig:DensityTemperaturePhaseDiagram}
\end{figure}

We run simulations using two different thermodynamic setups. The `isothermal' simulations used an isothermal equation of state
\begin{equation}
P = \cs^2 \rho,
\end{equation}
where $\cs$ is a constant that we set to either $\cs=1\kms$ or $\cs=10\kms$ (representative of `cold' and `warm' ISM). In the isothermal simulations, $\mathcal{L}=0$ in Eq.~\eqref{eq:energy}. We note that the isothermal approximation does not correspond to a physical situation where there is no heating and cooling. Instead, it means that cooling and heating always exactly balance in such a way that the gas temperature is kept constant \citep[e.g.][]{Klessen2016}. For example, the energy released in shocks is instantaneously radiated away in the isothermal approximation.

The `chemistry' simulations used the adiabatic equation of state
\begin{equation}
    \label{eq:eos}
    P = (\gamma - 1) \rho e_{\rm th},
\end{equation}
where $\gamma = 5/3$ is the adiabatic index. These simulations account for the chemical evolution of the gas using an updated version of the NL97 chemical network from \citet{gc12}, which is based on the work of \citet{gm07a,gm07b} and \citet{nl97}. This network solves for the non-equilibrium abundances of H, H$_{2}$, H$^{+}$, C$^{+}$, O, CO, and free electrons. The heating and cooling contained in the term $\mathcal{L}$ in Eq.~\eqref{eq:energy} are calculated on-the-fly by the network based on the instantaneous chemical composition of the gas and taking into account a number of processes, including radiative cooling, heat released by the formation of H$_2$ on dust grains, and an averaged interstellar radiation field (ISRF) and cosmic ray ionization rate (CRIR). The ISRF is set to the standard value $G_0$ measured in the solar neighbourhood \citep{draine78} diminished by a local attenuation factor which depends on the amount of gas present within 30~pc of each computational cell. This attenuation factor is introduced to account for the effects of dust extinction and H$_{2}$ self-shielding and is calculated using the {\sc Treecol} algorithm described in \cite{clark12}. The value was chosen originally as it was similar to the typical observed separation of OB stars in the Solar neighbourhood. Although in the dense CMZ environment the separation might be smaller, we choose to keep the same value here for consistency with previous simulations \citep{Tress2020}. The cosmic ray ionisation rate (CRIR) is fixed to $\zeta_{\rm H} = 3 \times 10^{-17} \: {\rm s^{-1}}$ \citep{gl78}. Although these values are typical for the Solar neighbourhood and likely too small for the CMZ \citep{Clark2013,Oka2019}, we expect this to have little effects on the dynamics of the gas discussed in this paper. Indeed, \citet{Sormani2018} (we refer also to the discussion in section~2.3 of \citealt{Tress2020}) have shown that the strength of the ISRF and CRIR do not affect the large-scale dynamics of the gas in the Galactic Centre region significantly. The main effect is to modify the amount of gas in different ISM phases since cosmic rays and UV photons dissociate and ionise molecular gas. Increasing the ISRF and CRIR has an effect on the dynamics (and on the MRI and inflow rates) that is similar to increasing the effective sound speed of the gas. To check this, we have run some exploratory simulations with higher ISRF and CRIR, and we found that the dynamical behaviour of these simulations is similar to the high-$c_{\rm s}$ simulations discussed below in Sect. \ref{sec:gasmorphology}. A more complete study of the effects of varying the ISRF and CRIR is outside the scope of this work.

Finally, we imposed a numerical temperature floor $T_{\rm floor} =20$~K on the simulated ISM. Without this floor, the code would occasionally produce anomalously low temperatures in cells close to the resolution limit undergoing strong adiabatic cooling, causing it to crash. The chemical network is the same as used in \cite{Sormani2018} and \cite{Tress2020}, and more details can be found in section 3.4 of the former or section 2.3 of the latter. Figure~\ref{fig:DensityTemperaturePhaseDiagram} shows the typical density-temperature phase diagram in our chemistry simulations.

The number density is defined as
\begin{equation}
 n = \frac{\rho}{\mu m_{\rm p}}\,,
\end{equation}
where $\mu$ is the mean molecular weight and $m_{\rm p}$ is the proton mass. As a reference, at the assumed solar metallicity the mean molecular weight is $\mu=0.67, 1.27, 2.23$ for fully ionised, neutral, and fully molecular gas respectively. The temperature of the gas in the chemistry simulations is defined as $T= P/(n k_{\rm b})$, where $k_{\rm b}$ is the Boltzmann constant. 

\subsection{Initial conditions} \label{sec:IC}

\begin{figure}
	\includegraphics[width=\columnwidth]{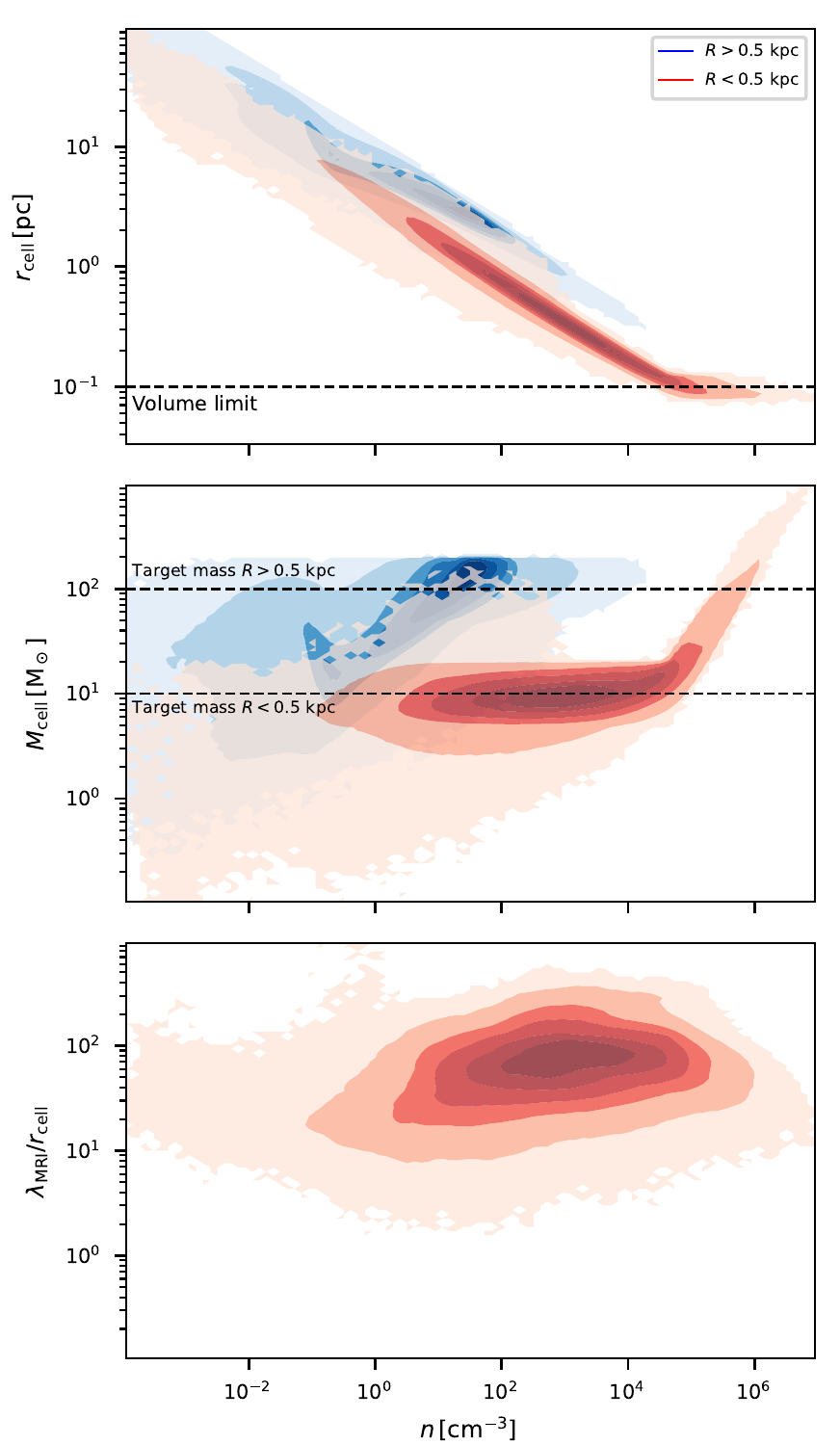}
        \caption{Resolution in our fiducial CHEM\_MHD simulation (Table~\ref{tab:sims}) at $t=196$~Myr. \emph{Top}: spatial resolution as a function of total gas density. $r_{\rm cell}$ is the radius of a sphere with the same volume as the cell. The horizontal black dashed line indicates the volume limit of the cells (Sect.~\ref{sec:methods:summary}). \emph{Middle}: mass of cells as a function of density. In both panels, blue contours are for the region at $R>0.5 \kpc$ with a target mass of $100 \Msun$, while red contours are for the $R<0.5 \kpc$ with higher resolution at a target mass of $10 \Msun$ (Sect.~\ref{sec:methods:summary}). \emph{Bottom}: $\lambda_{\rm MRI}/r_{\rm cell}$ as a function of density for cells in the CMZ ($R<0.5$~kpc), where $\lambda_{\rm MRI}=2\pi v_A / \Omega$ is the characteristic wavelength of the MRI, $v_A$ is the Alfven speed, and $\Omega=v_\phi/R$ is the angular velocity of each gas cell. We see that the MRI is well resolved in the CMZ. The contours contain [100, 99, 90, 75, 50, 25]\% of the total number of cells.}
    \label{fig:Resolution}
\end{figure}

We initialised the density according to the following axisymmetric density distribution:
\begin{equation}
\rho(R,z) = \frac{\Sigma_0}{4 z_{\rm d}}  \exp\left(- \frac{R_{\rm m}}{R} - \frac{R}{R_{\rm d}}\right) \sech^2\left(\frac{z}{2 z_{\rm d}}\right)\, ,
\end{equation}
where $(R,\phi,z)$ denote standard cylindrical coordinates, $z_{\rm d} = 85$~pc, $R_{\rm d} = 7$~kpc, $R_{\rm m} = 1.5$~kpc, $\Sigma_0 = 50$~${\rm M_\odot} \pc^{-2}$, and we cut the disc so that $\rho=10^{-28} \; \rm g / cm^3$ for $R\geq 5\kpc$. This profile roughly matches the observed radial distribution of gas in the Galaxy \citep{KalberlaDedes2008,HeyerDame2015} and is identical to the one used in \cite{Tress2020}. The total initial gas mass in the simulation is $\simeq 1.5 \times 10^9 \Msun$. The computational box has a total size of $24 \times 24 \times 24 \kpc$ with periodic boundary conditions. The box is sufficiently large that the outer boundary has a negligible effect on the evolution of the simulated galaxy.

In order to avoid transients, we introduced the bar gradually \citep[e.g.][]{Athan92b}. We started with gas in equilibrium on circular orbits in an axisymmetrised potential and then we turned on the non-axisymmetric part of the potential linearly during the first $146 \Myr$ (approximately one bar rotation) while keeping constant the total mass which generates the underlying external potential (not to be confused with the mass of the gas in the simulation). Therefore, only the simulation at $t \geq 146 \Myr$, when the bar is fully on, will be considered for the analysis in this paper. The simulations were run until $t=300$~Myr.

The initial temperature for the chemistry simulations is $T_0=10^4 \rm\, K$ everywhere. The precise value does not affect the outcome of the simulation since a new equilibrium is reached within a few megayears (and well before the bar is fully turned on) through the balance of heating and cooling processes.

Unless otherwise specified, we started with a purely poloidal uniform `seed' magnetic field of $\bfB_0 = 0.02 \,\mu\G \, \hat{\mathbf{z}}$. We have also experimented with different initial magnetic field strengths and with initial toroidal (rather than poloidal) geometry; the results of these experiments are briefly discussed in Appendix~\ref{sec:ICimpact}.

\subsection{Summary of simulation runs} \label{sec:methods:summary}

\begin{table}
	\centering
	\begin{tabular}{lccc} 
		name  & eq.~of state & sound speed ($\cs$) & physics \\
		\hline
         ISO\_01\_HD & isothermal & $1\kms$ & HD \\
         ISO\_01\_MHD & isothermal & $1\kms$ & MHD \\
         ISO\_10\_HD & isothermal & $10\kms$ & HD \\
         ISO\_10\_MHD & isothermal & $10\kms$ & MHD \\
         CHEM\_HD & chemistry & variable & HD \\
         CHEM\_MHD$^*$ & chemistry & variable & MHD \\
		\hline
        *(fiducial) \\
	\end{tabular}
    \caption{Summary of the main simulations.}
\label{tab:sims}
\end{table}

Table \ref{tab:sims} shows a summary of the main simulations presented in this paper. In addition to these simulations, we have run various tests in which we varied parameters such as the resolution, the initial magnetic field, or where we cut out the CMZ to isolate it from interaction with the large-scale environment. These additional simulations are introduced and discussed when appropriate throughout the paper.

\begin{figure*}
	\includegraphics[width=0.98\textwidth]{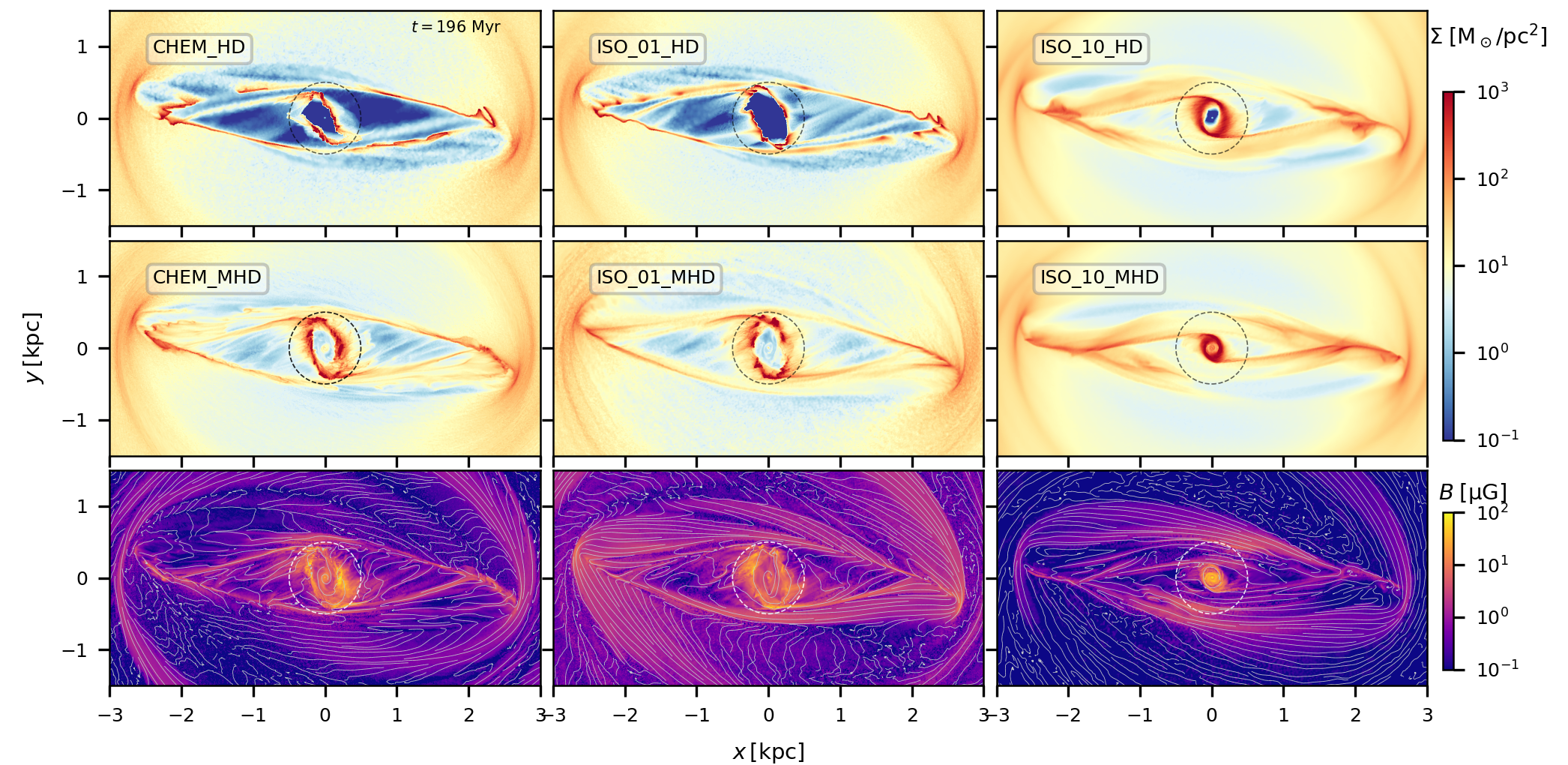}
	\caption{State of the system at $t=196$~Myr for our set of simulations. \emph{Top and middle}: face-on gas surface density for all the simulations listed in Table~\ref{tab:sims}. \emph{Bottom}: projected mass-weighted magnetic field $|\langle \bfB \rangle_z| = (\langle B_x \rangle_z^2 + \langle B_y \rangle_z^2 + \langle B_z \rangle_z^2)^{1/2}$ for the three magnetised simulations in Table~\ref{tab:sims}, where $\langle X \rangle_z = (\int \rho X \, \di z)/(\int \rho\, \di z)$ denotes the vertically integrated mass-weighted density of a quantity $X$. Contours denote the streamlines of the projected field $\langle B_x \rangle_z \hatex + \langle B_y \rangle_z \hatey$. The dashed circles indicate the region with higher numerical resolution at $R<500 \pc$ (Sect.~\ref{sec:methods:summary}). Rotation is clockwise. }
    \label{fig:Slices_large}
\end{figure*}

\begin{figure*}
	\includegraphics[width=0.98\textwidth]{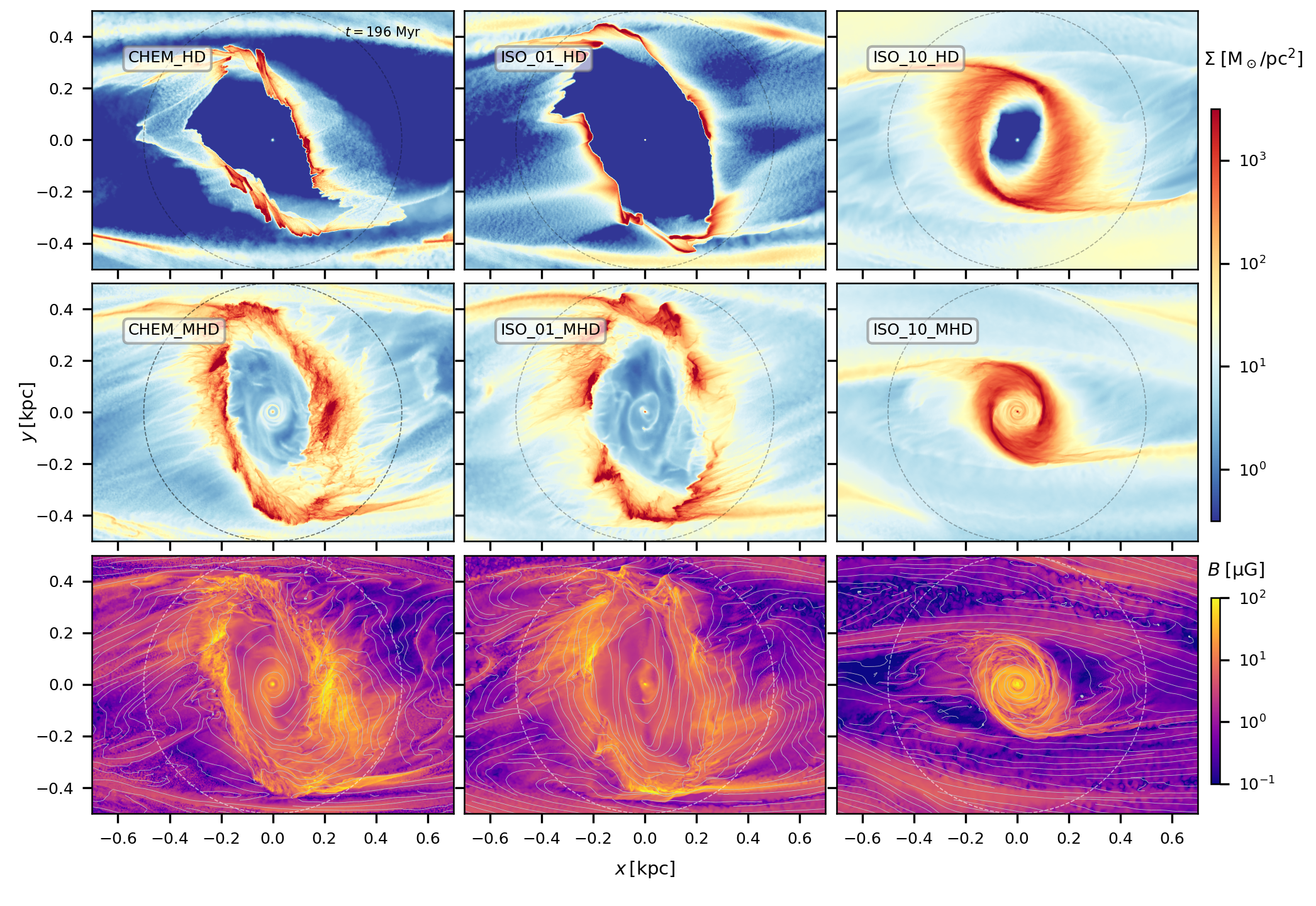}
        \caption{Same as Fig.~\ref{fig:Slices_large}, but zooming-in onto the Central Molecular Zone. }
    \label{fig:Slices}
\end{figure*}

The resolution of our simulations is specified by the target mass of each computational cell. The target mass for all the simulations listed in Table \ref{tab:sims} is $100 \Msun$ for $R>500\pc$ and $10 \Msun$ for $R<500\pc$. The resolution is therefore higher in the CMZ region than in the Galactic disc. The system of mass refinement in {\sc arepo} splits cells whose mass becomes greater than twice the target mass and merges cells whose mass becomes lower than half the target mass. Because this keeps the mass of the cells approximately constant, our spatial resolution varies as a function of the local gas density. We also implemented a minimum cell volume to prevent excessive refinement and computational slowdown in areas of high density: cells with an effective cell radius less than $r_{\rm cell}=[3 V_{\rm cell}/(4\pi)]^{1/3}=0.1 \pc$, where $V_{\rm cell}$ is the cell volume, were not permitted to divide into smaller cells. The typical number of cells in our simulations is around 25 million, of which approximately 10 million are located in the higher-resolution region at $R<500\pc$. Figure~\ref{fig:Resolution} shows the resolution as a function of density for our fiducial model CHEM\_MHD. 

\section{Gas morphology} \label{sec:gasmorphology}

\begin{figure*}
	\includegraphics[width=\textwidth]{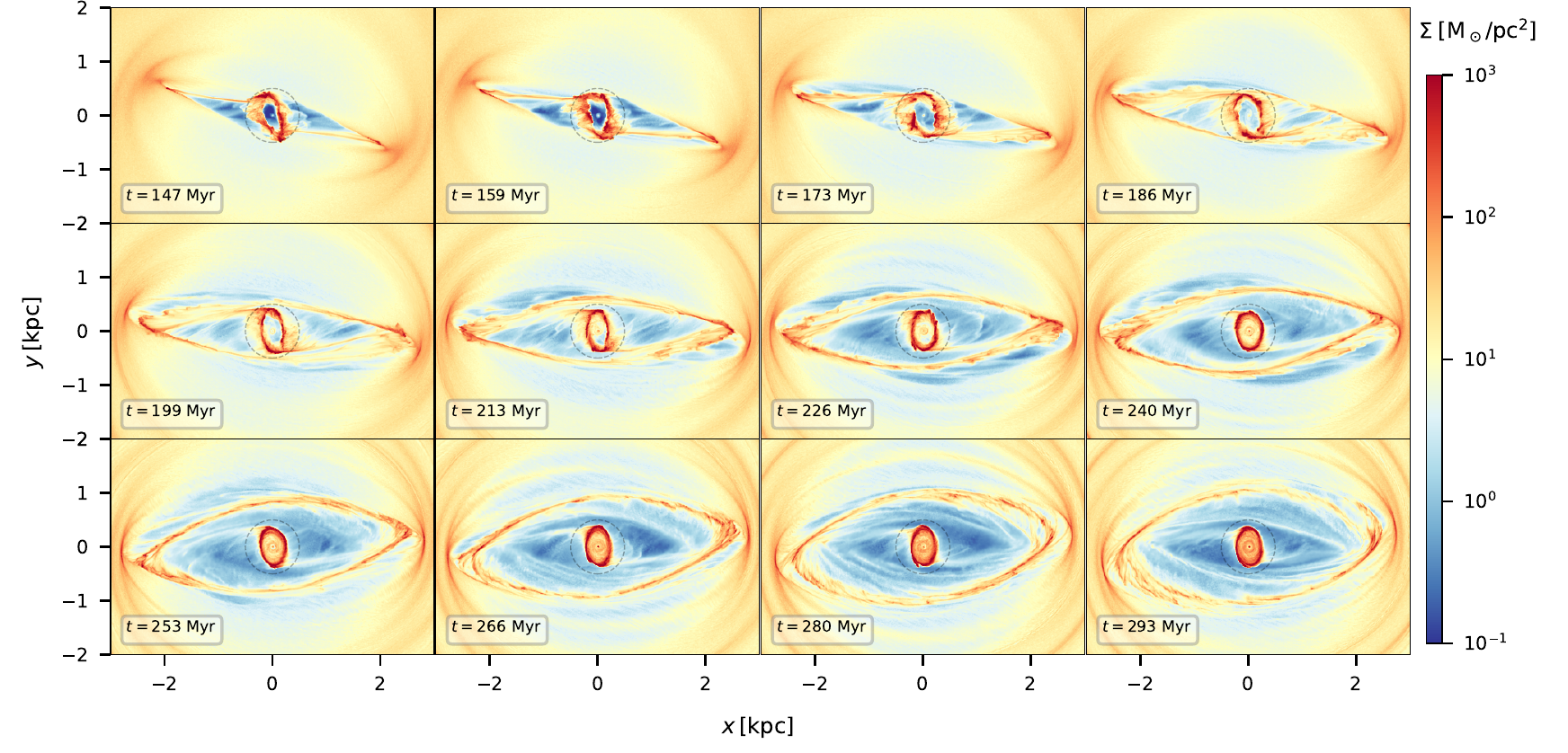}
        \caption{Time evolution of the surface gas density in the CHEM\_MHD simulation. Rotation is clockwise.} 
    \label{fig:rhoproj_vs_time}
\end{figure*}

\begin{figure*}
	\includegraphics[width=\textwidth]{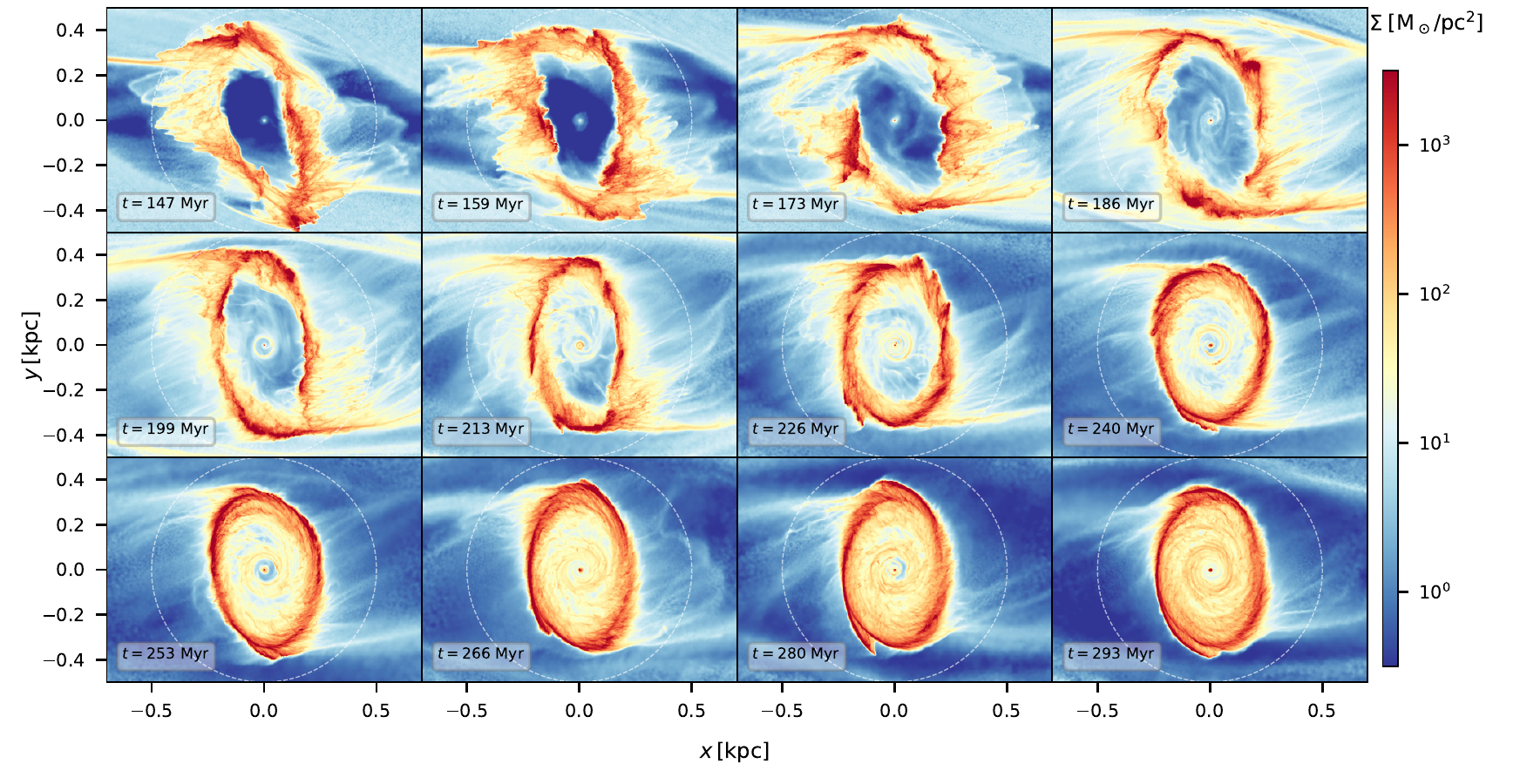}
        \caption{Time evolution of the gas density in the central regions of the CHEM\_MHD simulations. This is the same as Fig.~\ref{fig:rhoproj_vs_time}, but zooming-in in the central region. The interior of the CMZ gas ring is empty at $t=147\Myr$ and then is gradually filled with gas due to MRI-driven accretion (Sect.~\ref{sec:inflow}).} 
    \label{fig:rhoslice_vs_time}
\end{figure*}

Figures~\ref{fig:Slices_large} and \ref{fig:Slices} show the face-on gas surface density and magnetic fields for the six simulations listed in Table~\ref{tab:sims} as well as the magnetic field for the three magnetised simulations, while Figs.~\ref{fig:rhoproj_vs_time} and \ref{fig:rhoslice_vs_time} show the time evolution of our fiducial CHEM\_MHD simulation. The gas morphology and general flow pattern in all simulations have the typical characteristics of gas flow in barred potentials, such as the presence of large-scale shocks on the leading side of the bar that transport the gas towards the centre (also known as `bar dust lanes', e.g. \citealt{Athan92b}), and a central ring-like accumulation of gas, that in the Milky Way corresponds to the CMZ. The general characteristics of this flow have been extensively discussed in numerous papers to which we refer for more in-depth discussions \citep[e.g.][]{Athan92b,Sellwood1993,Fux1999,Kim2012a,Sormani2015a,Sormani2018,Tress2020}. Here, we focus only on the differences that appear when a magnetic field is introduced.

The first difference is that the magnetic fields tend to decrease the radius of the CMZ ring-like structure. This is noticeable if we compare the ISO\_10\_HD to the ISO\_10\_MHD simulations in the right column of Fig.~\ref{fig:Slices_large}. The CMZ in the magnetised simulation is slightly smaller than in the non-magnetised one. The explanation is likely the following. It is well-known that the radius of the nuclear ring in simulations is strongly dependent on the sound speed \citep[e.g.][]{Englmaier1997,Patsis2000,Li2015,Sormani2015a}. \cite{Sormani2024} argued that this dependence can be explained in terms of density waves excited by the bar potential. These density waves remove angular momentum, clear out a region around the inner Lindblad resonance, and transport the gas inwards where it accumulates into a ring. When the sound speed is larger, density waves are stronger and can be excited over a more extended region, and transport the gas into a ring of smaller radius. Magnetic fields increase the effective sound speed of the gas by exerting magnetic pressure, and therefore produce smaller rings. The amount by which the effective sound speed is increased by magnetic fields can be roughly estimated by adding in quadrature the Alfv\'en velocity defined as 
\begin{equation} \label{eq:valfven}
    \bfv_A=\frac{\bfB}{(4 \pi \rho)^{1/2}}.
\end{equation} 
In our simulations, the Alfv\'en velocity in the dense gas in the CMZ ring is typically of the order of $|\bfv_A|\simeq 5\kms$ (Sect.~\ref{sec:Bstrength}), and indeed the effect seen in Fig.~\ref{fig:Slices_large} is comparable to the effect seen in isothermal unmagnetised simulations when the sound speed is increased by roughly this amount in quadrature \citep{Sormani2024}.

\begin{figure}
	\includegraphics[width=\columnwidth]{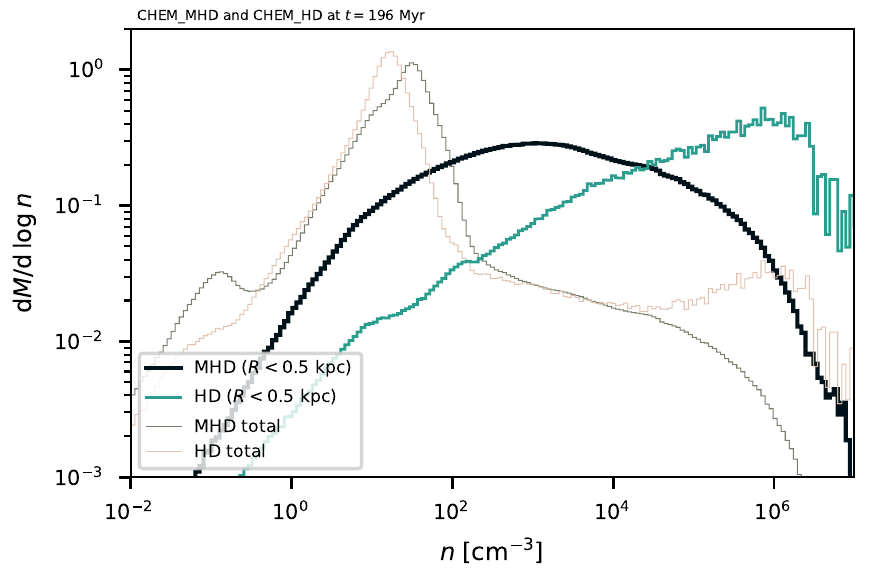}
        \caption{Mass-weighted probability density function (PDF) of the CHEM\_MHD and CHEM\_HD simulations (Table~\ref{tab:sims}) at $t=196\Myr$. The thick lines are for the CMZ ($R<500\pc$), while the thin lines are for the entire simulation. Each curve is normalised so that $\int (\di M / \di \log n) \, \di \log n = 1$. In the CHEM\_HD simulation, most of the gas in the ring accumulates at the highest densities purely due to orbital convergence. This is not due to the gas self-gravity, since it is switched off in our simulations. The presence of magnetic fields shifts the peak to lower densities by providing pressure support and driving random motions. The total distribution is bimodal, as expected from a two-phase medium \citep[e.g.][]{Seta2022}.}
    \label{fig:DensityPDF}
\end{figure}

A second difference is the probability density function (PDF). It is well known that magnetic fields can affect the density PDF \citep[e.g.][]{Federrath2013}. Figure~\ref{fig:DensityPDF} shows that in the unmagnetised simulations most of the gas mass in the CMZ (thick blue line) lies at the highest densities ($n \gtrsim 10^{6} \, \cm^{-3}$), because all the gas in the dense ring tends to occupy the same orbit and there is only the thermal pressure preventing further compression. In the magnetised simulation instead, the mass PDF has a peak at a density of $n \sim 10^{3} \, \cm^{-3}$. This is because the magnetic fields provide pressure support and also drive turbulence, which increases the random motions of the gas and prevents it from accumulating at too high densities \citep[see also][]{Molina2012}. Indeed, Fig.~\ref{fig:Slices} shows that the CMZ ring in the CHEM\_HD simulation is very thin and dense\footnote{We note that the accumulation of gas at very high density is not due to self-gravity here since this is switched off in our simulations. The confinement to a thin ring is entirely due to the dynamics in the bar potential.}, while in the CHEM\_MHD simulation it is puffed up by turbulence. Turbulence also puffs up the disc in the vertical direction and increases the disc vertical scale-height, which is known to be directly related to the amount of turbulence in galactic discs \citep[e.g.][]{Ostriker2022}. We will discuss turbulence and the mechanism driving it more in detail in Sect.~\ref{sec:turbulence}.

A third difference occurs in the region inside the dense ring. Figure~\ref{fig:rhoslice_vs_time} shows that the region inside the ring in the CHEM\_MHD simulation is devoid of gas at $t=147\Myr$, and then gradually gets filled with gas. By contrast, in the CHEM\_HD simulation the region inside the ring remains devoid of gas at all times. Figure~\ref{fig:Slices} illustrates this difference in the HD and MHD simulations by comparing snapshots at the same time $t=196$~Myr (compare the CHEM\_HD panel with the CHEM\_MHD panel). The filling up of the ring interior in the magnetised simulation occurs because magnetic fields drive inward accretion from the CMZ towards the central few parsecs. It is interesting to note that supernova feedback can also produce a similar effect of filling up the ring (fig.~9 in \citealt{Tress2020}). Thus, it will be important in the future to understand which effect is stronger, and what is the non-linear interaction between the two. We discuss further the inflows driven by the magnetic field in Sect.~\ref{sec:inflow}.

\begin{figure*}
	\includegraphics[width=\textwidth]{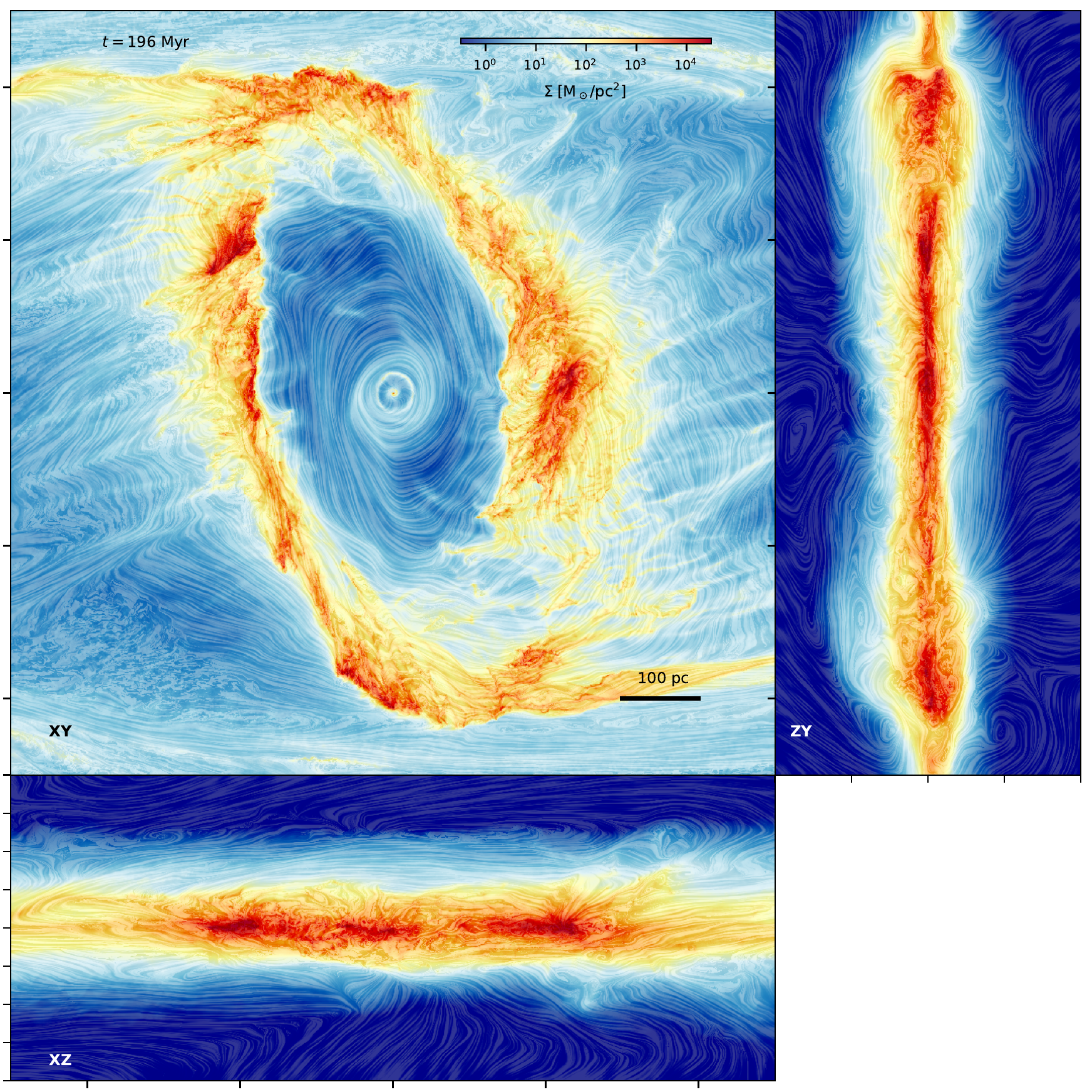}
        \caption{Visualisation of the magnetic field and column density in the CMZ in our fiducial CHEM\_MHD simulation at $t=196 \Myr$. The colour-scale represents the total gas column density. The line pattern indicates the orientation of the mass-weighted integrated magnetic field (i.e. $\langle B_x\rangle_z \hatex + \langle B_y \rangle_z \hatey$ where $\langle{X}\rangle_z = (\int \rho X \, \di z)/(\int \rho\, \di z)$ for the $xy$ panel, and analogous definitions for the $xz$ and $yz$ panels) obtained with the line integral convolution method. }
    \label{fig:MagnetiFieldMorph}
\end{figure*}

\begin{figure*}
	\includegraphics[width=\textwidth]{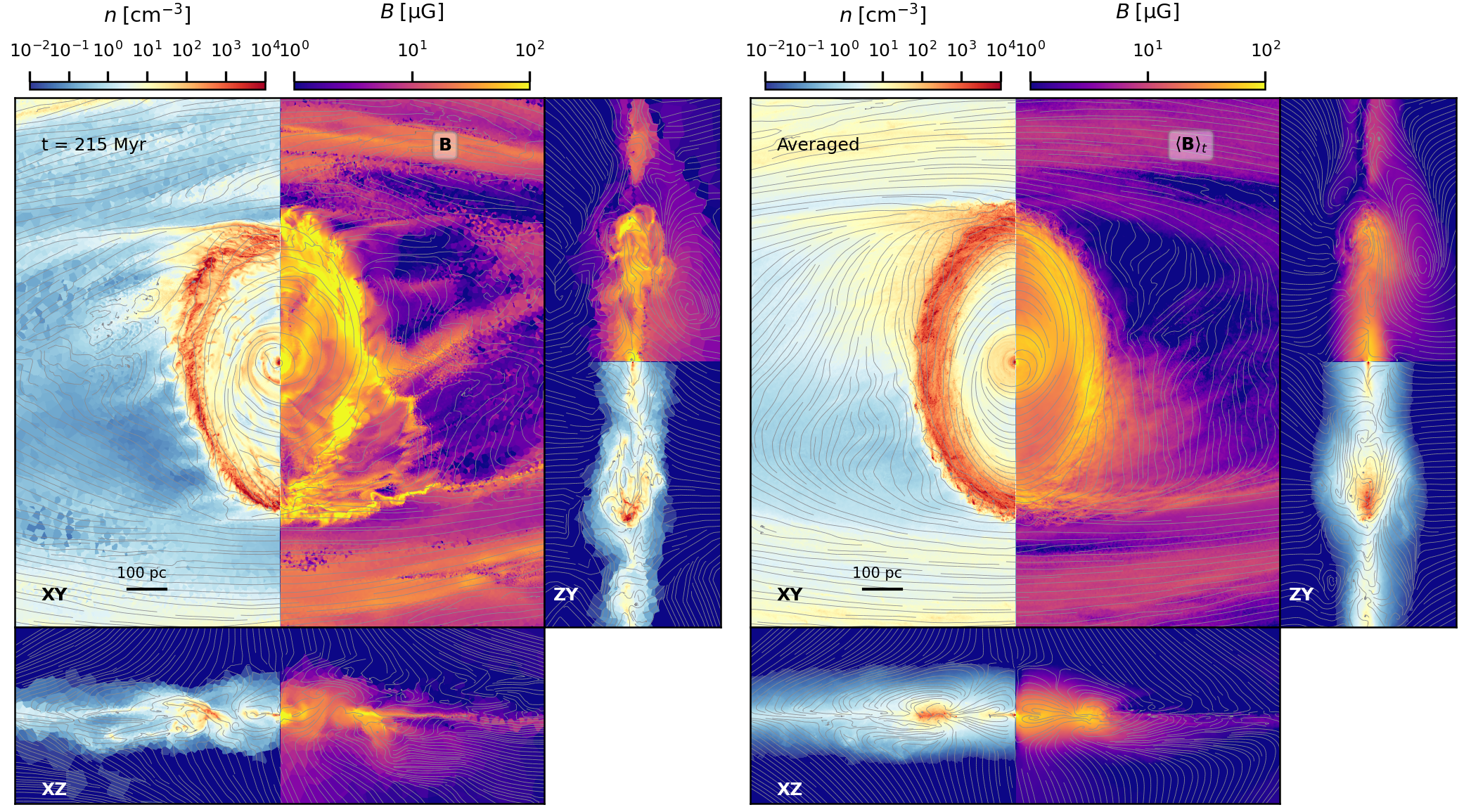}
        \caption{Instantaneous (left) and time-averaged (right, Eq.~\ref{eq:decomposition}) density and magnetic fields. Plotted are slices in the $z=0$, $y=0$, and $x=0$ planes (not integrated quantities). Lines show the magnetic vector field. The time-average is taken between $200-250$~Myr.} 
    \label{fig:BDecomposition}
\end{figure*}

\begin{figure}
    \begin{subfigure}{\columnwidth}
    	\includegraphics[width=\columnwidth]{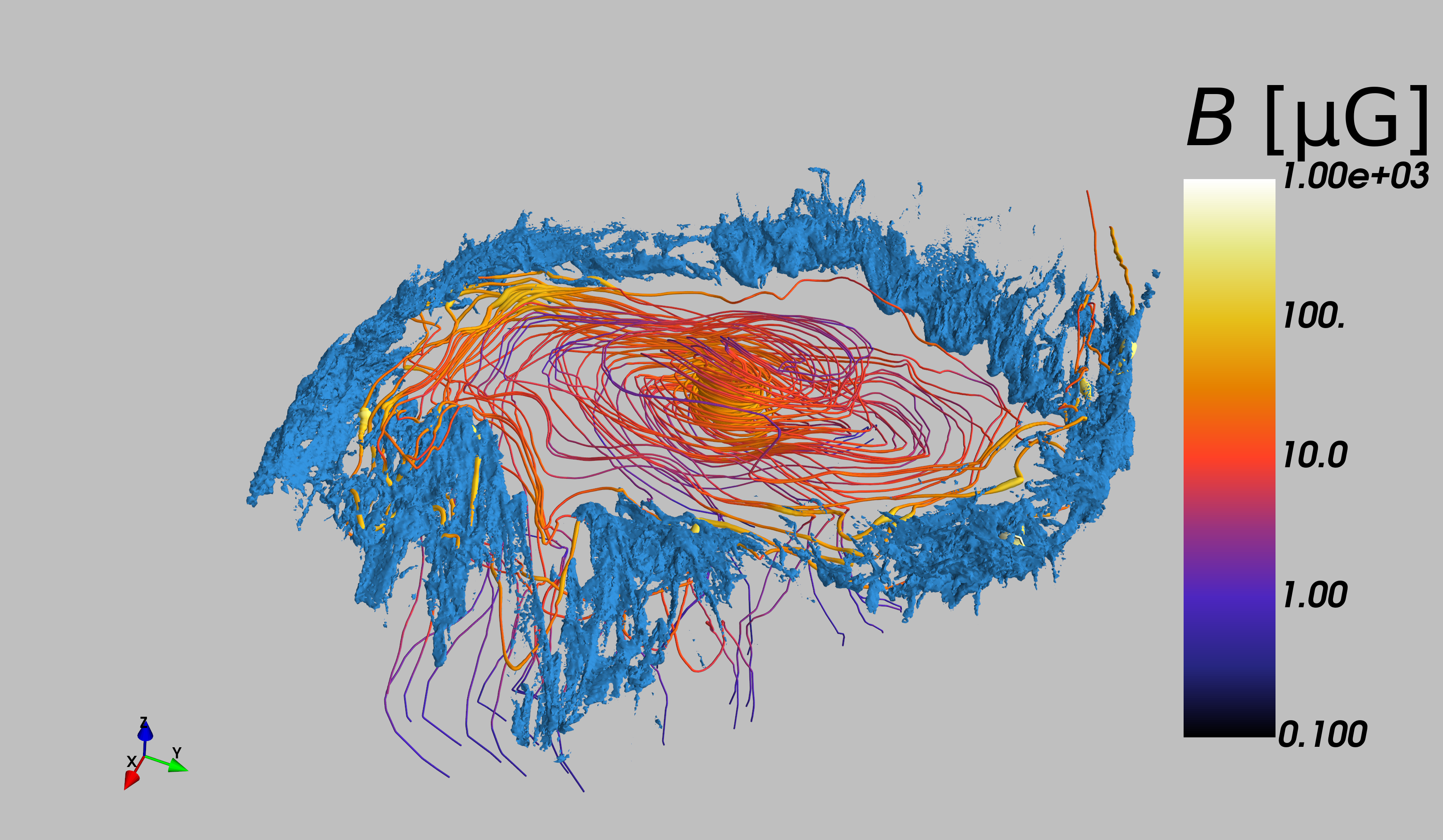}
    \end{subfigure}
    \begin{subfigure}{\columnwidth}
    	\includegraphics[width=\columnwidth]{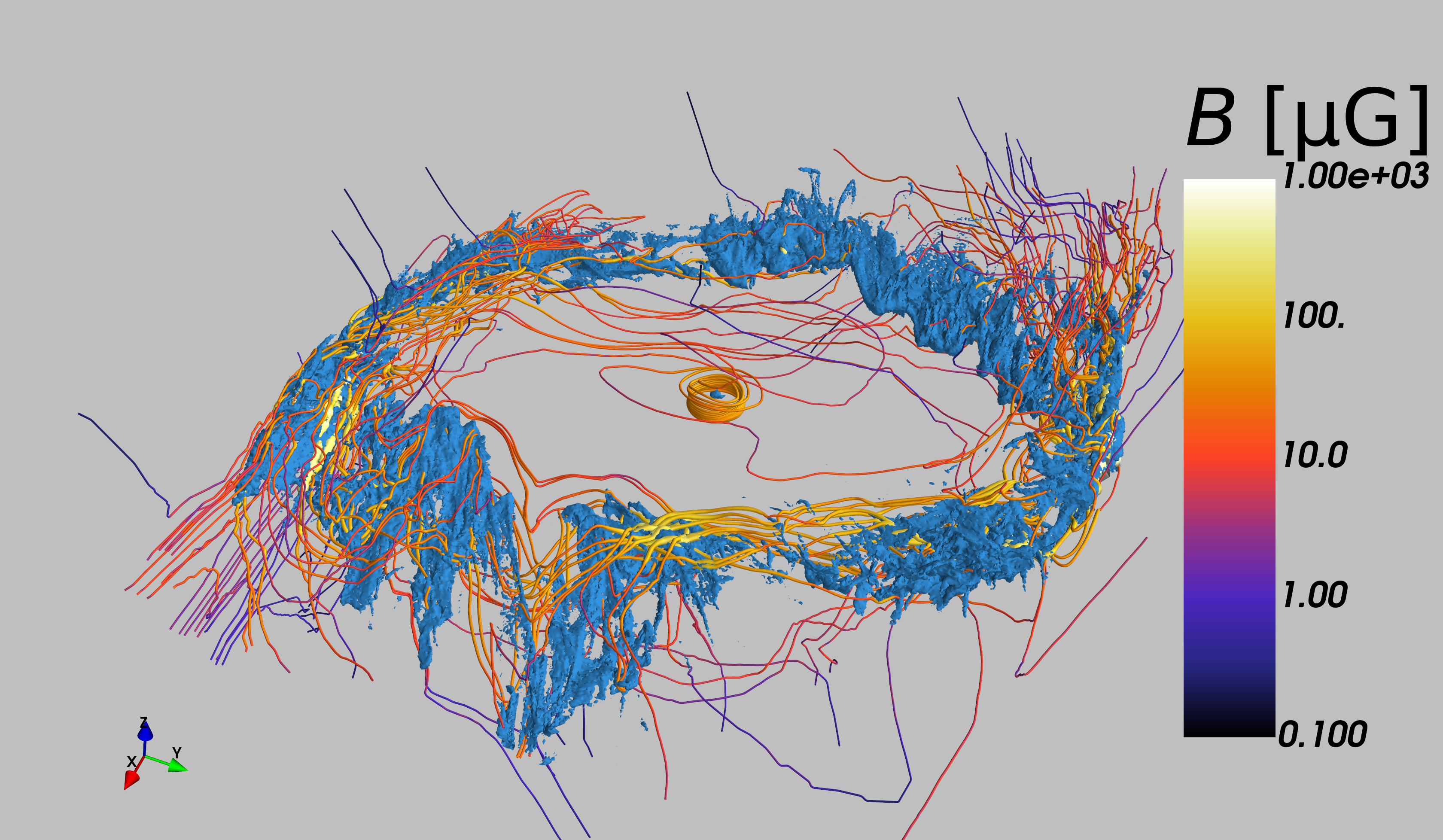}
    \end{subfigure}
    \caption{3D visualisation of the magnetic field lines in our fiducial CHEM\_MHD simulation at $t=176$~Myr. The field lines are constructed by starting at a given set of points, and following the field lines until they close on themselves or leave the domain to go to infinity (field lines cannot `end' within the domain because $\nabla\cdot\bfB=0$, i.e. they behave like velocity streamlines in an incompressible flow). In the top panel we use a set of starting points distributed on a hexagonal prism centred on the Galactic centre and whose faces are located inside the gas ring, roughly midway between the centre and the CMZ gas ring. In the bottom panel we use a set of starting points located near the end of the `bar lanes', just outside the CMZ ring. The blue solid surface represents an isodensity surface at $n=100$~cm$^{-3}$.}
    \label{fig:fieldlines}
\end{figure}

\begin{figure}
    \begin{subfigure}{\columnwidth}
    	\includegraphics[width=\columnwidth]{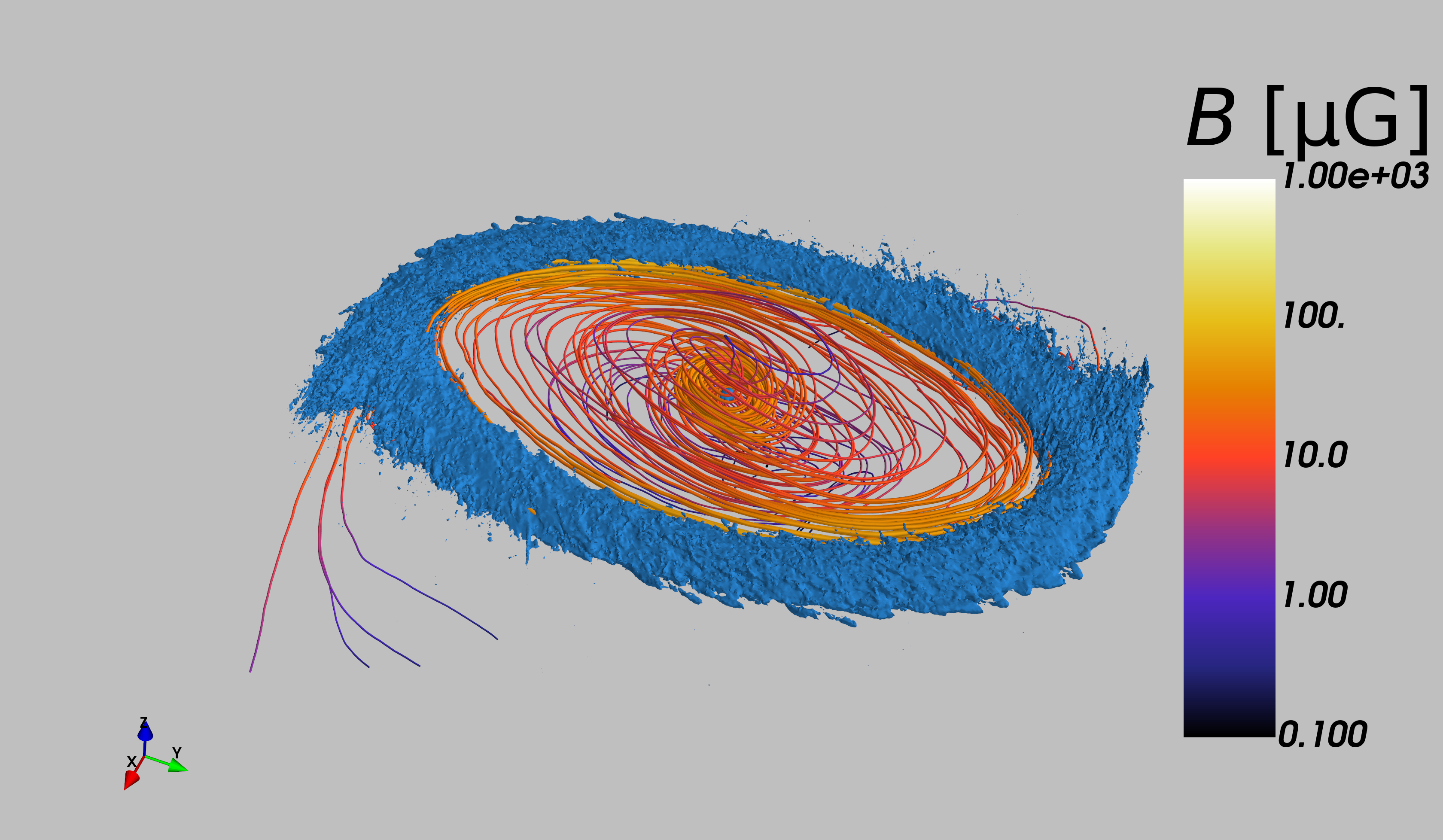}
    \end{subfigure}
    \begin{subfigure}{\columnwidth}
    	\includegraphics[width=\columnwidth]{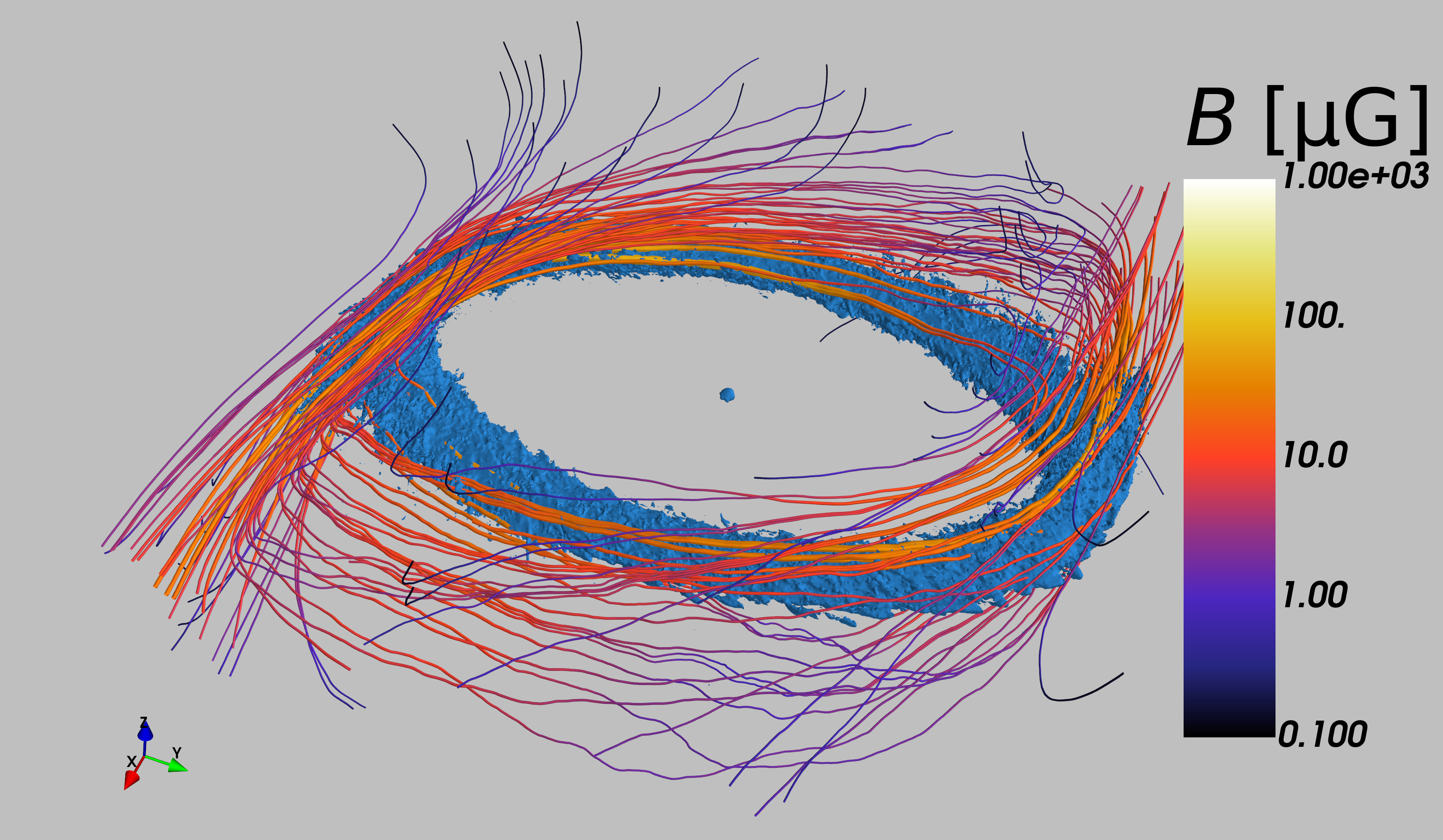}
    \end{subfigure}
    \caption{Same as Fig.~\ref{fig:fieldlines}, but for the time-averaged magnetic and density fields shown in Fig.~\ref{fig:BDecomposition}. This figure clearly illustrates that the streamlines are parallel to the bar dust lanes, and the regularity of the time-averaged field.}
    \label{fig:fieldlines_average}
\end{figure}

\begin{figure}
	\includegraphics[width=\columnwidth]{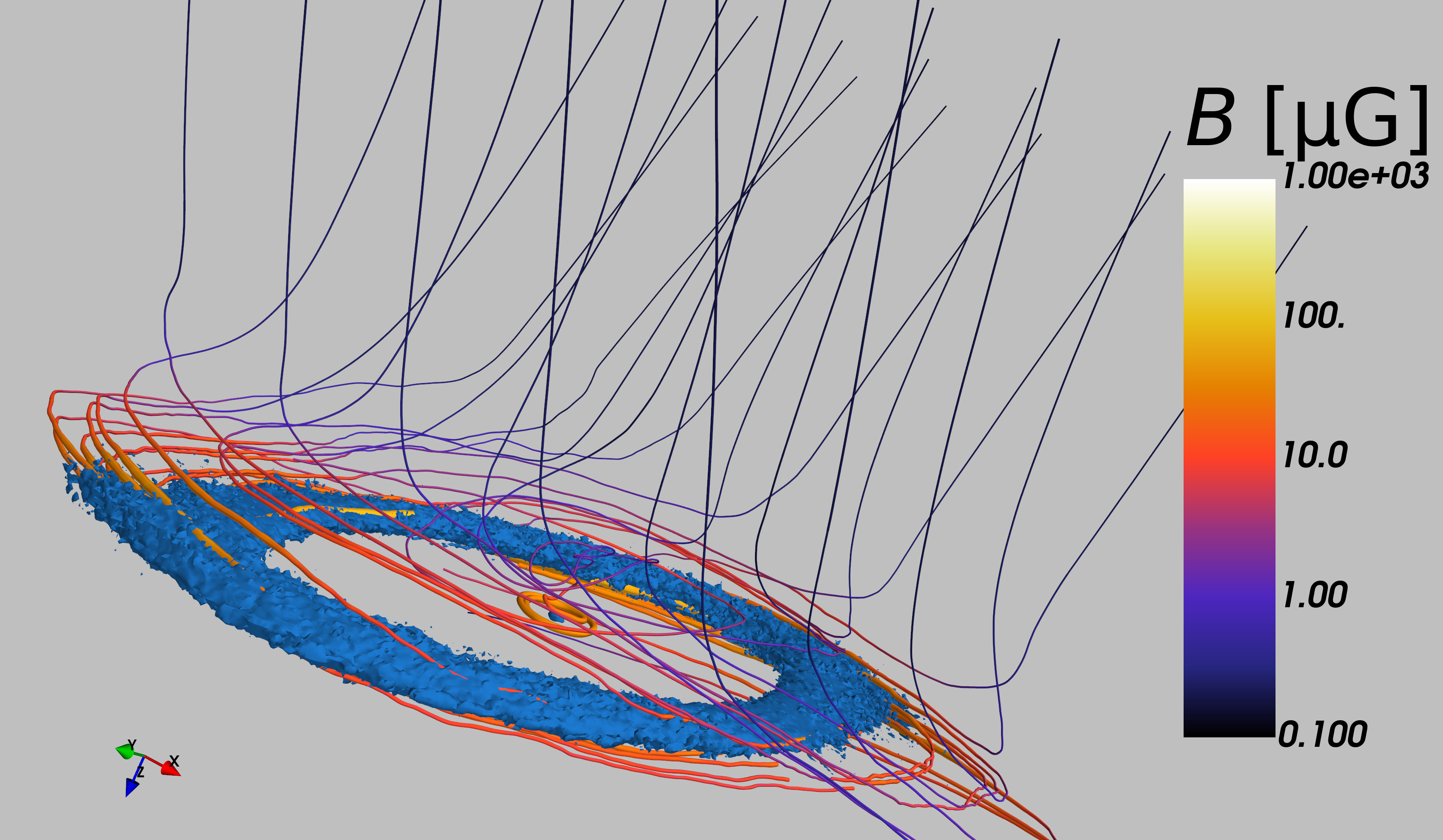}
        \caption{Similar to Fig.~\ref{fig:fieldlines}, but using the time-averaged magnetic field as in Fig.~\ref{fig:BDecomposition} and a set of starting points that are located approximately $30 \pc$ above the plane. The streamlines in this spiral up vertically, illustrating the transition from toroidal to poloidal magnetic field.} 
    \label{fig:fieldlines_average_above_plane}
\end{figure}

\begin{figure}
	\includegraphics[width=\columnwidth]{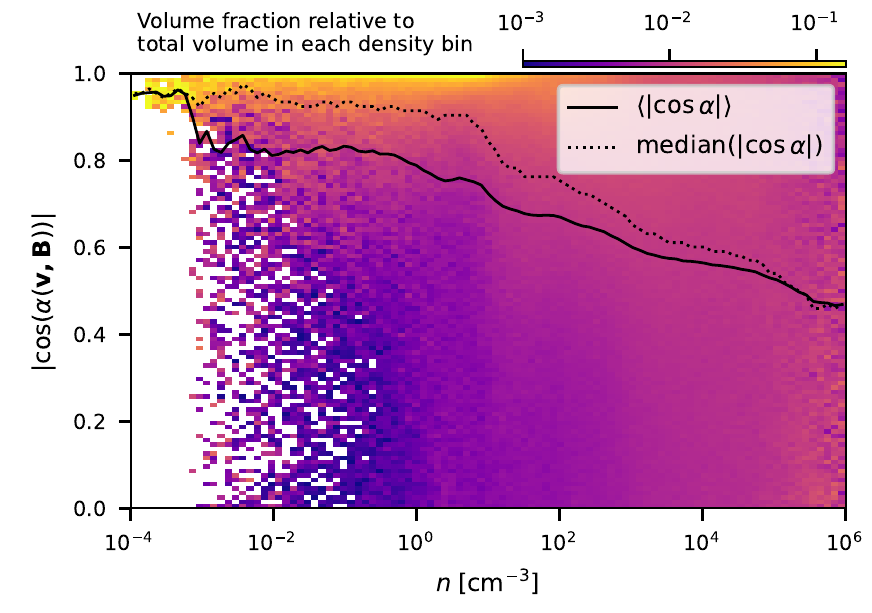}
        \caption{Relative orientation of the gas velocity and magnetic fields as a function of total gas density for our fiducial CHEM\_MHD simulation at $t=186$~Myr. Colours indicate the angle $|\cos \alpha|$ defined by Eq.~\eqref{eq:alpha}. Each slice at fixed $n$ is normalised separately for improved clarity. The black solid line shows the average value of $\cos \alpha$ at the given density, while the dotted one shows the median instead. If the orientation of $\bfv$ with respect to $\bfB$ were completely random (in the solid angle), the distribution would be uniform and the black lines would be horizontal at a value $|\cos\alpha|=0.5$. The plot shows that the velocity and magnetic fields become progressively more random as we move to higher densities.} 
    \label{fig:cosalphaPDF}
\end{figure}

\begin{figure}
	\includegraphics[width=\columnwidth]{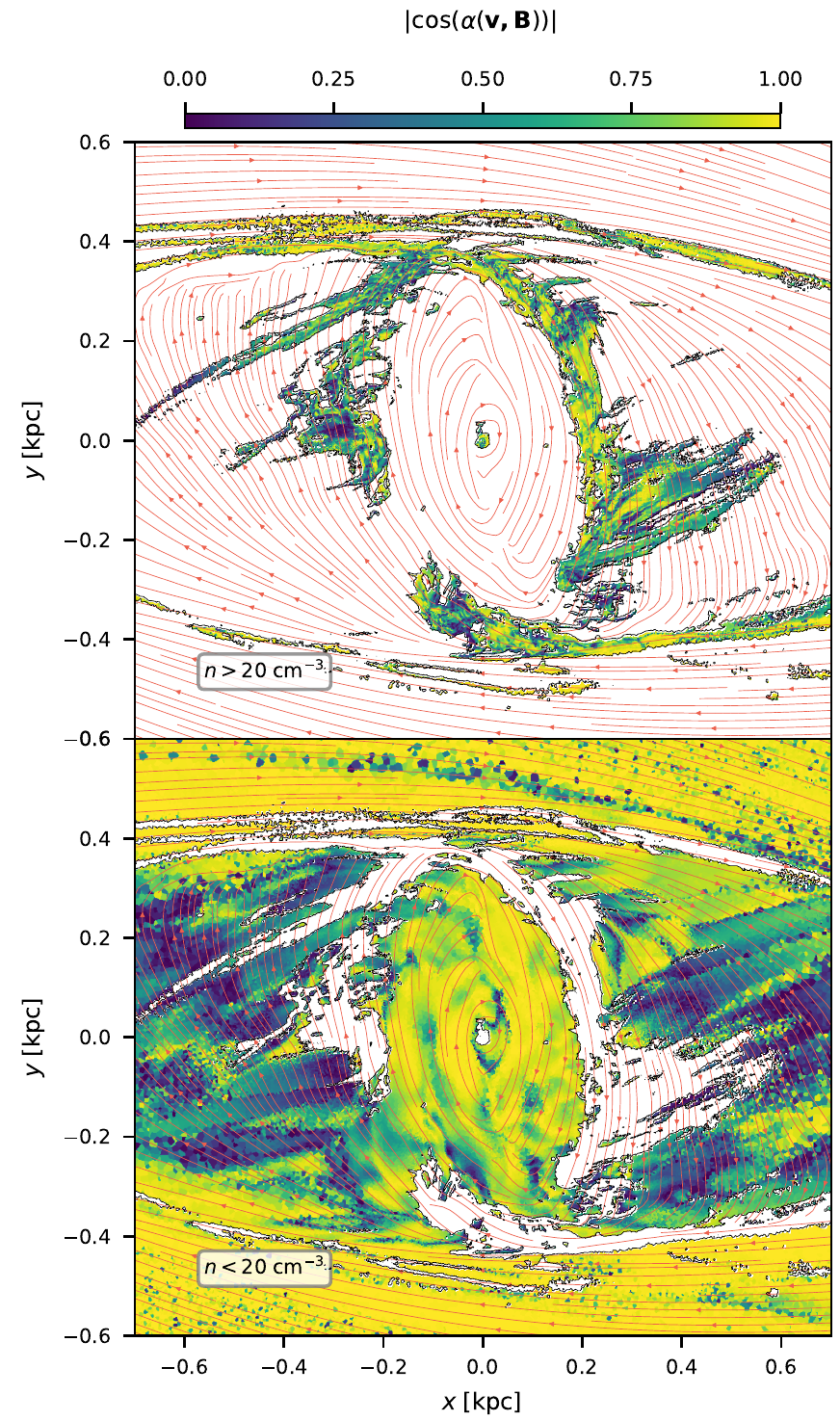}
        \caption{Relative orientation of the gas velocity and magnetic fields in a slice at $z=0$ in our fiducial CHEM\_MHD simulation at $t=186$~Myr. A snapshot at early time is chosen here to show the orientation in the dust lanes as well, which are depleted at later times. The colour-scale indicates $\cos \alpha$ defined in Eq.~\eqref{eq:alpha}, where $\cos\alpha=1$ and $\cos\alpha=0$ correspond to $\bfB$ and $\bfv$ being parallel and perpendicular respectively. Streamlines with red arrows trace gas velocity in the frame co-rotating with the bar. The plot is separated into two panels for densities $n>20 \cm^{-3}$ (top) and $n<20 \cm^{-3}$ (bottom) for clarity. Velocity and magnetic fields are well aligned in the less turbulent regions ($\cos \alpha$ close to 1), while their orientation becomes progressively more random in regions where the turbulence is dynamically more important.} 
    \label{fig:alphaslice}
\end{figure}

\begin{figure}
	\includegraphics[width=\columnwidth]{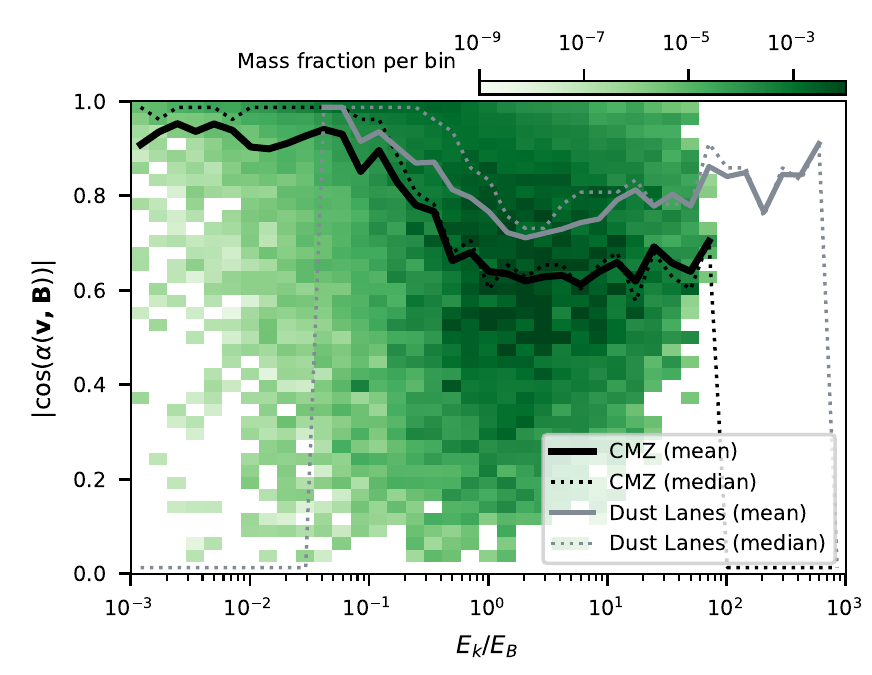}
        \caption{Relative orientation of the gas velocity and magnetic fields as a function of $E_k/E_B$ averaged over $20$~pc cubical bins. The green distribution in the background includes only gas in the CMZ and excludes the gas in the bar lanes. The black lines show the average (solid) and median (dotted) of the distribution. Once the turbulence becomes dynamically important, the relative orientation becomes more random. An exception is given by gas in the dust lanes, for which the average (solid grey line) and median (dotted grey line) are shown.} 
    \label{fig:cosalphavsTurbulence}
\end{figure}

\begin{figure*}
	\includegraphics[width=\textwidth]{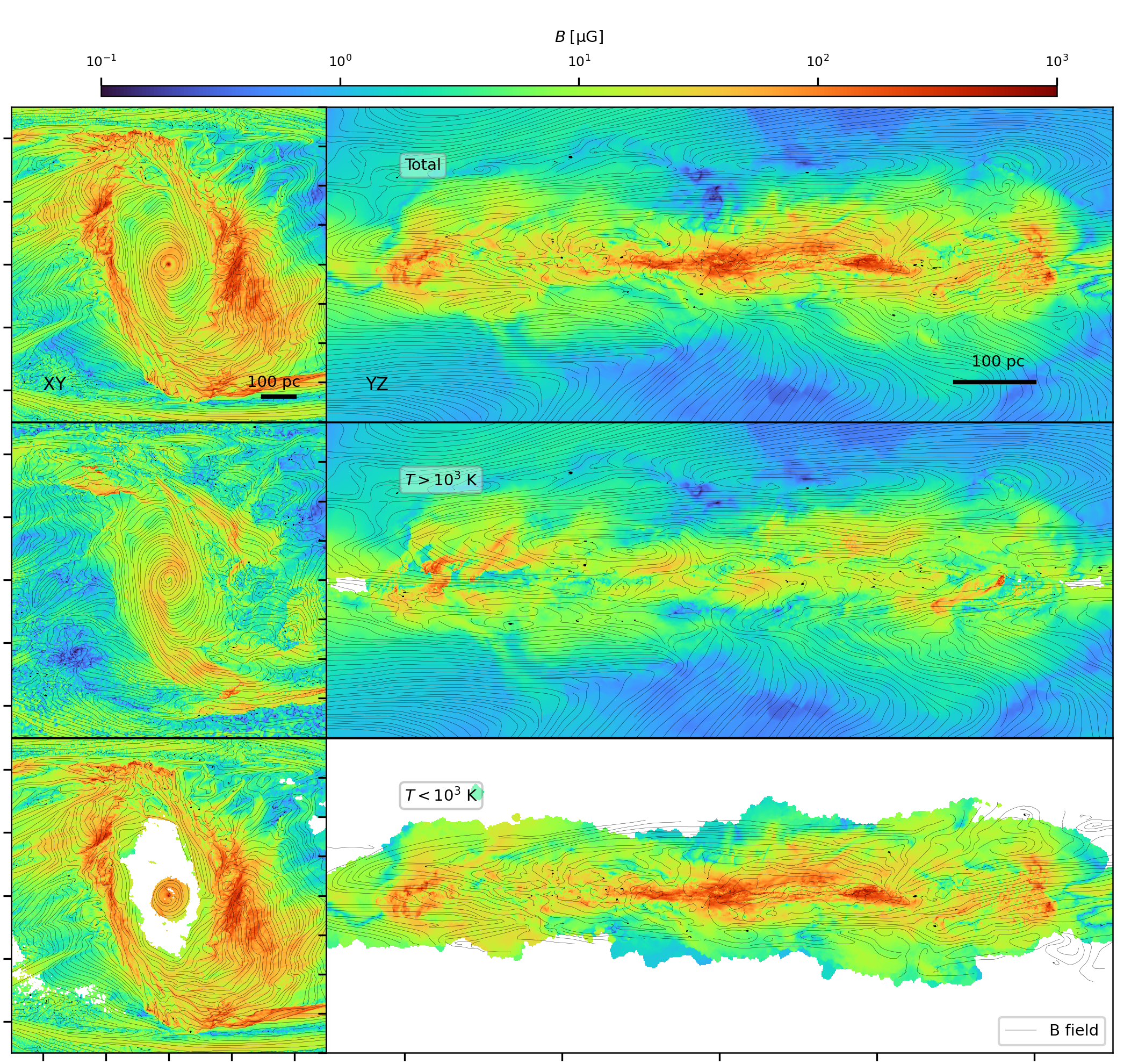}
        \caption{Magnetic field in the different ISM phases. Colours represent magnetic field intensity and lines are magnetic field lines obtained with the integral convolution method. \emph{Top}: all the gas. \emph{Middle}: warm phase ($T>10^3$~K). \emph{Bottom}: cold phase ($T<10^3$~K). The magnetic field in the cold gas is mostly parallel to the plane, while the vertical field above and below the plane belongs to the warm diffuse phase.}
    \label{fig:BDifferentPhases}
\end{figure*}

\begin{figure}
	\includegraphics[width=\columnwidth]{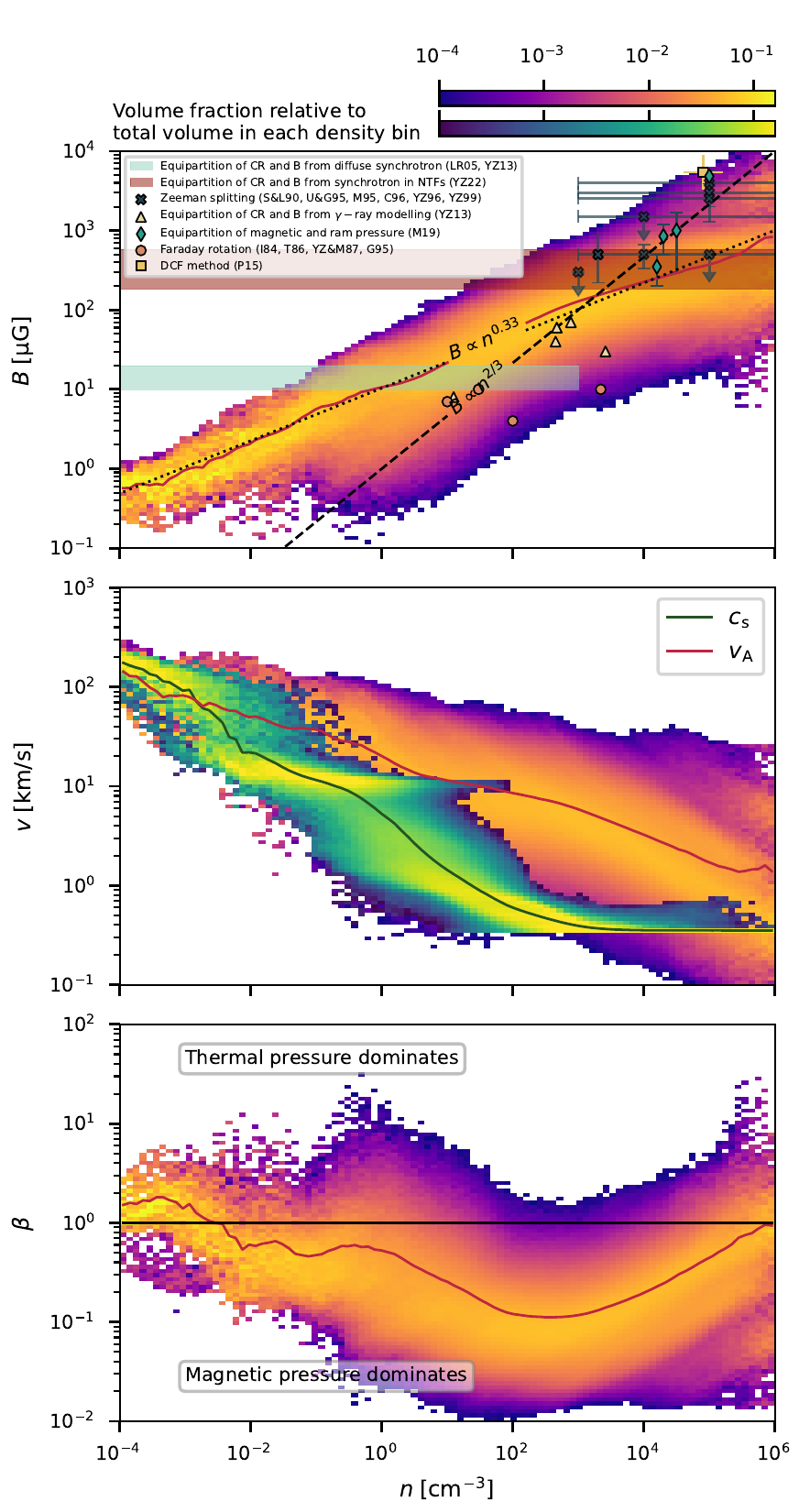}
        \caption[]{Magnetic field properties as a function of density at $R<500$~pc in our fiducial CHEM\_MHD simulation at $t=186$~Myr. \emph{Top}: magnetic field strength. The $B\propto n^{2/3}$ indicates the scaling expected from simple isotropic collapse with flux-freezing \citep{Mestel1965}. The dotted line $B \propto n^{0.33}$ is the fit to the average instead. Symbols denote various estimates of the magnetic field in the CMZ from the literature. References are abbreviated as follows: \citet[][LR05]{LaRosa2005}, \citet[][YZ13]{Yusef-Zadeh2013}, \citet[][YZ22]{Yusef-Zadeh2022}, \citet[][S\&L90]{Schwarz1990}, \citet[][U\&G95]{Uchida1995}, \citet[][M95]{Marshall1995}, \citet[][C96]{Crutcher1996}, \citet[][YZ96]{Yusef-Zadeh1996}, \citet[][YZ99]{Yusef-Zadeh1999}, \citet[][M19]{Mangilli2019}, \citet[][I84]{Inoue1984}, \citet[][T86]{Tsuboi1986}, \citet[][YZ\&M87]{Yusef-Zadeh1987}, \citet[][G95]{Gray1995}, \citet[][P15]{Pillai2015}. \emph{Middle}: Alfv\'en velocity (Eq.~\ref{eq:valfven}) and thermal sound speed $\cs=\gamma P/\rho$ where $P$ is the thermal pressure (Eq.~\ref{eq:eos}). \emph{Bottom}: plasma $\beta=P/P_{\rm mag}$, where $P_{\rm mag}=B^2/(8\pi)$ is the magnetic pressure. In all panels, each density slice is normalised separately for clarity of visualisation, and the solid lines show the average value at the given density.}
    \label{fig:BvsRho}
\end{figure}

\section{Magnetic fields in the bar and CMZ regions} \label{sec:BfieldsCMZbar}

A general impression of the magnetic field geometry and strength in our simulations can be obtained from the bottom rows in Figs.~\ref{fig:Slices_large} and \ref{fig:Slices}. An alternative visualisation of the magnetic field in the CMZ for our fiducial CHEM\_MHD simulation is shown in Fig.~\ref{fig:MagnetiFieldMorph}. When we consider the magnetic field properties in more detail, we find the following general characteristics, which are explored in the dedicated subsections below:
\begin{enumerate}
    \item The magnetic field can be understood as the sum of a `regular' time-averaged component and a fluctuating `turbulent' component (Sect.~\ref{sec:decomposition}).
    \item The magnetic field is generally aligned with the gas velocity vectors, and the magnetic field geometry changes from toroidal near the $z=0$ plane to poloidal at $|z|>0$ (Sect.~\ref{sec:geometry}).
    \item The magnetic field strength scales as a function of gas density roughly as $B\propto n^{0.33}$(Sect~\ref{sec:Bstrength}).
\end{enumerate}

\subsection{Decomposition into regular and turbulent components} \label{sec:decomposition}

It is instructive to decompose the magnetic field as
\begin{equation} \label{eq:decomposition}
    \bfB(R,\phi,z,t) = \langle \bfB \rangle_t (R,\phi,z) + \Delta \bfB(R,\phi,z,t)\,
\end{equation}
where $\langle \bfB \rangle_t$ is a time-averaged `regular' component, $\Delta \bfB$ is an irregular instantaneous `turbulent' fluctuation, and the time average of a quantity $X$ over an interval $\Delta t$ is defined as:
\begin{equation}
    \langle X\rangle_t = \frac{\int_t^{t+\Delta t} X(R,\phi,z,t)\, \di t}{\Delta t}\,.
\end{equation}
We note that $\langle \Delta B_i\rangle_t=0$ and $\langle B_i^2 \rangle_t = \langle B_i \rangle_t^2 + \langle \Delta B_i^2 \rangle_t$ by definition. Moreover, although the flow reaches an approximate steady-state at $t>150\Myr$, some slow changes in the global gas and magnetic field configuration continue to happen as a consequence of the continuous gas inflow towards the centre. However, these changes are slow enough (typical timescale $\sim$100~Myr) that the decomposition into time-averaged and instantaneous components proves to be useful over timescales of tens of megayears (corresponding to a few rotations in the gas ring, where the orbital period is $\sim$10~Myr). Here we considered time averages over $t=200\mhyphen250\Myr$. The conclusions of this subsection are not significantly affected by this choice.

Figure~\ref{fig:BDecomposition} compares the instantaneous field $\bfB$ with the time-averaged $\langle \bfB \rangle_t$ for our fiducial CHEM\_MHD simulation. The time-averaged magnetic field has a very regular structure which resembles the velocity streamlines of gas flowing in a bar potential. Albeit regular, the time-averaged field is far from simple, and exhibits a complex `butterfly' morphology in the $xz$ and $yz$ planes, which we discuss in more detail in Sect.~\ref{sec:geometry}. The difference $\Delta \bfB = \bfB - \langle \bfB \rangle_t$ is larger where the gas is more turbulent, for example in the nuclear ring, indicating that the turbulence causes magnetic field fluctuations.

We quantified the strength of the regular and irregular magnetic fields using the following mass-weighted and volume-weighted averages:
\begin{align}
 B_{{\rm tot},\rho} & = \left< \frac{\int_V  |\bfB| \rho  \, \di V}{\int_V  \rho  \, \di V} \right> \label{eq:Btot} \\
 B_{{\rm reg}, \rho} & = \frac{\int_V  |\langle \bfB\rangle_t |\langle \rho \rangle_t \, \di V}{\int_V \langle \rho \rangle_t \, \di V} \,\\
 B_{{\rm trb},\rho} & = \left< \frac{\int_V  |\Delta\bfB| \rho  \, \di V}{\int_V  \rho  \, \di V} \right>
\end{align}
and
\begin{align}
 B_{{\rm tot},V} & = \left< \frac{\int_V  |\bfB|  \, \di V}{\int_V   \, \di V} \right> \label{eq:Bvtot} \\
 B_{{\rm reg}, V} & = \frac{\int_V  |\langle \bfB\rangle_t | \, \di V}{\int_V  \, \di V} \,\\
 B_{{\rm trb},V} & = \left< \frac{\int_V  |\Delta\bfB|  \, \di V}{\int_V    \, \di V} \right>
\end{align}
where the integrals are carried out over a volume $V$. Taking $V$ as the region where $R<500\pc$ and the time-averaged density is $\langle n \rangle_t > 10^2$~cm$^{-3}$, we find $B_{{\rm tot},\rho} \simeq 305$~$\mu$G, $B_{{\rm reg},\rho} \simeq 34$~$\mu$G, $B_{{\rm trb},\rho} \simeq 298$~$\mu$G, and  $B_{{\rm tot},V} \simeq 33$~$\mu$G, $B_{{\rm reg},V} \simeq 16$~$\mu$G, $B_{{\rm trb},V} \simeq 29$~$\mu$G. The mass-weighted turbulent component is larger than the volume-weighted turbulent component because denser gas (where most of the mass is) is more turbulent than the diffuse gas on average.

Figures~\ref{fig:fieldlines}, \ref{fig:fieldlines_average}, and \ref{fig:fieldlines_average_above_plane} offer a 3D visualisation of the instantaneous and time-averaged magnetic field, confirming the complex nature of the instantaneous field and the regular structure of the time-average component discussed above.

\subsection{Magnetic field geometry and relation with the velocity field} \label{sec:geometry}

Figures~\ref{fig:MagnetiFieldMorph} and~\ref{fig:BDecomposition} suggest that the magnetic field in the $z=0$ plane is mostly parallel to the plane, and tends to be oriented parallel to the velocity vectors of the gas flowing in the barred potential. 

We explore further the alignment between velocity and magnetic fields using their relative angle, defined as
\begin{equation} \label{eq:alpha}
    \cos\alpha = \frac{\bfv \cdot \bfB}{|\bfv| |\bfB|} \,.
\end{equation}
Figure~\ref{fig:cosalphaPDF} shows the distribution of $|\cos\alpha|$ as a function of density, where each slice at a given density was normalised separately for clarity. When $\bfB$ and $\bfv$ are perfectly aligned, $|\cos\alpha|=1$. When the orientation between $\bfB$ and $\bfv$ is completely random (i.e. $\bfB$ is uniformly distributed on the surface of a sphere around $\bfv$), the distribution is uniform in $|\cos\alpha|$ in the interval [0,1], with an average value of $|\cos\alpha|=0.5$. The figure therefore shows that the orientation becomes progressively more random as the gas gets denser. The distribution is shown for a single snapshot, but is qualitatively similar throughout the simulation. 

Figure~\ref{fig:alphaslice} plots an $xy$ map of $|\cos\alpha|$ for a slice in the plane $z=0$. This shows that in regions of comparable densities, the alignment becomes more random where there is more turbulence. For example, the bottom panel shows significant disalignment in the intra-lane region outside the CMZ ring where gas is more turbulent, because no closed orbits exist for the gas to flow on, than in regions where the gas can smoothly flow on $x_1/x_2$ orbits (fig.~5 in \citealt{Sormani2015a}). This suggests that it is the turbulence that tends to disalign the fields. 

This point is supported by Fig.~\ref{fig:cosalphavsTurbulence}, where we plotted the distribution of $|\cos\alpha|$ as a function of the ratio of the kinetic energy in turbulent motions to the magnetic energy $E_K/E_B$, which is a measure of the dynamical importance of the turbulence, in $20$~pc bins (Sect.~\ref{sec:turbulence}). In the CMZ, the magnetic field is aligned with the velocity in regions where $E_B > E_K$, but the distribution becomes progressively more random as the turbulence becomes more dynamically important. This is similar to the finding of \cite{Iffrig2017}, with the difference that in their case the turbulence was driven by supernova feedback, while in our case it is driven by the magnetorotational instability (Sect.~\ref{sec:turbulence}). The dust lanes behave differently than the CMZ in this respect and were excluded form the distribution shown in Fig.~\ref{fig:cosalphavsTurbulence}. In the dust lanes, the magnetic field stays aligned even in regions with extremely high $E_K/E_B$ ratios. This is likely because these regions are large-scale galactic shocks, and thus have an exceptionally high shear and sharp velocity discontinuities.

Since it is the turbulence that `disaligns' the velocity and magnetic fields, we would expect the regular time-averaged $\langle \bfB \rangle_t$, in which turbulent fluctuations are averaged out, to follow even more closely the velocity streamlines. Indeed, Figs.~\ref{fig:BDecomposition} and \ref{fig:fieldlines_average} confirm this. In particular, the regular magnetic field in the CMZ ring is nearly parallel to the ring itself (Figs.~\ref{fig:BDecomposition} and \ref{fig:fieldlines_average}), and the magnetic field in the `bar lanes' is roughly parallel to the lanes themselves (Figs.~\ref{fig:Slices_large}, \ref{fig:BDecomposition}, and \ref{fig:alphaslice}), and therefore to the velocity, since the latter is approximately parallel to the lanes in the frame co-rotating with the bar.

The $xz$ and $zy$ slices of the time-averaged regular field in Fig.~\ref{fig:BDecomposition} illustrate an interesting characteristic of the magnetic field geometry: the field transitions from mostly toroidal (i.e. along $\hatephi$) near the $z=0$ plane to mostly poloidal (i.e. along $\hateR$ and $\hatez$) at $|z|>0$. The transition happens through a complex `butterfly' pattern, in which the field wraps around the dense gas in the ring (xz projection of the time-averaged field in Fig.~\ref{fig:BDecomposition}). The transition can also be appreciated in the 3D visualisation of Fig.~\ref{fig:fieldlines_average_above_plane}, which follows field lines as they change from toroidal just above the midplane to nearly vertical away from the plane. Figure~\ref{fig:BDifferentPhases} separates the magnetic field geometry into the cold ($T<10^3$~K) and warm ($T>10^4$~K) phases, illustrating that the vertical field above and below the plane mostly belongs to the warm diffuse phase.

\subsubsection{Implications for the Milky Way}

The transition from horizontal (parallel to the plane) to vertical field as we move away from the mid-plane that we see in our simulations is reminiscent of the similar transition observed in the Milky Way as we move from diffuse to denser gas mentioned in Sect.~\ref{sec:intro}. However, we must be careful in drawing a comparison. The geometry of the magnetic field in the CMZ is probably affected by the presence of a Galactic outflow \citep[e.g.][]{Ponti2021,Heywood2022}, which is absent in our simulations due to the lack of stellar feedback and cosmic ray physics \citep[e.g.][]{Girichidis2024}. Nevertheless, it is interesting to note that a transition to a perpendicular field as we leave the plane also happens independent of a Galactic outflow.

Based on our finding that the magnetic field vectors tend to be aligned with gas velocity vectors, especially in the diffuse phase, we might speculate that the vertical magnetic field lines observed in the Milky Way diffuse gas at latitudes $|b|>0.4^\circ$ are tracing vertical streaming of the gas associated with the multi-phase Galactic outflow \citep{Ponti2021}. We might expect that potential de-alignment due to turbulence is not happening in the diffuse gas above and below the plane, as it is sufficiently far from the midplane for turbulence driven by magnetic fields (Sect.~\ref{sec:turbulence}) and/or stellar feedback (which predominantly occurs in the dense gas) to be ineffective.

\subsubsection{Implications for external barred galaxies}

Our finding that the magnetic field on large (kiloparsec) scales in the bar region is approximately aligned with the gas velocity streamlines is consistent with observations of polarised radio continuum emission of nearby barred galaxies such as NGC 1097 and NGC 1365 \citep{Moss2001,Beck2005}. The comparison of the orientation of the magnetic field in the nuclear ring is more tricky. The observed pitch angle of the magnetic field inferred from synchrotron-emitting gas in NGC 1097 is rather large, $\theta\sim 40^\circ$. The pitch angle of the regular $\langle \bfB \rangle_t$ component in our simulations is much smaller (Fig.~\ref{fig:BDecomposition}). However, (i) the pitch angle of the instantaneous magnetic field $\bfB$ often appears much larger due to the presence of fluctuations perpendicular to the ring (Figs.~\ref{fig:Slices} and \ref{fig:MagnetiFieldMorph}); (ii) it is not clear to what extent our figures, which display  the magnetic field in all gas components, are representative of synchrotron-emitting gas. A proper comparison will require synthetic observations of the synchrotron-emitting gas and a more careful comparison with observations, which is outside the scope of this paper.

The fact that the magnetic field is parallel to the bar lanes emerges spontaneously from the global flow in our simulations, and justifies the assumption of \cite{Moon2023}, who injected the magnetic field parallel to the velocity vectors into the computational box in their semi-global simulations.

\subsection{Magnetic field strength as a function of density} \label{sec:Bstrength}

The top panel in Fig.~\ref{fig:BvsRho} shows the strength of the $\bfB$ field as a function of total gas density in our fiducial CHEM\_MHD simulation. We find that the magnetic field scales approximately as
\begin{equation} \label{eq:Bscaling}
B = 102 \, \mu {\rm G} \left(\frac{n}{10^3 \, \rm cm^{-3}}\right)^{k} \,,
\end{equation}
where $k=0.33$. This scaling can be approximately understood as follows. Consider expansion or contraction of gas under the assumption of flux freezing. We can distinguish the following three limiting cases (described for example in section~3.3.1 in the book of \citealt{Kulsrud2005}):
\begin{align}
    B \propto \rho^1 & \quad \text{if the expansion or contraction is perpendicular to } \bfB; \\
    B \propto \rho^0 & \quad \text{if the expansion/contraction is parallel to } \bfB; \\
    B \propto \rho^\frac{2}{3} & \quad \text{if the expansion or contraction is isotropic}. 
\end{align}

When $\bfB$ is dynamically dominant (compared to the turbulent motions that cause expansion or contraction), we expect the gas to flow preferentially parallel to $\bfB$, and therefore we expect $k<2/3$. In our simulations, as we will see in Sect.~\ref{sec:turbulence}, the magnetic energy density is typically $20-40\%$ of the turbulent kinetic energy density, and the Alfv\'en speed is comparable to the turbulent velocity dispersion. Thus, we are in a trans-Alfv\'enic non-self gravitating regime, in which magnetic fields play a non-negligible dynamical role. We therefore expect the gas to flow more readily in the direction parallel to $\bfB$ than perpendicular to it, and therefore we expect $k<2/3$.

The exponent exhibits a small secular evolution in the simulation, changing from $k\simeq0.4$ at $t=150$~Myr to $k\simeq0.28$ at the end of the simulation, the mean being $k\simeq0.33$. This variation is somewhat smaller than what is seen in galaxy-scale simulations including supernova feedback \citep{Konstantinou2024}, which are likely a source of higher time variation of the exponent. 

The magnetic field strength as a function of density in our simulations is consistent with the rather sparse and uncertain measurements of the magnetic field strengths in the literature reported in the top panel of Fig.~\ref{fig:BvsRho}. It is also interesting to note that the power-law index of $0.33$ in Eq.~\eqref{eq:Bscaling} is similar to the value of $0.4$ reported by \cite{Liu2022}, which was obtained by compiling polarised dust emission observations of star forming regions from the literature and computing the magnetic field strength using the Davis-Chandrasekhar-Fermi (DCF) method, although one should note that the estimated power law index has large variations depending on how the magnetic field strength was estimated and the same authors also report a larger value of $0.57$ when they estimate the field differently (their section.~3.2.1 and their fig.~3). Finally, it is worth noting that we expect the introduction of the gas self-gravity, star formation and of the associated stellar feedback, that are switched off in our simulations, to likely affect the scaling of the magnetic field strength with density \citep{Girichidis2018}.

\section{Turbulence} \label{sec:turbulence}

We have already noted in Sect.~\ref{sec:gasmorphology} that the introduction of magnetic fields causes turbulence that changes the density PDF and `puffs up' the gas in the ring. To quantify the magnetic-driven turbulence in more detail we covered the $R<500\pc$ region with non-overlapping cubical bins 20~pc on-a-side and calculated the vertical velocity dispersion in each bin, defined as
\begin{equation}
\sigma_z^2 = \frac{1}{N} \sum_i \Delta v_{z,i}^2  \,,   
\end{equation}
where $N$ is the total number of cells in the 20~pc bin, 
\begin{equation}
    \Delta \bfv_i=\bfv_i - \bfv_{{\rm CM},i} \,
\end{equation}
is the velocity of the $i$-th cell relative to the centre of mass of the 20~pc bin (we subtracted the centre of mass velocity, since bulk motions do not contribute to turbulent kinetic energy, e.g. \citealt{Stewart2022}), and the sum extends over all cells in the bin. We used the dispersion in the $z$ direction to quantify the turbulence as it is less affected by streaming and rotational motions than the dispersion in other directions. We have also checked that once the streaming motions are taken into account the velocity dispersion is roughly isotropic in our simulations, which we find to be true within a factor of $\sim 1.5$.\footnote{In particular we find the dispersion $\sigma_x$ in the direction of the bar minor axis is almost identical to $\sigma_z$, while the dispersion $\sigma_y$ in the direction of the bar major axis is generally slightly larger, which is partly because streaming motions and velocity gradients are greater in this direction and therefore more difficult to subtract.}

\begin{figure}
	\includegraphics[width=\columnwidth]{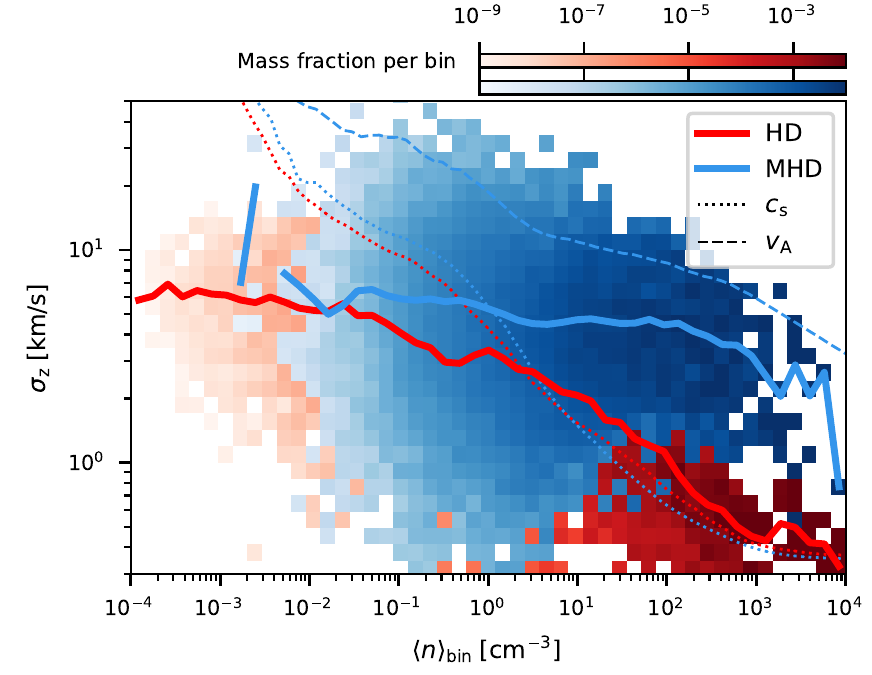}
        \caption{Velocity dispersion $\sigma_z$ in the $z$ direction in 20~pc on-a-side cubical bins for the magnetised CHEM\_ MHD (blue) and unmagnetised CHEM\_HD (red) simulations at $R<500\pc$ and $t=186$~Myr. The simulations are identical except that the one has magnetic fields and the other does not. On the $x$ axis is the average density within the bin. Red and blue solid lines indicate the average $\sigma_z$ at the given density. Dashed lines indicate the mass-weighted averaged Alfv\'en speed, and dotted lines the thermal sound speed. The magnetised simulation is significantly more turbulent. The turbulence is driven by the magneto-rotational instability (Sect.~\ref{sec:turbulence}).}  
    \label{fig:TurbulentEnergy_B_NoB}
\end{figure}

\begin{figure}
	\includegraphics[width=\columnwidth]{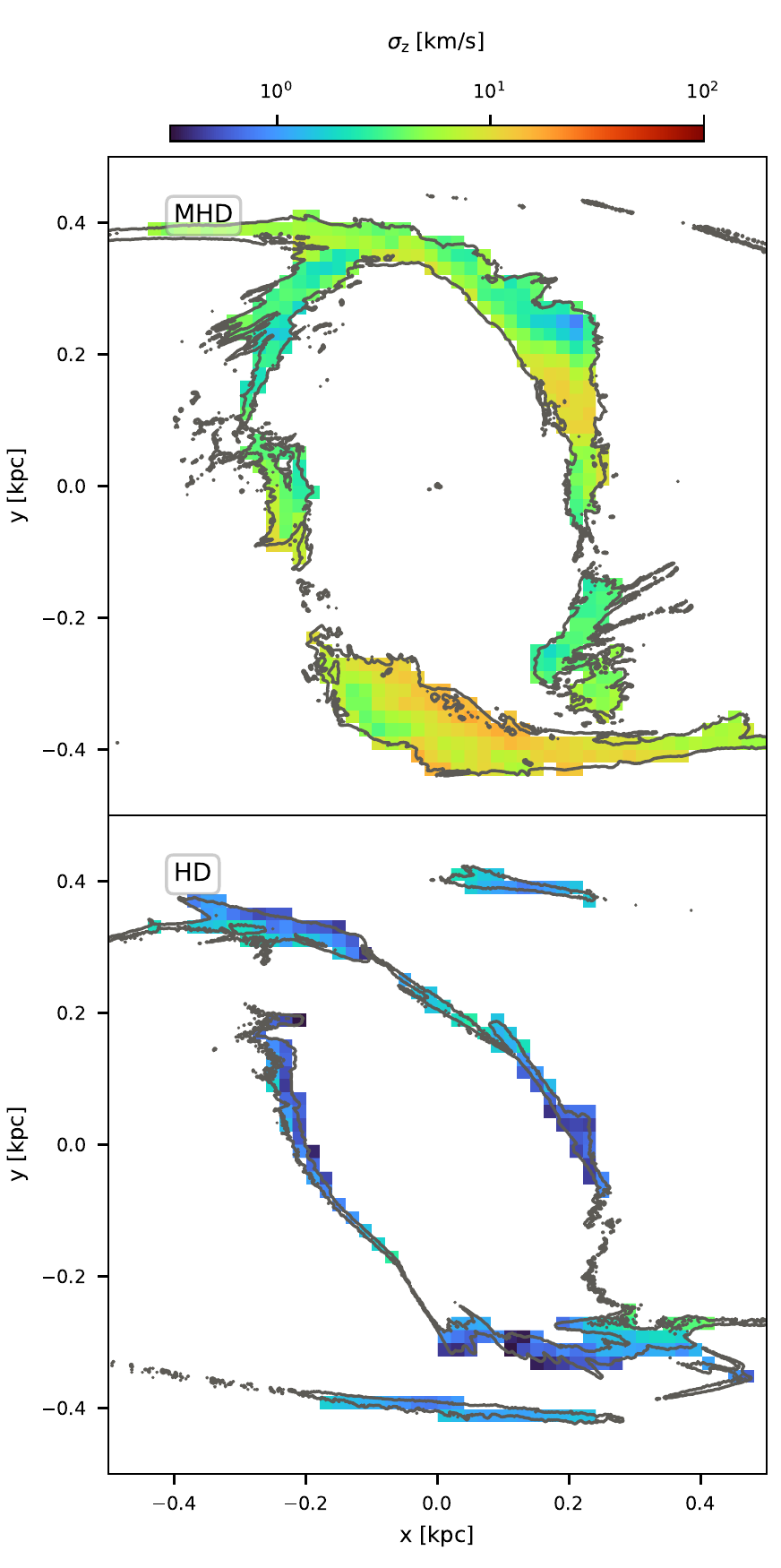}
        \caption{Vertical velocity dispersion $\sigma_z$ in the CMZ ring in the CHEM\_MHD (top) and CHEM\_HD (bottom) simulations at $t=186\Myr$. The velocity dispersion in the dense gas ring is much larger in the magnetised simulation than in the unmagnetised one due to the MRI-driven turbulence. Only regions with surface density $\Sigma \geq 100 \rm \, \Msun \pc^{-2}$ are shown. The velocity dispersion is calculated using cubical bins 20~pc on-a-side as in Fig.~\ref{fig:TurbulentEnergy_B_NoB}.} 
    \label{fig:sigmaz}
\end{figure}

Figure~\ref{fig:TurbulentEnergy_B_NoB} compares $\sigma_z$ in the magnetised CHEM\_MHD simulation and in its unmagnetised version CHEM\_HD (Table~\ref{tab:sims}) at $t=186$~Myr. Similarly, in Fig.~\ref{fig:sigmaz} we show a map of $\sigma_z$ in the CMZ for the two simulations. It is clear that the introduction of magnetic fields causes a significant increase of the velocity dispersion and of the turbulent kinetic energy. The increment is more significant in the high-density gas. For example, at bin-averaged densities of $\langle n \rangle_{\rm bin}=10^2 \cm^{-3}$ the velocity dispersion increases from $\sigma_z\simeq1\kms$ to $\sigma_z\simeq5\kms$. Comparing these numbers to the sound and Alfv\'en speeds (dashed and dotted lines in Fig.~\ref{fig:TurbulentEnergy_B_NoB}) shows that the turbulence is supersonic and trans-Alfv\'enic. 

The question arises as to what physical mechanism drives the turbulence in these simulations. We included neither the gas self-gravity nor star formation, so self-gravity and stellar feedback are ruled out as possible sources of the turbulence. The bar-driven inflow onto the CMZ can drive turbulence by converting bulk kinetic energy into turbulent motions \citep{Kruijssen2014,Sormani2019,Henshaw2023}. However, the unmagnetised CHEM\_HD simulation, which also includes the bar-driven inflow, displays a much lower level of turbulence than the magnetised CHEM\_MHD simulation. To quantify the relative importance of the bar-driven inflow on the turbulence in the CHEM\_MHD simulation, we calculated the velocity dispersion in this simulation at much later times ($t>250\Myr)$, after the bar-driven inflow effectively shuts down because it runs out of gas (Sect.~\ref{sec:inflow}). We find that after the bar inflow shuts down the turbulence decreases until it settles to an intermediate value of $\sigma_z \simeq 3 \kms$ at bin-averaged densities of $\langle n \rangle_{\rm bin}=10^2 \cm^{-3}$. This value is then maintained for very long times, well beyond the turbulence decaying times (or vertical crossing time). This suggests that during the time in which the bar inflow is active (as at  $t=186$~Myr in Fig.~\ref{fig:TurbulentEnergy_B_NoB}) the turbulence is driven by a combination of bar inflow and magnetic processes, while the turbulence at later times (after the bar-driven inflow shuts off) is maintained by purely magnetic processes. To further investigate this we have done the following experiment (Appendix~\ref{sec:Cutout}): we have stopped the simulation at $t=166$~Myr, removed all the gas at $R>500\pc$ so that we are left only with the CMZ gas ring, and then restarted the simulation. In this way, we remove the large-scale bar inflow and continue the simulation with only the CMZ ring evolving `in isolation'. We find that the turbulence settles to the same intermediate value that we find in the CHEM\_MHD simulations at late times, which results in $\sigma_z \simeq 3 \kms$ at bin-averaged densities of $\langle n \rangle_{\rm bin}=10^2 \cm^{-3}$. We repeated this test using an axisymmetrised potential after restarting the simulation, to exclude any possible influence of the non-axisymmetric gravitational potential, and find again that turbulence is maintained at the same level. These tests confirm that the turbulence at $t=186$~Myr in Fig.~\ref{fig:TurbulentEnergy_B_NoB} is driven by a combination of bar inflow and magnetic processes, while the turbulence at later times (after the bar inflow shuts down) is purely due to magnetic processes.

A well-known and effective mechanism to generate turbulence in astrophysical accretion discs is the magneto-rotational instability \citep[MRI;][]{Velikhov1959,Balbus1991,Hawley1991,Balbus1998}. This instability occurs in every (even weakly) magnetised disc in differential rotation in which the angular speed $\Omega(R)$ decreases as a function of radius, and has been shown to work in the $\beta<1$ limit that is relevant here (for example \citealt{Kim2000}, \citealt{Piontek2007}, and Appendix C of \citealt{Jacquemin-Ide2021}). The MRI generates turbulence by extracting the energy stored in differential rotation and converting it into turbulent fluid motions. Thus, in MRI-driven turbulence the magnetic stresses act as a mediator, allowing the turbulence to tap into the differential rotation that would otherwise not be converted into turbulent motions. 

Our simulated CMZ satisfies the conditions for the onset of the MRI, and Fig.~\ref{fig:Resolution} shows that the MRI is well-resolved in our simulations. The MRI is therefore the most natural candidate to drive turbulence. The MHD code {\sc arepo} that we are using has been already tested to correctly reproduce the linear phase of the MRI \citep{Pakmor2013}. We therefore conclude that the MRI (in its saturated state) is driving the turbulence in our magnetised simulations at late times (after the bar inflows shuts off). 

To compare turbulent and other types of energy, we computed the turbulent kinetic, magnetic, and thermal energies in each 20~pc bin as follows: 
\begin{align}
    E_{\rm k} & = \sum_i \frac{1}{2} m_i |\Delta \bfv_i|^2 \,, \label{eq:Ek} \\
    E_B & = \sum_i V_i \frac{|\bfB|_i^2}{8\pi}\,, \label{eq:EB} \\
    E_{\rm th} & = \sum_i m_i e_{{\rm th}, i}\,, \label{eq:Eth}
\end{align}
where $m_i$ is the mass of the $i$-th cell, $\bfB_i$ is its magnetic field, $e_{{\rm th}, i}$ its thermal energy per unit mass, $V_i$ its volume, and the sum extends over all cells in the bin.

\begin{figure}
    \includegraphics{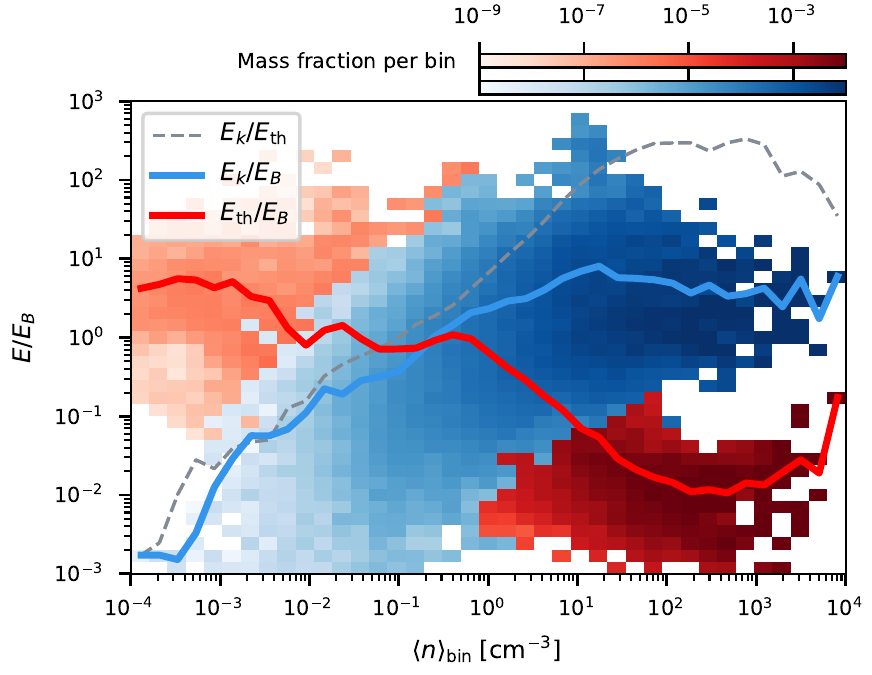}
    \caption{Energy ratios in 20~pc on-a-side cubical bins in our CHEM\_HD (red) and CHEM\_MHD (blue) simulations at $R<500\pc$ and $t=186$~Myr. Blue is the ratio $E_{\rm k}/E_B$ between turbulent kinetic energy (Eq.~\ref{eq:Ek}) and magnetic energy (Eq.~\ref{eq:EB}). Red is the ratio $E_{\rm th}/E_{\rm B}$ between thermal energy $E_{\rm th}$ (Eq.~\ref{eq:Eth}) and magnetic energy. On the $x$ axis is the average density within the respective bin. The solid lines represent the average values at the given density. The average of the turbulent kinetic energy and thermal energy is also shown as the grey dashed line. At high densities ($\langle n \rangle_{\rm bin}>1 \, {\rm cm}^{-3}$), the kinetic energy is a few times the magnetic energy ($E_{\rm k}/E_B\sim 2\mhyphen5$), while the thermal energy is negligible compared to the magnetic energy. At low densities ($\langle n \rangle_{\rm bin}<1 \, {\rm cm}^{-3}$) the situation is inverted, with the thermal energy surpassing the magnetic energy and the turbulent kinetic energy becoming negligible at very low densities.}
    \label{fig:EPEB}
\end{figure}

Figure~\ref{fig:EPEB} plots the energy ratios at $t=186$~Myr, when turbulence is driven by a combination of bar-driven inflow and MRI. We find that (i) in high-density gas (bin-averaged density $\langle n\rangle_{\rm bin}>1 \cm^{-3}$), the magnetic energy is $20\mhyphen 40\%$ of the turbulent kinetic energy; (ii) in low-density gas ($\langle n \rangle_{\rm bin}<0.1 \cm^{-3}$), the magnetic energy is approximately in equipartition with the thermal energy, and the turbulent energy is small. These ratios are similar to those found in studies of feedback-driven and gravity-driven turbulence, which suggest that in general $E_k/E_B \gtrsim 2$ \citep[e.g.][]{Federrath2011,Rieder2017,Gent2021,Higashi2024}. In contrast, at later times, after the bar inflow shuts off, the ratio between the turbulent kinetic energy and the magnetic energy decreases and reaches mass-weighted average values $E_k/E_B<1$, which is typical of purely MRI-driven turbulence \citep[e.g.][]{Balbus1998,Kim2003,Sano2004,Minoshima2015}. The decrease in ratio is mainly driven by a decrease in $E_k$, while $E_B$ remains approximately constant (Sect.\ref{sec:Bevolution}). These findings corroborate the idea that while the bar inflow is active the turbulence is driven by a combination of the inflow and the MRI, while when the bar inflow shuts off the turbulence is purely MRI-driven. 

In conclusion, we have found that the combination of bar-driven inflow and MRI turbulence sustains vertical velocity dispersions that on scales of 20~pc are of the order of $\sigma_z \sim 5\kms$, while the MRI alone sustains $\sigma_z \sim 3\kms$. Both of these numbers are smaller than the $\sigma_{\rm los}\sim 10\kms$ line-of-sight dispersion observed in the CMZ on the same scales \citep[e.g.][]{Shetty2012, Henshaw2016}. This suggests that a further ingredient, likely stellar feedback such as supernovae and/or stellar winds, is necessary to explain the observed levels of turbulence in the CMZ (for example \citet{Tassis2022} and section~4.3.4 in the review of \citealt{Henshaw2023}). However, we note that both the bar-driven turbulence and MRI-driven turbulence are expected to be primarily solenoidal \citep{Gong2020}, so they might be at the origin of the solenoidal driving of turbulence observed in the `Brick' cloud \citep{Federrath2016}.

\section{Growth of the magnetic field} \label{sec:Bevolution}

\begin{figure}
	\includegraphics[width=\columnwidth]{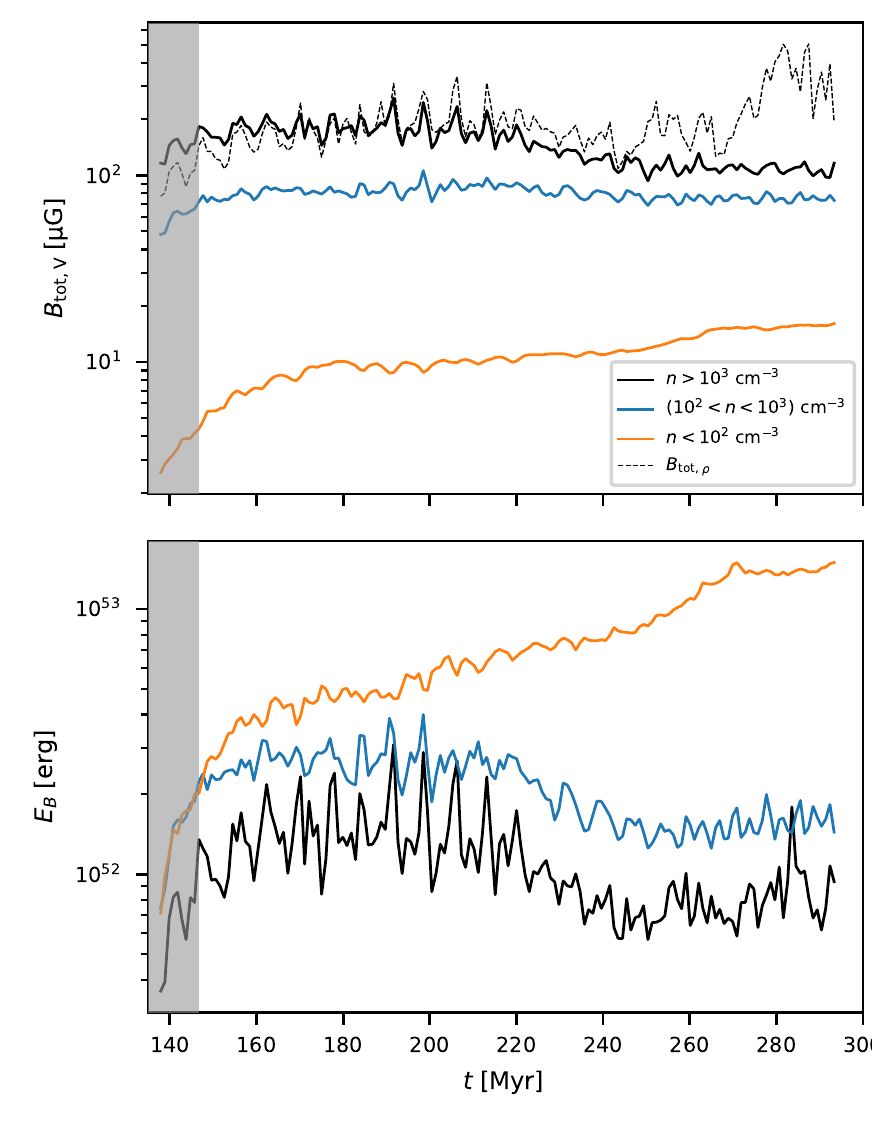}
        \caption{Time evolution of the magnetic field in our fiducial CHEM\_MHD simulation at $R<500\pc$ and $|z|<100$~pc. \emph{Top}: The full lines are the volume-weighted magnetic fields (Eq.~\ref{eq:Btot}) averaged over the indicated density range. The black dashed line is the mass-weighted magnetic field averaged over all densities (Eq.~\ref{eq:Btot}). The figure shows that the field grows in time starting from the initial seed as a result of dynamo action (Sect.~\ref{sec:Bevolution}). \emph{Bottom}: integrated magnetic energy in the same density regimes. We note that while the denser part of the ISM has higher magnetic field values, most of the magnetic energy is actually in the low density regime.} 
    \label{fig:BEB}
\end{figure}

Figure~\ref{fig:BEB} plots the time evolution of the volume- and mass-weighted magnetic fields in the region $R<500\pc$ in our fiducial simulation as a function of time. We started with an initially uniform seed field of $\bfB_0 = 0.02 \,\mu\G \, \hat{\mathbf{z}}$. The field grows until it saturates at typical values $B_{\rm tot}\sim 200$~$\mu\G$. We split the plot into different density ranges, the cutoff density chosen at typical values for which the ISM becomes molecular ($>10^2$~cm$^{-3}$). We find that saturation is reached more quickly in the dense gas, where turbulence is dynamically more important, and more slowly in the diffuse gas (top panel in Fig.~\ref{fig:BEB}). In Appendix \ref{sec:resolutionstudy} we show that the saturation field strength does not depend on the numerical resolution, while in Appendix \ref{sec:ICimpact} we show that it does not depend on the strength and orientation of the magnetic fields in the initial conditions.

\citet{Chandran2000} proposed that the magnetic field in the CMZ grows by accumulation of magnetic flux that is frozen into the bar-driven inflow and is advected into the CMZ. However, a key assumption of their model is that the magnetic field in the inflow is vertical (i.e. in the $\hatez$ direction), so that magnetic field lines get squeezed together as they move inwards, leading to the B field amplification. This assumption is invalid in our simulations because, as discussed in Sect.~\ref{sec:geometry}, the magnetic field in the bar lanes that transport the inflow is parallel to the velocity vectors, which mostly lie in the plane $z=0$. This implies that the magnetic field does not grow by magnetic flux accumulation via the mechanism envisioned by \citet{Chandran2000} in our simulations. To confirm this, we have performed an experiment similar to the one described in Sect.~\ref{sec:turbulence}: we stopped the simulation at $t=166$~Myr, removed all the gas at $R>500\pc$ so that we are left only with the CMZ ring, reset the magnetic field to the initial value $\bfB_0 = 0.02 \,\mu\G \, \hat{\mathbf{z}}$ everywhere, and then restarted the simulation. In this way, we remove the bar-driven inflow and any related magnetic flux accumulation. We find that the magnetic field still grows and reaches the same saturation level as in the `full' simulation (Appendix~\ref{sec:Cutout}). Repeating the test using an axisymmetrised potential after restarting the simulation leads to the same result. We conclude that the magnetic field in the CMZ does not grow by magnetic flux accumulation advected with the bar-driven inflow.

It is likely that magnetic fields in our simulations grow by dynamo action. Dynamo action can be defined as the process by which motions in the fluid amplify the magnetic fields over time. Differential rotation can amplify a toroidal magnetic field by shearing and stretching the radial field \citep[the so-called $\Omega$ effect, for example][]{Parker1955,Moffatt1978,Mestel2012}. Turbulent motions can lift the gas upwards in the plane and create and/or amplify a poloidal component from the toroidal component by inducing stretch-twist-fold motions \citep[an example of such motions is the so called $\alpha$ effect, e.g.][]{Parker1955,Parker1971,Parker1992,Childress1995}. The two effects together can produce a cycle that leads to a net increase of the magnetic field intensity over time.

Differential rotation is naturally present in our simulations. Turbulence in our simulations is mostly driven by the MRI as we discussed in Sect.~\ref{sec:turbulence}. It is therefore likely that the magnetic field in our simulations grows by a combination of $\Omega$-dynamo induced by the differential rotation and an MRI-driven dynamo. Indeed, it is well-known that the MRI can drive dynamo action \citep[e.g.][]{Brandenburg1995,Stone1996,Ziegler2000,Vishniac2009,Guan2011,Hawley2013,Bodo2014,Dhang2020}. In an MRI-driven dynamo, magnetic fields are not only amplified by the turbulent velocity fluctuations, but they also produce the turbulent velocity field itself via the MRI \citep{Balbus1998}. In this aspect, MRI-driven dynamo action is different from dynamos driven by supernova feedback for example (or mean-field dynamos), where the magnetic field responds to velocity perturbations induced by something external (in this case, the supernovae). This is reflected by the fact that ratios between turbulent kinetic energy and magnetic energy are typically lower in MRI-driven dynamos than in stellar feedback-driven dynamos (discussion in Sect.~\ref{sec:turbulence}).

In summary, dynamo action via differential rotation and MRI-driven turbulent motions is likely responsible for the growth of magnetic fields in our simulation. A more complete investigation of the dynamo action in these simulations is out of the scope of this paper but is a worthwhile direction for future studies. In particular, one could analyse the turbulence in the contexts of mean-field and MRI-driven dynamos to see which framework better describes the simulations. Whether quantities such as kinetic helicity, which is a prerequisite for the mean-field $\alpha$-$\Omega$ mechanism, are the main driver for magnetic field growth here \citep[e.g.][]{Ntormousi2020} or if small-scale dynamo powered by MRI-driven turbulence are more important. By following high-turbulence regions as they evolve in their orbit around the Galactic centre we could understand how the energy is transferred between turbulent motion and $B$ and vice-versa.

\section{Inflow} \label{sec:inflow}
\begin{figure}
	\includegraphics[width=\columnwidth]{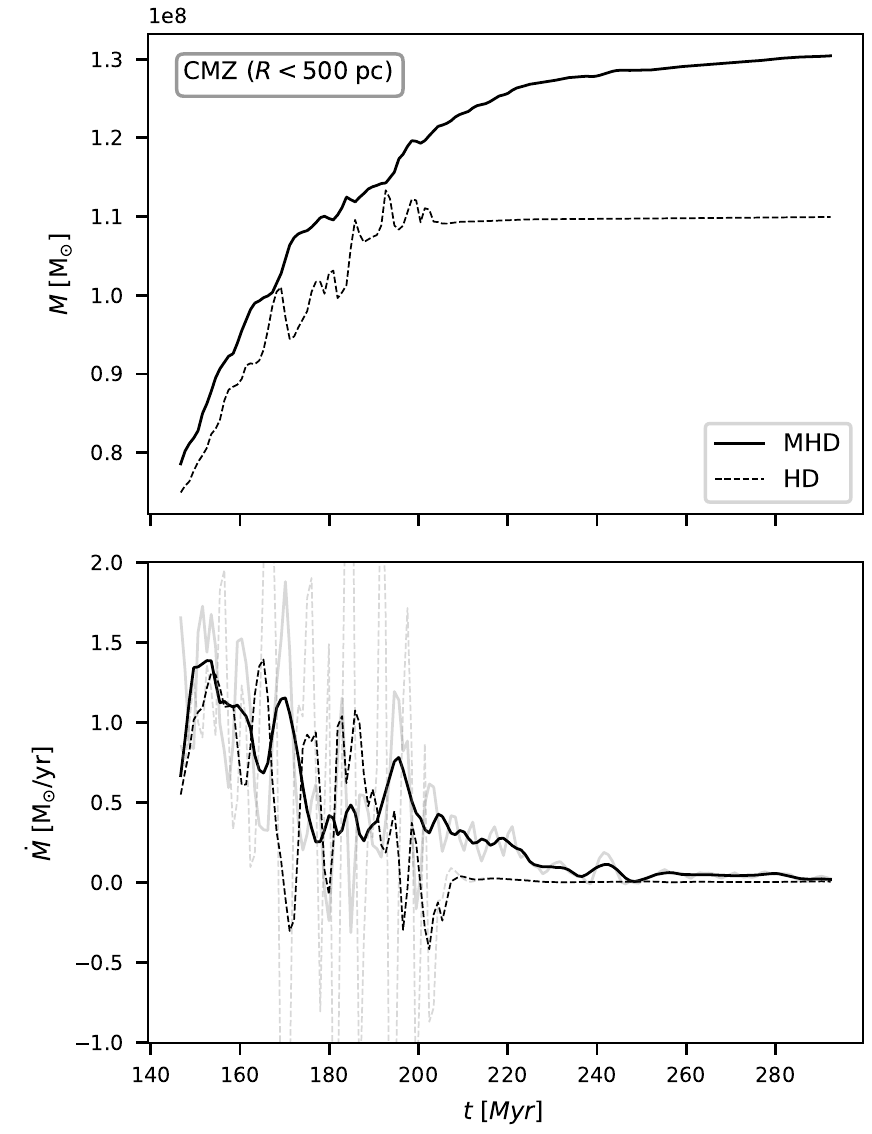}
        \caption{Mass accretion from the disc to the CMZ. \emph{Top}: Mass contained in the cylindrical volume $R<500\pc$ as a function of time for the CHEM\_MHD and CHEM\_HD simulations. \emph{Bottom}: inflow rate into the same volume, which represents the inflow onto the CMZ ring. The black lines represent the running average (over $7$~Myr) of the inflow rate, to smooth out the high time variability of $\dot{M}$. The instantaneous (non-averaged) inflow rate is shown as the light grey lines. The inflow rate in the MHD and HD simulation has a similar order of magnitude, but it lasts longer in the MHD simulation (roughly until $t\simeq200 \Myr$ in the HD simulation versus $t\simeq240\Myr$ in the MHD simulation). This is because magnetic fields transport gas radially inwards within the disc at $R>3\kpc$ and replenish the gas reservoir that supplies the bar-driven inflow at the outskirts of the bar, which instead runs out of gas in the purely HD simulations.  Inflow at these scales is mainly driven by the gravitational torques of the Galactic bar (Sect.~\ref{sec:inflow}).} 
    \label{fig:MassInflowCMZ}
\end{figure}

\begin{figure}
	\includegraphics[width=\columnwidth]{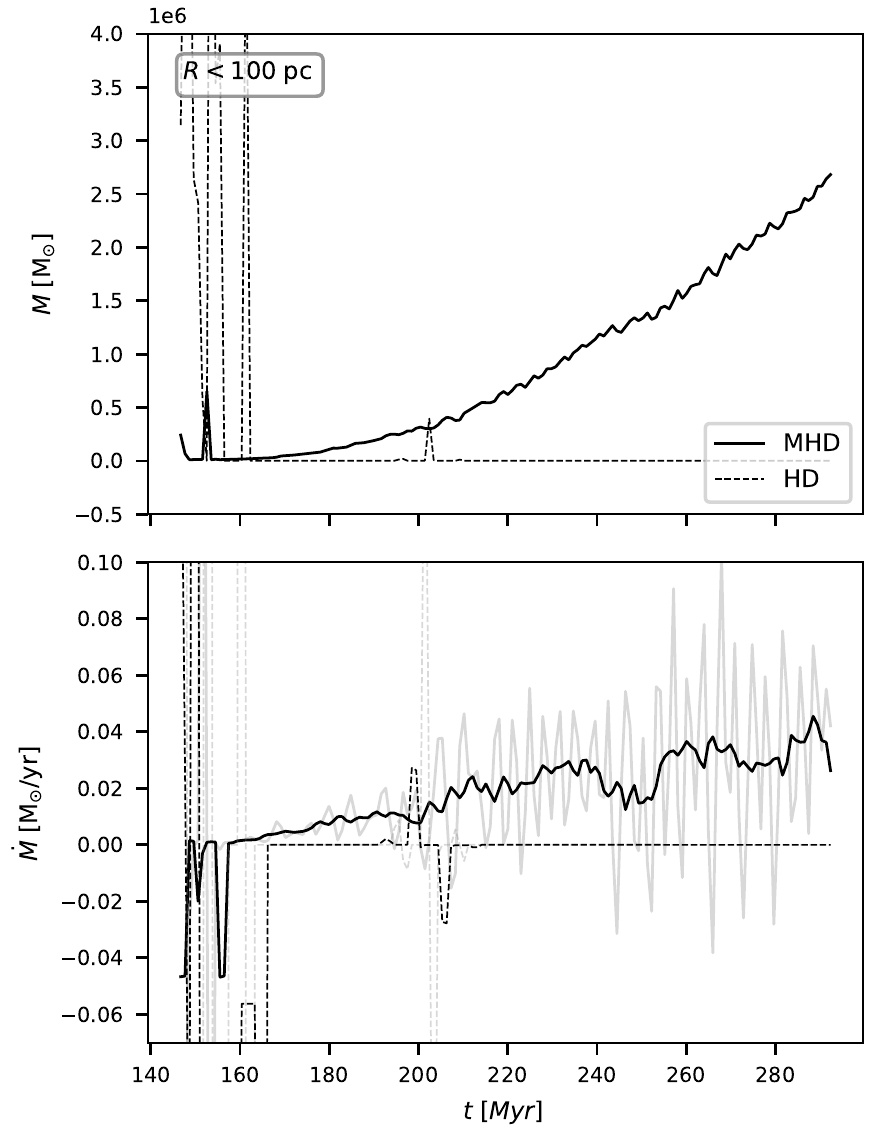}
        \caption{Mass accretion from the CMZ inwards. \emph{Top:} Mass contained in the cylindrical volume $R<150\pc$ as a function of time for the CHEM\_MHD and CHEM\_HD simulations. \emph{Bottom}: inflow rate into the same volume, which represents the inflow from the CMZ ring inwards. Similarly to Fig.~\ref{fig:MassInflowCMZ}, we show the time averaged inflow rate (black lines) as well as the non-averaged one (grey lines). The inflow is zero in the HD simulation, while it is significant in the MHD simulation. The spikes in the top panel for the HD simulation are caused by orbiting clouds that temporarily enter the $R=150\pc$ region, but that eventually leave ($M$ at the end of the HD simulation is zero as can be seen in the top panel). Inflow at these scales is mainly driven by the magnetorotational instability (Sect.~\ref{sec:inflow}). } 
    \label{fig:MassInflowCentre}
\end{figure}

The inflow of gas towards the centre in our simulations can be schematically divided into two regimes operating in different radial ranges which correspond to different physical driving mechanisms. These two regimes are:
\begin{enumerate}
    \item The bar-driven inflow: from the outskirts of the bar ($R\simeq3\kpc$) down to the CMZ gas ring ($R\simeq200\mhyphen300\pc$).
    \item The nuclear inflow: from the CMZ gas ring to the central few parsecs.
\end{enumerate}
Figures~\ref{fig:MassInflowCMZ} and \ref{fig:MassInflowCentre} quantify the mass inflow rates at two radii that correspond to these two regimes. Figure~\ref{fig:MassInflowCMZ} shows that the bar-driven inflow rate has a similar magnitude ($\dot{M}\simeq 1 \Msunyr$) in both the MHD and HD simulations during the first $\simeq 30\Myr$ after the bar is fully on ($t=146\Myr$, Sect.~\ref{sec:IC}), and then steadily declines. The reason for the decline is primarily that the gas reservoir located at the outskirts of the bar, which supplies the bar-driven inflow, runs out of gas. In other words, once the bar has cleared out all the inner disc region within its reach at $R\lesssim 3\kpc$, the inflow stops. The inflow lasts a bit longer in the MHD simulation than in the purely HD simulation because the MRI-driven turbulence transports some additional gas from the outer disc at $R>3\kpc$ down to $R\simeq 3\kpc$ where it can be `captured' by the bar. Eventually, the bar-driven inflow runs out of gas in the MHD simulation too. This is expected since our simulations do not include the most efficient mechanisms that are believed to replenish the gas supply at the outskirts of the bar, such as raining of gas
with low angular momentum from the circumgalactic medium or interactions between bar and spiral arms \citep[e.g.][]{Lacey1985,Bilitewski2012}. Magnetic stress in the bar lanes can also enhance the bar inflow by removing angular momentum \citep{Kim2012b}, but the torques analysis below shows that this is a secondary effect and that gravitational torques dominate over Maxwell torques in this regime. Thus, the bar-driven inflow is only marginally affected by the presence of the magnetic fields.

Figure~\ref{fig:MassInflowCentre} shows that the nuclear inflow is practically zero in the purely HD simulation. In this simulation, all the mass is accumulating in the ring and no gas is moving further in. In contrast, the MHD simulation has a significant inflow rate of order $\dot{M}\simeq 0.01\mhyphen0.1 \Msunyr$ that is time-varying with a general trend upwards with time. Thus, in contrast to the bar-driven inflow that is almost unaffected by the presence of magnetic fields, the nuclear inflow changes dramatically when magnetic fields are introduced. As we shall see below, the mass transport in the nuclear inflow is driven by the MRI. The nuclear inflow is one to two orders of magnitude smaller than the bar-driven inflow, so there is a net mass accumulation in the CMZ ring. However, the numerical values depend on the numerical resolution and do not appear to be converged at the maximum resolution we can afford. In Appendix \ref{sec:resolutionstudy} we perform a resolution study and we show that the nuclear inflow tends to decrease as the resolution is increased. Indeed, it is well known that convergence is hard to achieve in global simulations of MRI-driven accretion discs \citep{Hawley2013}. Therefore, the nuclear inflow rates derived here should be considered upper limits.


We now analyse in more detail the physical mechanisms that drive the inflows. We start by looking at the transport of angular momentum in our simulations. Consider the cylindrical region within $R=R_0$, with volume $V$ and surface $S$. Combining Eqs.~\eqref{eq:Continuity} and \eqref{eq:euler} the rate of change of the total angular momentum contained within this cylindrical volume can be expressed as (Appendix \ref{sec:appendixA}):
\begin{equation} \label{eq:Lzdot}
    \frac{\pa L_z}{\pa t} = F_{\rm Rey} + F_{\rm Max} + F_{\rm Grav} \,,
\end{equation}
where
\begin{equation} \label{eq:Lzdef}
    L_z = \int_V \rho R v_\phi \, \di V \,,
\end{equation}
is the total $z$ angular momentum contained inside the cylinder, and 
\begin{align}
F_{\rm Rey} & = -\int_S R \rho v_\phi v_R \, \di S\, , \label{eq:FR} \\
F_{\rm Max} & = -\int_S R T_{R\phi} \, \di S\, , \label{eq:FM} \\
F_{\rm Grav} & = - \int_V \rho \frac{\pa \Phi}{\pa \phi}\, \di V\,, \label{eq:FG}
\end{align}
are the fluxes of angular momentum in and out of the cylinder, $T_{R \phi} = - {B_\phi B_R}/({4 \pi})$ is the component of the Maxwell stress tensor defined in Sect.~\ref{sec:methods}, $\di V$ denotes the volume element, $\di S$ is the surface area element, and $(R,\phi,z)$ denote Galactocentric cylindrical coordinates

\begin{figure}
    \includegraphics[width=\columnwidth]{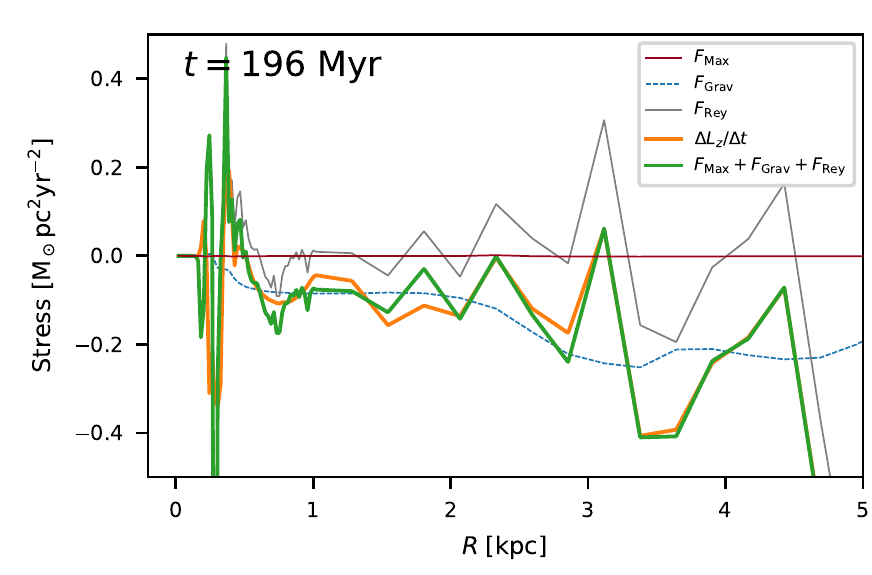}
    \caption{Rate of change of the total angular momentum contained within cylindrical radius $R$ (Eq.~\ref{eq:Lzdot}) for the snapshot $t=196 \Myr$ of the CHEM\_MHD simulation. $\Delta L_z/\Delta t$ is calculated by summing the angular momentum of all gas cells (Eq.~\ref{eq:Lzdef}) in two different snapshots separated by $\Delta t \simeq 1 \Myr$, taking the difference and dividing by $\Delta t$. The other contributions are calculated using Eqs.~\eqref{eq:FR}-\eqref{eq:FG}.} 
    \label{fig:sumcheck}
\end{figure}

Equation \eqref{eq:Lzdot} states that the change in the total angular momentum of the gas contained within the cylinder is the sum of three contributions: (i) the Reynolds flux $F_{\rm Rey}$ due to bulk motions of the gas entering or leaving the cylinder; (ii) the Maxwell flux $F_{\rm Max}$ due to magnetic forces; and (iii) the gravitational term $F_{\rm Grav}$ due to gravitational torques from the external bar potential. In Fig.~\ref{fig:sumcheck} we performed a sanity check by calculating separately the left-hand-side (LHS) and right-hand-side (RHS) of Eq.~\ref{eq:Lzdef}. The two agree well as a function of $R$, which gives us confidence that our code is working correctly.

To explore the mass transport, first we decomposed the velocity field as
\begin{equation}
\bfv = v_0 \hatephi + \bfu\,, \label{eq:v0}
\end{equation}
where
\begin{equation}
v_0(R,t) = \frac{ \langle \rho v_\phi \rangle_{\phi z} }{\langle \rho \rangle}\,,
\end{equation}
is the average mass-weighted rotation velocity, $\bfu$ represents the deviations and 
\begin{equation}
\langle X \rangle_{\phi z} = \frac{1}{2\pi} \int_0^{2 \pi} \di \phi \, \int X \, \di z \,,
\end{equation}
denotes the vertical and azimuthal average of a physical quantity $X$. We note that $u_R = v_R$ by definition. We also decomposed the Reynolds flux (Eq.~\eqref{eq:FR}) into an average and turbulent component:
\begin{align} \label{eq:Frey}
F_{\rm Rey} = F^{\rm ave}_{\rm Rey} + F^{\rm trb}_{\rm Rey}
\end{align}
where
\begin{align}
F^{\rm ave}_{\rm Rey} & = -\int_S R \rho v_0 v_R \, \di S\, , \\
F^{\rm trb}_{\rm Rey} & = -\int_S R \rho u_\phi v_R \, \di S\, . \label{eq:Ftrb}
\end{align}

In Appendix \ref{sec:appendixA} we show that mass accretion rates of a quasi-steady state can be understood as the sum of three contributions: turbulent Reynolds stresses, Maxwell stresses, and gravitational torques from the bar.\footnote{As can be seen from Eq~\eqref{eq:Mdot2}, it is only the turbulent part $F^{\rm trb}_{\rm Rey}$ of the Reynolds flux \eqref{eq:Frey} that is related to a net mass inflow rate. Indeed, a large angular momentum flux $F_{\rm Rey}$ does not necessarily imply a large mass inflow rate. This is because fluid elements can oscillate radially carrying more angular momentum on their outward radial journey than on their return, with individual fluid elements neither gaining nor losing angular momentum on average. This type of transport has been named `lorry transport' by \citet{Lynden-Bell1972}, who explained how fluid elements can `transport angular momentum just as a system of lorries can transport coal without accumulating a growing store on the lorries themselves'. The term $F^{\rm ave}_{\rm Rey}$, which is typically much larger than $F^{\rm trb}_{\rm Rey}$, is related to a transport of angular momentum without a corresponding transport of mass.} By determining which contribution dominates at each radius in our simulation, we can identify the physical mechanism driving the inflow.
\begin{figure}
    \begin{subfigure}{\columnwidth}
	    \includegraphics[width=\columnwidth]{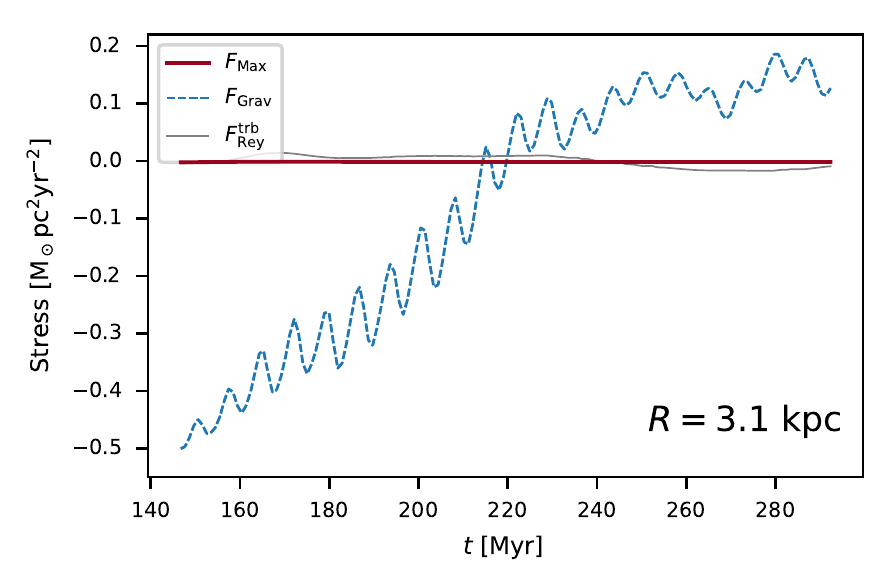}
    \end{subfigure}
    \begin{subfigure}{\columnwidth}
    	\includegraphics[width=\columnwidth]{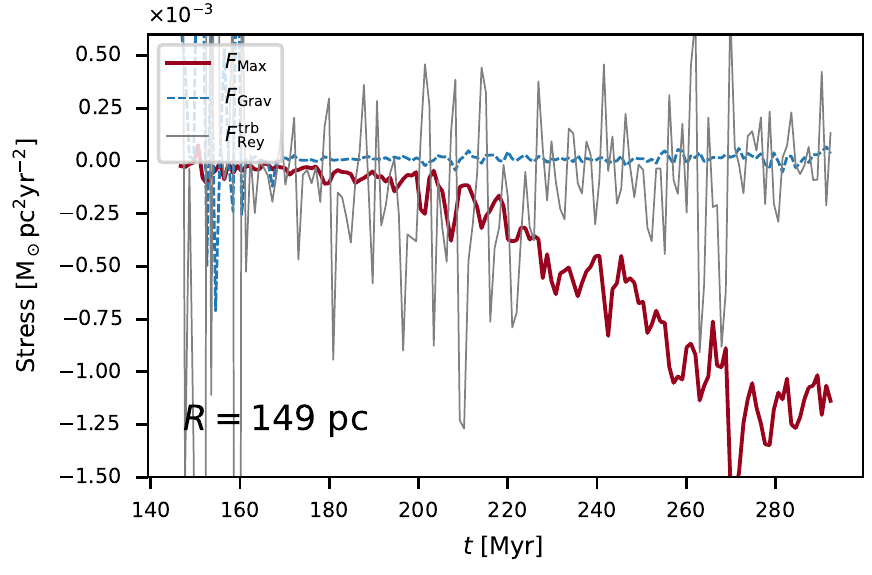}
    \end{subfigure}
    \caption{Turbulent Reynolds, Maxwell, and gravitational stresses at a given radius as a function of time in the CHEM\_MHD simulation. \emph{Top:} at $R=3.1\kpc$, which tracks the entire bar region. The oscillations on timescales $t\simeq 7 \Myr$ represent a torque-weighted orbital period of clouds orbiting in the bar region (Fig.~\ref{fig:GravTorquesMap} and recalling that angular momentum is not conserved in a non-axisymmetric bar potential). The overall upward trend is because angular momentum is removed by the bar at a decreasing rate as the gas reservoir that supplies the bar-driven inflow is depleted and the bar-driven inflow stops (Fig.~\ref{fig:MassInflowCMZ} and Sect.~\ref{sec:inflow}). \emph{Bottom:} $R=149\pc$, which tracks what happens \emph{inside} the CMZ gas ring. Maxwell stresses become dominant at $t>200 \Myr$, and show that magnetic stresses are responsible for the nuclear inflow from the CMZ ring inwards.} 
    \label{fig:Lzdot_vs_t}
\end{figure}

\begin{figure}
    \begin{subfigure}{\columnwidth}
    	\includegraphics[width=\columnwidth]{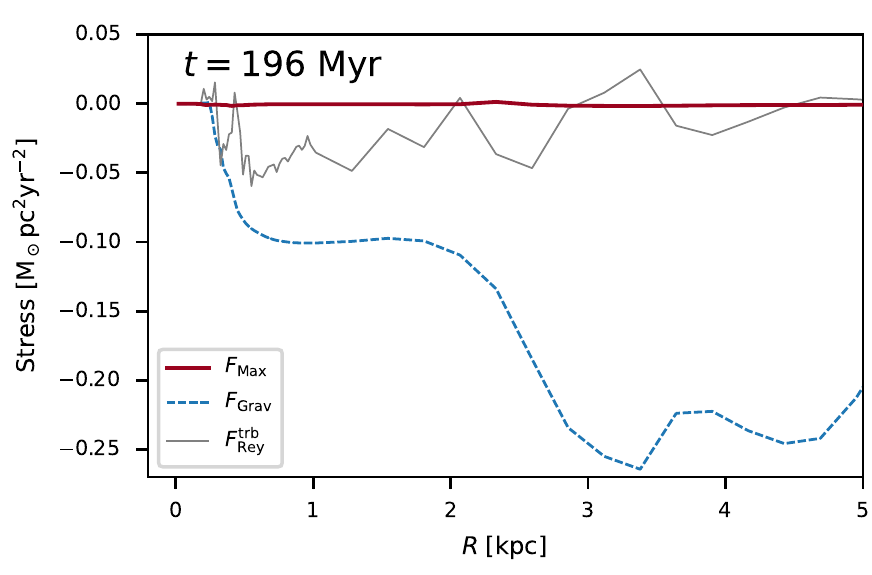}
    \begin{subfigure}{\columnwidth}
	    \includegraphics[width=\columnwidth]{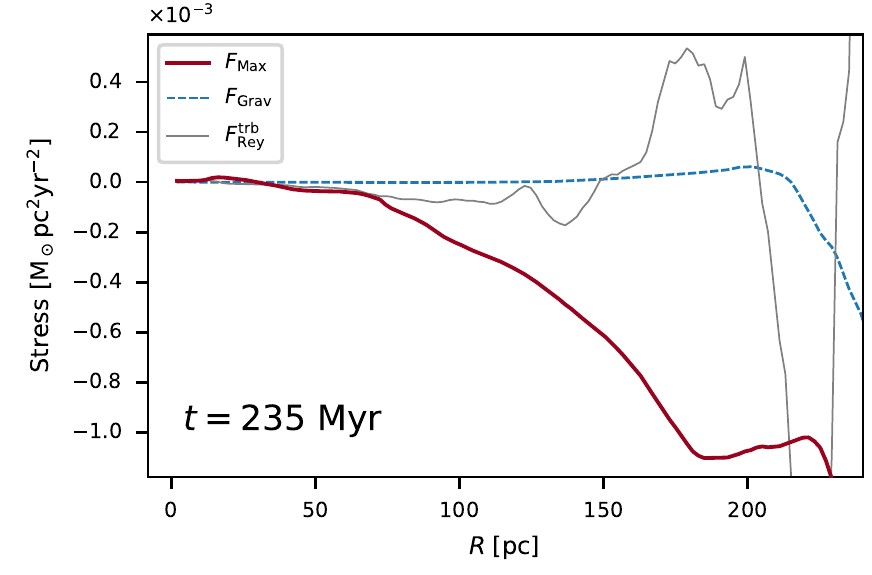}
    \end{subfigure}
    \end{subfigure}
        \caption{Turbulent Reynolds, Maxwell, and gravitational stresses at a given time as a function of radius for the CHEM\_MHD simulation. \emph{Top}: for $R<5\kpc$. \emph{Bottom}: for $R<200\pc$. At $200<R<3\kpc$ the gravitational stresses dominate, and therefore the bar-driven inflow is driven by the gravitational torques of the Galactic bar (as the name would imply). At $0<R<200 \pc$ the Maxwell stress dominate, and therefore the nuclear inflow is driven by magnetic fields. We note that the top and bottom panels are for different snapshots, which are selected to better capture the moments the bar-driven and nuclear inflows take place.} 
    \label{fig:Lzdot_vs_R}
\end{figure}

\begin{figure}
    \includegraphics[width=\columnwidth]{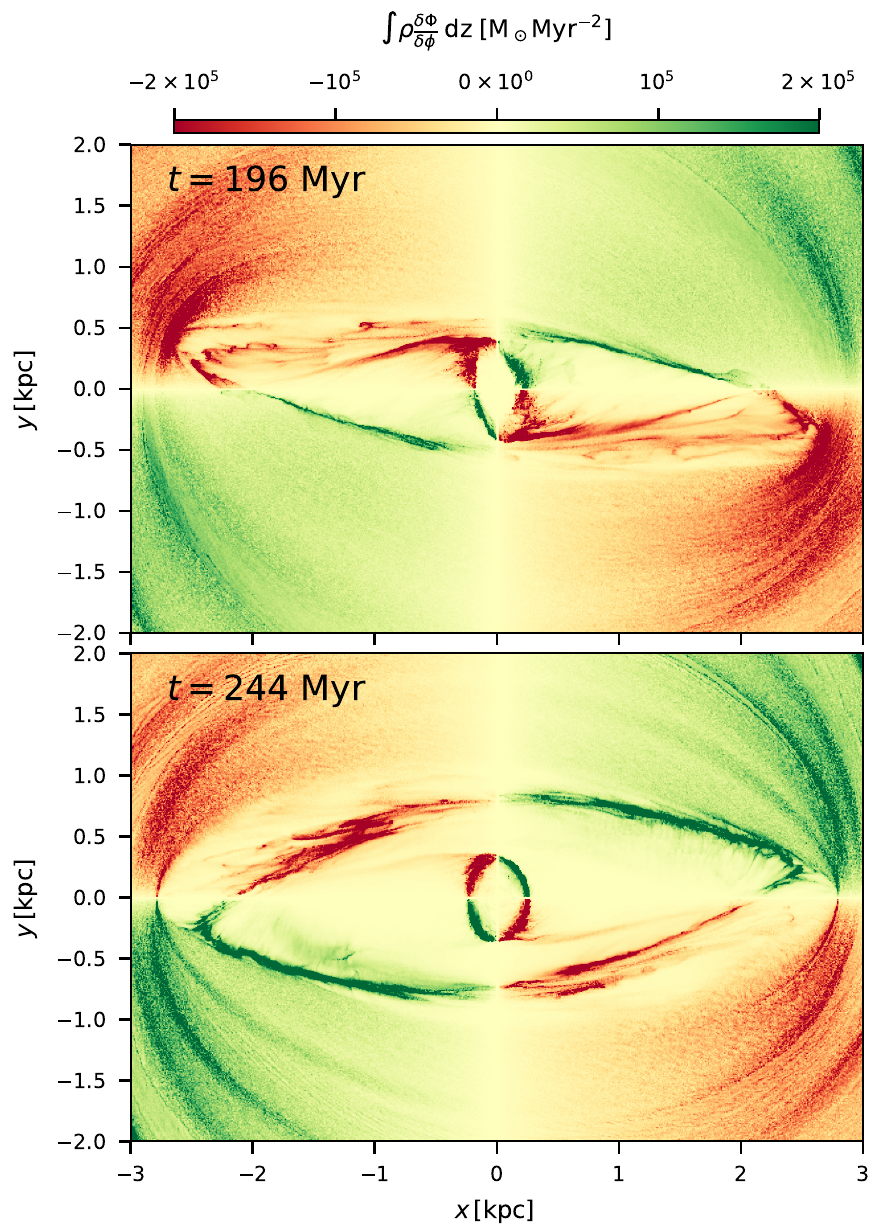}
    \caption{Mass-weighted gravitational torques map for two selected snapshots of our CHEM\_MHD simulation. The bar removes (adds) angular momentum in the two red (green) quadrants.} 
    \label{fig:GravTorquesMap}
\end{figure}

Figure~\ref{fig:Lzdot_vs_t} shows the turbulent Reynolds, Maxwell, and gravitational stresses as a function of time for two selected radii. The top panel is for $R=3.1\kpc$. On these scales, the gravitational torques dominate at all times, demonstrating that they are the ones driving the bar-driven inflow. The gravitational torques display regular oscillations on a timescale of $t\simeq 7\Myr$, which arise for the following reason. Let us consider a particle on a closed periodic elongated orbit in a non-axisymmetric bar potential, for example an $x_1$ orbit \citep{Contopoulos1989}. The angular momentum is not conserved in a bar potential, so the angular momentum of the particle will oscillate with a period equal to the orbital period. During half of the orbit, the particle loses angular momentum to the bar, for half of the orbit it gains angular momentum from the bar, while there is no net gain in the long term since the orbit is periodic. The oscillations in Fig.~\ref{fig:Lzdot_vs_t} are simply a torque-weighted version of this type of orbital oscillations, averaged over all clouds at $R<3.1\kpc$. These oscillations are visible because the gas is not symmetrically distributed with respect to the Galactic centre as shown in Fig.~\ref{fig:GravTorquesMap} (otherwise contributions on opposite sides would cancel out). These considerations explain the oscillations, but they do not explain why the average value of $F_{\rm Grav}$ is less than zero nor the average upward trend of the $F_{\rm Grav}$. These two are explained by the fact that the gas orbits are not exactly periodic, and there is a net inflow. The upward trend in the $F_{\rm Grav}$ is because the bar-driven inflow decreases over time (Fig.~\ref{fig:MassInflowCMZ}). Eventually the gravitational torques even become slightly positive at $t>220\Myr$, so the gas is gaining angular momentum from the bar. This happens because, after the bar-driven inflow shuts down, there are residual large-scale oscillations of the overall gas distribution at the outskirts of the bar from the initial gradual turn-on of the bar, which albeit slow, is not completely `adiabatic'.  This can be seen in the bottom panel of Fig.~\ref{fig:GravTorquesMap}, which shows that the $x_1$ ring is tilted so that the tips at $x\simeq\pm 2.5\kpc$ are in the positive-torques quadrants.

The bottom panel in Fig.~\ref{fig:Lzdot_vs_t} is for $R=149\pc$, which corresponds to regions inside the CMZ gas ring. Here, the Maxwell stresses dominate at $t>200\Myr$, when the nuclear inflow is significant (bottom panel of Fig.~\ref{fig:MassInflowCentre}). Gravitational stresses are negligible at all times at this radius. This shows that the magnetic stresses are the mechanism driving the nuclear inflow in our simulations. 

Figure~\ref{fig:Lzdot_vs_R} corroborates the same conclusions by showing the stresses as a function of radius. At radii larger than that of the CMZ gas ring, the gravitational torques dominate (when the bar-driven inflow is active). Inside the CMZ ring, the magnetic stresses dominate (when the nuclear inflow is active). The turbulent Reynolds stresses are always negligible.

A natural question to ask is whether MRI-driven accretion would be the dominant inflow mechanism when further processes that are not included in the present simulations are taken into account. \cite{Tress2020} quantified the contribution of supernova feedback in driving a nuclear inflow from the CMZ inwards in simulations that included star formation and self-gravity but not magnetic fields. They found that supernova-driven turbulence can drive a nuclear inflow of approximately $\dot{M}\sim0.03\Msunyr$, which is highly variable in time. This is of the same order of magnitude of the MRI-driven inflow that we found here ($\dot{M}=0.01\mhyphen0.1\Msunyr$). Thus, it is not obvious which mechanism dominates, or even if there is a single dominant mechanism. Furthermore, supernova-driven feedback and magnetic fields could interact in a non-linear way when they are both present. \cite{Moon2023} run semi-global simulations that included both supernova feedback and magnetic fields, and found that the latter can significantly enhance nuclear accretion flows compared to the supernova-only case. Understanding the dominant mechanism for the nuclear inflow will require a careful comparison that explores all relevant physical processes under a computational setup that allows a clear comparison (same gravitational potential, resolution, code).

In summary, we can clearly distinguish between two regimes in our simulations. The transport of gas from the Galactic disc to the CMZ gas ring ($R \simeq 3\kpc \to 300 \pc$) is dominated by the gravitational torques of the Galactic bar, and it is only mildly affected by the presence of magnetic fields. This is what we refer to as the `bar-driven inflow'. The transport of gas from the CMZ towards the innermost few parsec is entirely due to MRI-driven turbulence and is mediated by magnetic stresses. We refer to this as the `nuclear inflow'.

\section{Summary and conclusions} \label{sec:conclusion}

We investigated the impact and properties of the magnetic fields in the central regions of the Milky Way using 3D magnetohydrodynamical simulations of non-self gravitating gas flowing in an externally imposed barred potential. We found the following results:

\subsection{Gas morphology}

\begin{itemize}
    \item Magnetic pressure tends to increase the effective sound speed of the gas, and to decrease the radius of the nuclear ring (Sect.~\ref{sec:gasmorphology}).
    \item Magnetically-driven turbulence puffs the gas up and increases the scale-height compared to non-magnetised simulations (Sect.~\ref{sec:gasmorphology}).
    \item Magnetically-driven accretion fills the region inside the CMZ gas ring with gas, which would remain devoid of gas in the absence of magnetic fields or other physical processes not included in the present simulations such as supernova feedback (Sect.~\ref{sec:gasmorphology}).
\end{itemize}

\subsection{Magnetic field properties}
\begin{itemize}
    \item The magnetic field can be conveniently decomposed into a regular time-averaged component and a random turbulent component (Sect.~\ref{sec:decomposition}).
    \item The regular component is generally well aligned with the gas velocity vectors of gas flowing in a bar potential. In particular, the magnetic field in the bar lanes that transport the gas from the Galactic disc to the CMZ is parallel to the lanes. Turbulence tends to `disalign' the magnetic and velocity fields (Sect.~\ref{sec:geometry}).
    \item The field geometry transitions from toroidal near the $z=0$ plane to poloidal at $|z|>0$. The transition happens through a complex `butterfly' pattern (Sect.~\ref{sec:geometry}).
    \item The magnetic field scales as a function of density as $B \simeq 102 \, (n/10^3 \, {\rm cm^{-3}})^{0.33} \, {\rm \mu G}$. This can be explained by the magnetic field playing a non-negligible dynamical role (Sect.~\ref{sec:Bstrength}).
\end{itemize}

\subsection{Turbulence}
\begin{itemize}
    \item The combination of bar inflow and magneto-rotational instability (MRI) drives turbulence in the CMZ and can maintain a velocity dispersion of the order $\sigma \sim 5\kms$ on a scale of 20~pc. The MRI alone in the absence of bar inflow maintains $\sigma_z \sim 3\kms$. Both these values are lower than the velocity dispersion observed in the CMZ on the same scale, indicating that magnetic fields and bar driven inflow alone cannot drive the full amount of turbulence observed in the CMZ. Stellar feedback is likely the missing ingredient that is necessary to fully explain the observed turbulence (Sect.~\ref{sec:turbulence}).
    \item When turbulence in the CMZ is driven by both the bar inflow and the MRI, the ratio between the turbulent kinetic and magnetic energy is $E_k/E_B \gtrsim 2$, similar to the value found in studies of stellar feedback-driven and gravity-driven turbulence. When the turbulence is driven by the MRI alone, this value decreases to $E_k/E_B < 1$, similar to studies of purely MRI-driven turbulence (Sect.~\ref{sec:turbulence}).
\end{itemize}

\subsection{Growth of the magnetic field}
\begin{itemize}
    \item Magnetic fields grow in our simulations because of dynamo action driven by a combination of MRI and differential rotation, until they saturate at a mass-weighted value in the CMZ ring of approximately $B_{\rm tot}\simeq 200~\mu\G$. 
    \item The saturation value is not sensitive to the initial strength and orientation of the magnetic field, and does not change significantly if we increase the resolution of the simulations (Sect.~\ref{sec:Bevolution})
\end{itemize}

\subsection{Inflows}
\begin{itemize}
    \item We can clearly distinguish two inflow regimes acting in different radial ranges in our simulation: (1) The bar-driven inflow that transports the gas from the Galactic disc ($R\simeq3\kpc$) down to the CMZ gas ring ($R=200\mhyphen300\pc$); (2) The nuclear inflow that transports the gas from the CMZ inwards towards the central few parsec (Sect.~\ref{sec:inflow}).
    \item The bar-driven inflow is driven by the gravitational torques of the Galactic bar and is only marginally influenced by the presence of magnetic fields. The inflow rate is of the order $\dot{M}\simeq 1 \Msunyr$ (Sect.~\ref{sec:inflow}).
    \item The nuclear inflow is driven by magnetic stresses in our simulations. The inflow rate is of the order of $\dot{M} \simeq 0.01\mhyphen0.1\Msunyr$. This suggests that MRI-driven transport is a viable mechanism to transport gas to the Nuclear Star Cluster (NSC) that will contribute to its in-situ star formation. A resolution study shows that the nuclear inflow rate decreases with increasing numerical resolution, and our simulations do not appear to be converged at the maximum resolution we can afford. The above values should therefore be considered as upper limits (Sect.~\ref{sec:inflow}).
\end{itemize}
\section*{Data availability}
Movies of the simulations can be found at the following link \url{https://youtube.com/playlist?list=PLlsb6ZGKWbI4yV0kB1kOvIVqm-Ag1oNW8&si=BMyEGd7FkZeNtRdr}. 
\begin{acknowledgements}
 \ackHelix \ackMCS \ackJDH \ackAMS \ackPG \ackVMR
 \ackLC
 \ackRSK \ackES \ackJMDK \ackJG \ackAG \ackCB
\end{acknowledgements}



\bibliographystyle{aa}
\bibliography{bibliography}




\clearpage

\begin{appendix}

\section{Conservation of angular momentum and mass transport} \label{sec:appendixA}

In this appendix we derive equations for the transport of angular momentum and mass. Our treatment mostly follows \citet{Balbus1999} and \citet{Moon2023}.

An equation for the angular momentum transport can be obtained from Eq.~\eqref{eq:euler}. Multiplying the azimuthal component of Eq.~\eqref{eq:euler} by $R$ using standard cylindrical coordinates $(R,\phi,z)$ and rearranging gives:
%
%

\begin{equation} \label{eq:angularmomentum}
\frac{\pa ( l_z )}{\pa t} + \nabla \cdot \mathbf{F}_J  = - \rho \frac{\partial \Phi}{\partial \phi} \,,
\end{equation}
where
\begin{align}
& l_z = \rho R v_\phi \,,\\ 
& \mathbf{F}_J = R \left( \rho v_\phi \bfv + P \hatephi - \frac{B_\phi}{4\pi} \bfB + \frac{B^2}{8 \pi} \hatephi \right)\,.
\end{align}
The quantity $l_z$ is the angular momentum per unit volume, while $\mathbf{F}_J$ is the flux of angular momentum, which is the sum of contributions due to bulk motions of the gas, pressure forces, and magnetic forces. The term $\rho \pa \Phi/\pa\phi$ is a source term representing the changes in angular momentum due to torques from the external potential. When $\rho \pa \Phi/\pa\phi=0$, Eq.~\eqref{eq:angularmomentum} implies that the total angular momentum is conserved. Indeed, the only agent that can change the total angular momentum in our simulations is the external bar potential. We note that the angular momentum per unit volume $l_z$ does not contain any contribution due to magnetic fields. This is because, although electromagnetic fields can in general carry angular momentum, this contribution is neglected in ideal MHD since it is of order $(v/c)^2$ with respect to the angular momentum contained in the gas.

The angular momentum flux in the radial direction is:
\begin{align}
F_{JR}  & = R \left(\rho v_\phi v_R + T_{R \phi}\right) \, ,\label{eq:radialflux}
\end{align}
where
\begin{align} \label{eq:maxwellRphi}
T_{R \phi} = - \frac{B_\phi B_R}{4 \pi}\,,
\end{align}
is the component of the Maxwell stress tensor defined in Sect.~\ref{sec:methods}. When \eqref{eq:radialflux} is integrated over the surface of a cylinder of radius $R$, it gives the flux of angular momentum through the surface of the cylinder. The term $\rho v_\phi v_R$ quantifies the angular momentum flux carried by gas that is physically crossing the surface $R$, while ${T}_{R\phi}$ is the contribution due to magnetic torques.
Equation~\eqref{eq:Lzdot} in the main text is obtained by integrating Eq.~\eqref{eq:angularmomentum} over the volume of a cylinder of radius $R_0$ and using the divergence theorem. 
 
We now relate these quantities to the mass accretion rate. Expanding Eq.~\eqref{eq:angularmomentum} in cylindrical coordinates, integrating in both the vertical ($z$) and azimuthal ($\phi$) direction, and assuming that the boundary terms in the vertical direction vanish, we obtain:
\begin{equation}  \label{eq:averaged}
\frac{ \partial \langle l_z \rangle}{\partial t} + \frac{1}{R} \frac{\partial \langle R F_{JR} \rangle}{\partial R} = - \langle \rho \frac{\partial \Phi}{\partial \phi}\rangle \,,
\end{equation}
where
\begin{equation}
\langle X \rangle = \frac{1}{2\pi} \int_0^{2 \pi} \di \phi \, \int X \, \di z \,,
\end{equation}
denotes the vertical and azimuthal average of a physical quantity $X$. 
Integrating and averaging the continuity equation \eqref{eq:Continuity} in the same way gives:
\begin{equation} \label{eq:continuityaveraged}
\frac{\pa \langle \rho \rangle }{\pa t} + \frac{1}{R} \frac{\pa}{\pa R}\left[ \langle R \rho v_R \rangle \right] = 0\,.
\end{equation}
Next, we decompose the velocity as
\begin{equation}
\bfv = v_0 \hatephi + \bfu\,, 
\end{equation}
where
\begin{equation}
v_0(R,t) = \frac{ \langle \rho v_\phi \rangle }{\langle \rho \rangle}\,,
\end{equation}
is the average rotation velocity and $\bfu$ represents the fluctuations. 
Substituting \eqref{eq:v0} into \eqref{eq:averaged}, using \eqref{eq:radialflux}, \eqref{eq:angularmomentum}, and the averaged continuity equation \eqref{eq:continuityaveraged}, and using the fact that $\langle \rho u_\phi \rangle=0$ by our definition of $v_0$, we obtain
\begin{equation} \label{eq:fluctuations}
R \langle \rho \rangle \frac{\pa v_0 }{\pa t}  +
\langle \rho u_R \rangle \frac{\partial (R v_0) }{\partial R} + \frac{1}{R} \frac{\pa \langle R^2 \rho u_R u_\phi \rangle}{\pa R} + \frac{1}{R} \frac{\pa \langle R^2 T_{R\phi} \rangle}{\pa R}  = - \langle \rho \frac{\partial \Phi}{\partial \phi}\rangle\,, 
\end{equation}
Equation \eqref{eq:fluctuations} can be rewritten in a more illuminating form as:
\begin{equation} \label{eq:Mdot2}
\dot{M} = \dot{M}_{\rm M} + \dot{M}_{\rm R} + \dot{M}_{\rm G} + \dot{M}_t \,,
\end{equation}
where
\begin{align}
\dot{M} = - 2 \pi R \langle \rho u_R \rangle \,,
\end{align}
is the total mass accretion rate at radius $R$, and
\begin{align}
& \dot{M}_{\rm R} = 2\pi \left[ \frac{\pa(R v_0)}{\pa R} \right]^{-1}  \frac{\pa \langle R^2 \rho u_R u_\phi \rangle}{\pa R} \,, \\
& \dot{M}_{\rm M} = 2\pi \left[ \frac{\pa(R v_0)}{\pa R} \right]^{-1} \frac{\pa \langle R^2 T_{R\phi} \rangle}{\pa R} \,, \\
& \dot{M}_{\rm G} = 2 \pi R \left[ \frac{\pa(R v_0)}{\pa R} \right]^{-1} \langle \rho \frac{\partial \Phi}{\partial \phi}\rangle \,, \\
& \dot{M}_{t} = 2 \pi R^2 \left[ \frac{\pa(R v_0)}{\pa R} \right]^{-1} \langle \rho \rangle \frac{\pa v_0}{\pa t}  \,.
\end{align} \\
The first three terms are the contributions due to Reynolds, Maxwell, and gravitational stresses respectively, while the last term is related to changes in the average rotation velocity with time, which are typically small and can be neglected in a quasi-steady state.

\section{Resolution study} \label{sec:resolutionstudy}

\begin{figure}
	\includegraphics[width=\columnwidth]{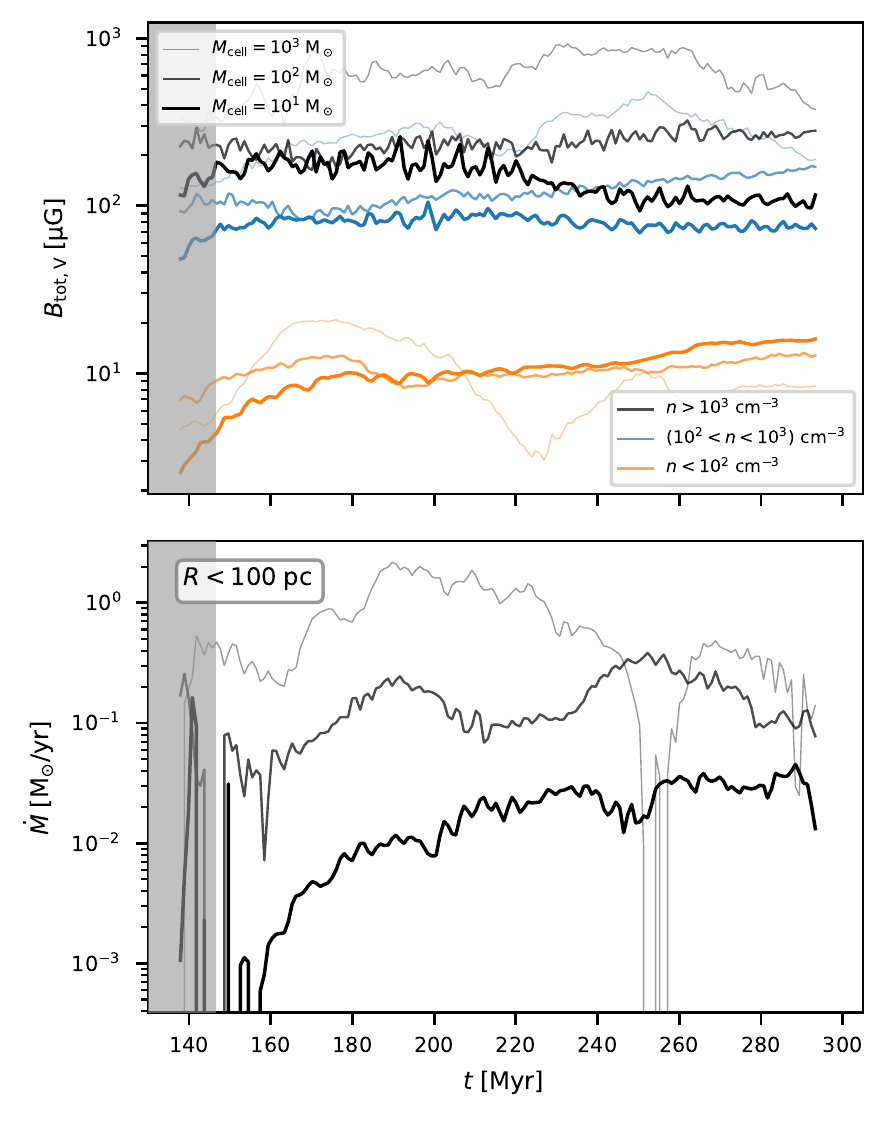}
        \caption{Impact of resolution on magnetic fields and inflow. \emph{Top}: volume-weighted average magnetic field intensity in the CMZ ($R<500$~pc, $|z|<100$~pc) in three different density regimes at different resolutions as a function of time. \emph{Bottom}:  mass accretion as a function of time from the CMZ inwards ($R < 100$~pc) for the three simulations at different resolutions. The saturation levels of the magnetic field are relatively insensitive to the resolution, while the inflow rate is highly sensitive to it.} 
    \label{fig:ResolutionStudy}
\end{figure}

We perform a resolution study to see how the magnetic field strength at saturation and the inflow rate depend on the numerical resolution. Figure~\ref{fig:ResolutionStudy} shows that the saturation magnetic field strengths are relatively insensitive to numerical resolution. In contrast, the MRI-driven inflow rate decreases as the resolution is increased, and does not appear to be converged even at the highest resolution we can afford. We can therefore only put an upper limit to the MRI-driven inflow rate.

\section{Impact of initial conditions} \label{sec:ICimpact}

\begin{figure}
	\includegraphics[width=\columnwidth]{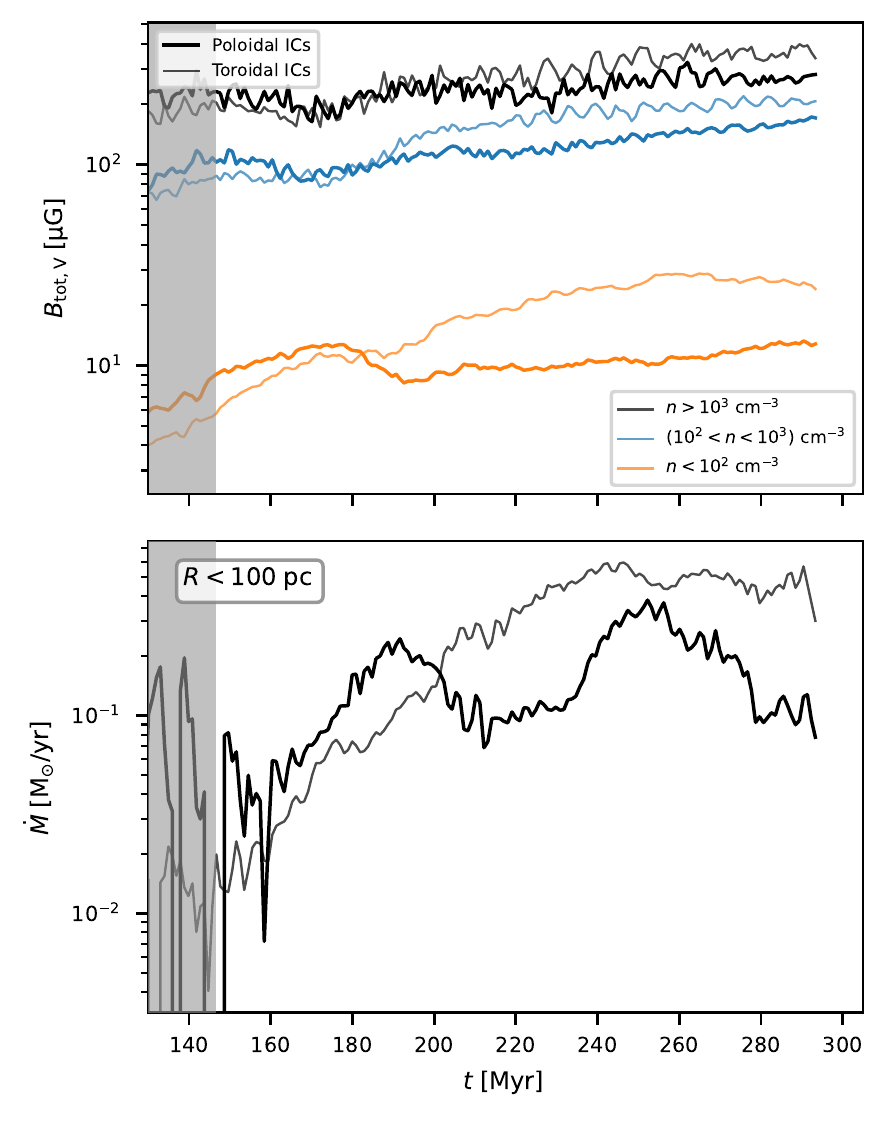}
        \caption{Impact of the initial conditions on magnetic fields and inflow. \emph{Top}: volume-weighted average magnetic field intensity in the CMZ ($R<500$~pc, $|z|<100$~pc) in three different density regimes for poloidal initial magnetic field and toroidal initial magnetic field. Both tests are run at a resolution of $100$~M$_\odot$ per cell. \emph{Bottom}: mass inflow rate as a function of time from the CMZ inwards ($R<100$~pc) in the two simulations. The figure shows that the magnetic field intensity and inflow rates are relatively insensitive to the initial orientation of the magnetic field.} 
    \label{fig:ToroidalvsPoloidal}
\end{figure}

We tested the impact of the initial conditions on the magnetic field evolution. Figure~\ref{fig:ToroidalvsPoloidal} compares the magnetic field strengths and inflow rates in the CMZ as a function of time for simulations that start with a uniform Poloidal seed magnetic field $\bfB_0=0.02\, \mu {\rm G} \, \hatez$ (as in the main text) versus a purely Toroidal magnetic field $\bfB_0=0.02\, \mu {\rm G}\, \hatephi$. We find that the magnetic field configuration and inflow rates are relatively insensitive to the initial orientation of the magnetic field. This is likely because the CMZ builds its gas and magnetic field reservoirs by bar-driven accretion during the course of the simulation, and memory of the initial conditions is erased in the process.

\section{Cutout tests} \label{sec:Cutout}

\begin{figure}
	\includegraphics[width=\columnwidth]{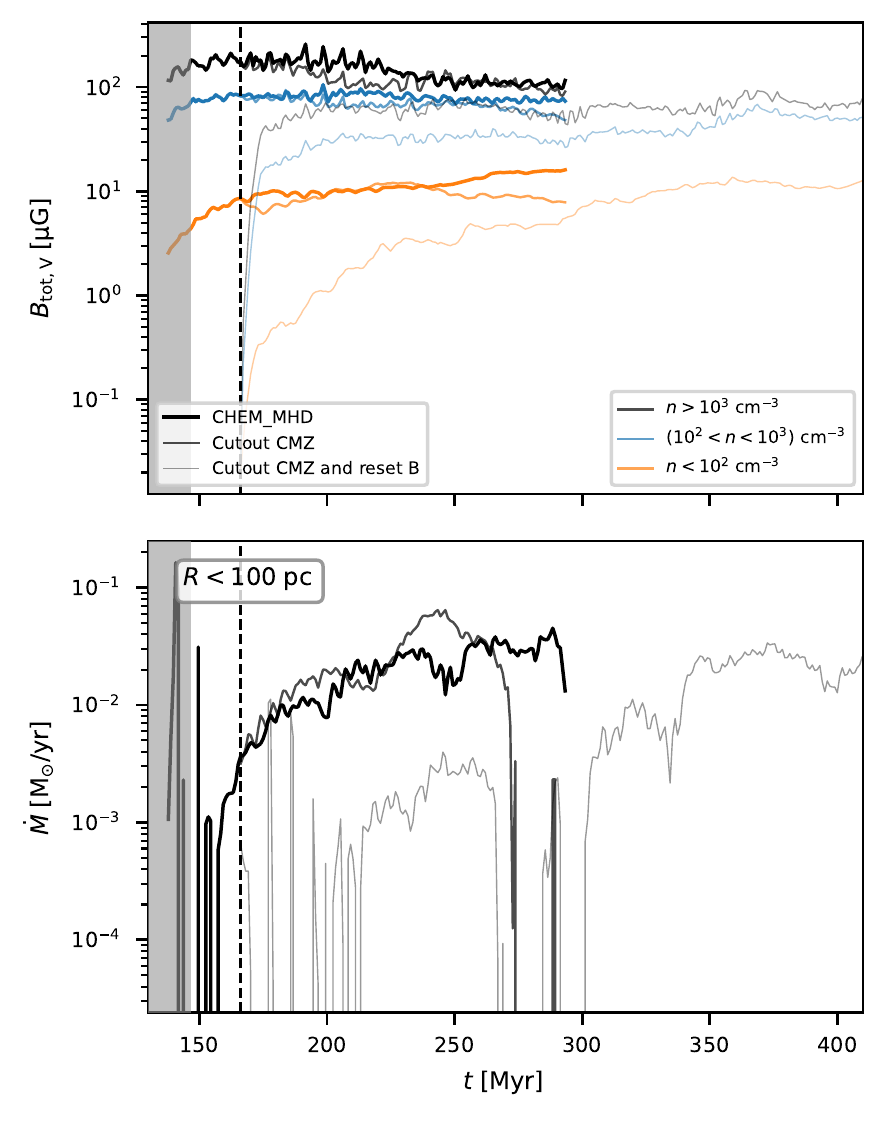}
        \caption{Effect of the cutout test on magnetic fields and inflow. \emph{Top}: volume-weighted average magnetic field intensity in the CMZ in three different density regimes for three simulaitions: (i) the fiducial CHEM\_MHD simulation; (ii) the test where we cut out the CMZ; (iii) and the test where we also reset the $B$ field (Appendix~\ref{sec:Cutout} for description). The latter was run until $400$~Myr to give time for dynamo processes and MRI to develop again. \emph{Bottom}: mass accretion as a function of time from the CMZ inwards ($R<100$~pc) for the three simulations. The vertical dashed line indicates the time at which we performed the cut-out.} 
    \label{fig:CutOut}
\end{figure}

To assess the importance of the bar-driven inflow on the turbulence and magnetic field properties in the CMZ we performed the following numerical experiments. In the first experiment we stop the simulation at $t=166$~Myr, remove all the gas at $R>500\pc$ so that we are left only with the CMZ gas ring, and then restart the simulation. In this way, we remove the large-scale bar inflow and continue the simulation with only the CMZ ring evolving `in isolation'. In the second experiment, we do the same and in addition we also axisymmetrised the gravitational potential and reset the magnetic field to the initial seed value $\bfB_0 = 0.02 \, \mu \rm G\, \hatez$. Figure~\ref{fig:CutOut} plots the evolution of the magnetic field and the inflow rates in these experiments. We find that the magnetic field intensity and inflow rates are not affected in the first experiment, and in the second both the magnetic field intensity and inflow rate eventually ramp up to the same values they had in the other simulations. We therefore conclude that the bar-driven inflow has little or no influence on the magnetic field intensity levels at saturation in the CMZ, nor on the inflow rate. These tests are consistent with our suggestion that the MRI is primarily responsible for driving the nuclear inflow within the CMZ ring in our simulations (Sect.~\ref{sec:inflow}).

\section{Stokes parameters}

The interstellar magnetic field is mostly measured indirectly. Dust polarisation observations can constrain the average magnetic field orientation along the line-of-sight, but does not tell us much about the magnetic field strength. Moreover the direction of the polarised emission does not follow simple rules of vector addition along the line-of-sight as the magnetic field does, but instead it is necessary to compute and integrate the Stokes parameters. 

\begin{figure*}
	\includegraphics[width=\textwidth]{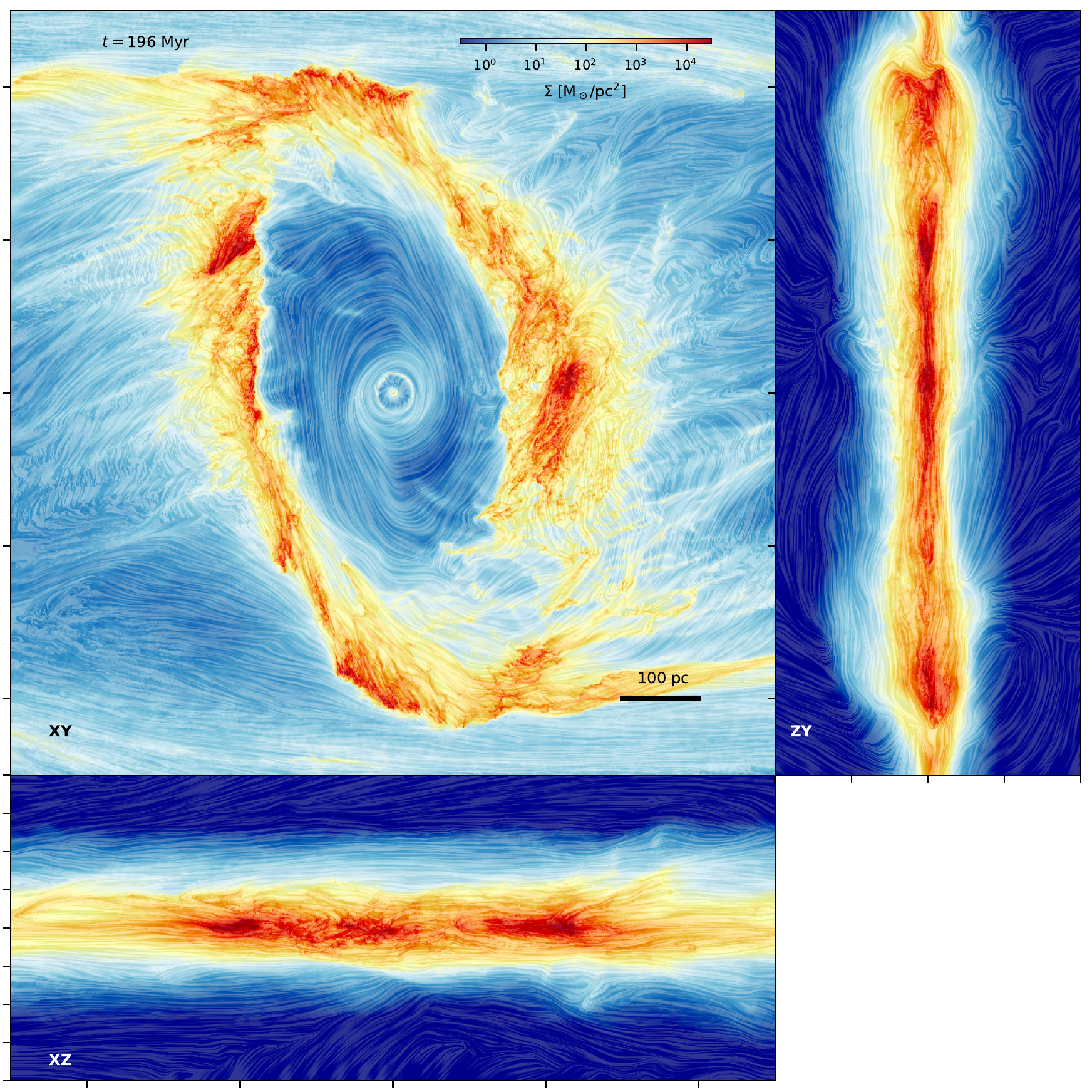}
        \caption{Same as Fig.~\ref{fig:MagnetiFieldMorph} but here we integrate the Stokes parameters along the line-of-sight instead of directly integrating the magnetic field.} 
    \label{fig:FOEO_Stokes}
\end{figure*}

To get a projected map of the magnetic field orientation that can be more easily related to observed dust polarisation maps, we computed mass-weighted normalised stokes parameters (for instance eqs. 15, 16, and 17 of \citealt{Soler2013}):

\begin{equation}
    Q_{xy} = \frac{B_x^2 - B_y^2} {B_x^2 + B_y^2},
\end{equation}
\begin{equation}
    U_{xy} = \frac{2 B_x B_y} {B_x^2 + B_y^2}.
\end{equation}
The mass weighted average magnetic field strength and its orientation are then 
\begin{equation}
    \langle | \mathbf{B}^{'} | \rangle_z = \sqrt{\langle Q_{xy} \rangle_z^2 + \langle U_{xy} \rangle_z^2}
\end{equation}
and
\begin{equation}
    \Phi_{\langle | \mathbf{B}^{'} | \rangle_z} = \frac{1}{2} \arctan \left( \frac{\langle U_{xy} \rangle_z}{\langle Q_{xy} \rangle_z} \right),
\end{equation}
where $\langle{X}\rangle_z = \left( \int \rho X \, \di z\right)/\left(\int \rho\, \di z \right)$. A Stokes parameter map of our simulations is shown in Fig.~\ref{fig:FOEO_Stokes}. We find only relatively minor differences with the direct $B$ field maps in Fig.~\ref{fig:MagnetiFieldMorph}.

\section{Divergence of the magnetic field} \label{sec:divB}

Discretization of the MHD equations introduces divergence errors of the magnetic field which are obviously spurious. These can quickly grow and dominate the dynamics of the gas. Various methods have been developed to keep them at bay. We use the Powell divergence cleaning technique \citep{Powell1999}, implemented in {\sc arepo} by \citet[][]{Pakmor2013}. Here, source terms proportional to $\nabla \cdot B$ are introduced in the equations such that the generated divergence is advected away, preventing it from growing further. This scheme is robust, local, and does not require additional restrictions on the time step. Moreover, it can be easily implemented on the unstructured moving mesh of the {\sc arepo} code. However, since the scheme is not divergence-free by construction, it is recommended to check the impact of these errors in the simulations. The implementation did not produce any obvious artefacts in test problems as well as in more complex and generic astrophysical simulations \citep{Pakmor2013, Pakmor2020} and the relative divergence error was contained provided high enough resolution. 

We define the relative error of the divergence as 
\begin{equation}
    \frac{\left (\nabla \cdot \mathbf{B}\right) r}{|\mathbf{B}|},
\end{equation}
where $r$ is the radius of the cell for which the relative error was computed. 

\begin{figure}
	\includegraphics[width=\columnwidth]{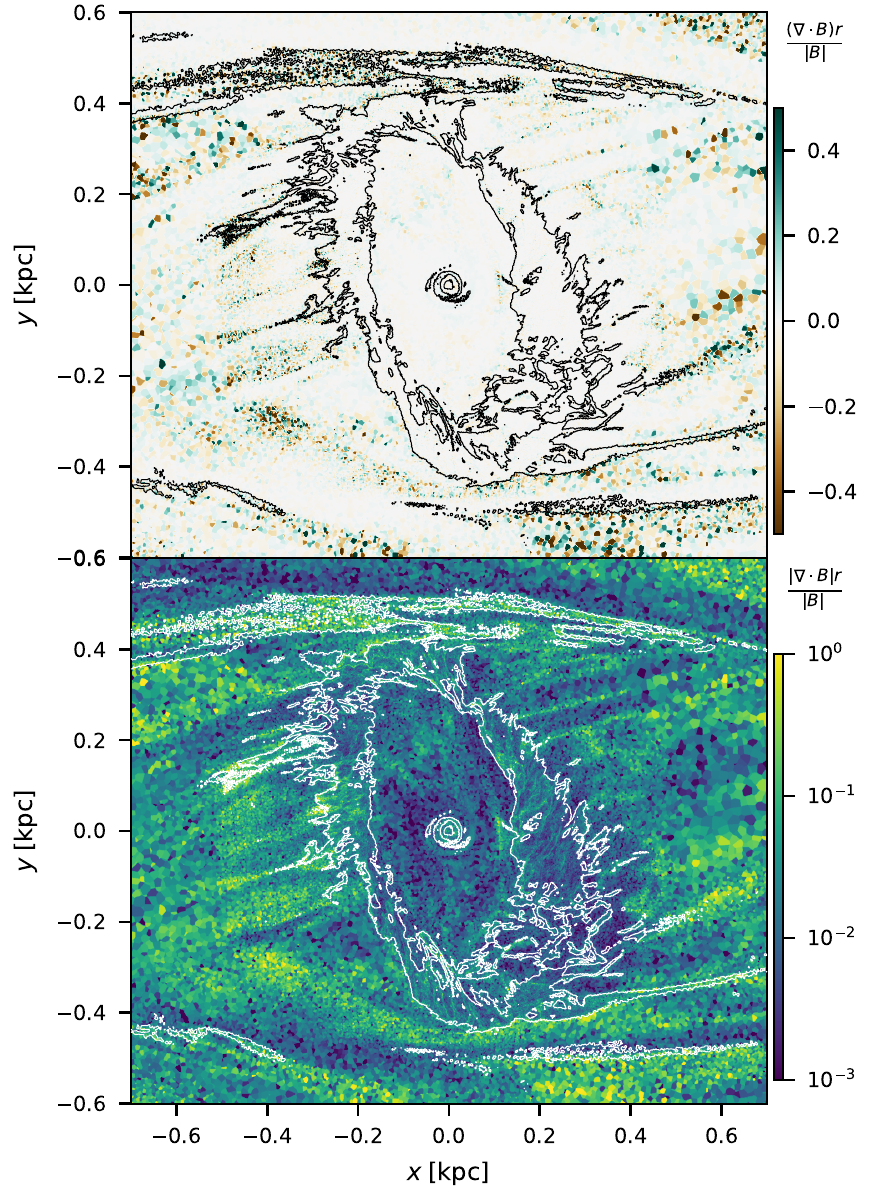}
        \caption{Map of the relative divergence error of the magnetic field for a slice through the $z=0$ plane at $t=196$~Myr. \emph{Top}: the divergence error map is shown considering its sign as well. \emph{Bottom}: map of the absolute value of the divergence error in $\log$ scale. The contour corresponds to an isodensity contour of $n=20$~cm$^{-3}$.} 
    \label{fig:DivBMap}
\end{figure}

In Fig.~\ref{fig:DivBMap} we show the relative divergence error map for a slice through the midplane at $t=196$~Myr. In the CMZ the divergence error is always small (of the order of a few percent). Cells with higher divergence errors are present mostly in the low density region surrounding the CMZ or at the interfaces between low and high density regions. Here the magnetic field gradients are highest and greater divergence errors are to be expected. 

In the top panel, the divergence error is shown including its sign. There, we see that no large patches of only positive (or negative) divergence errors are present. This means that even in regions with higher divergence errors, the error mostly averages out on larger scales, such that the dynamics of scales including more than just a few cells are properly followed. 

In the bottom panel, the absolute value in log scales is shown instead. We see that even for regions of relatively high divergence error, the error never grows excessively. 

\begin{figure}
	\includegraphics[width=\columnwidth]{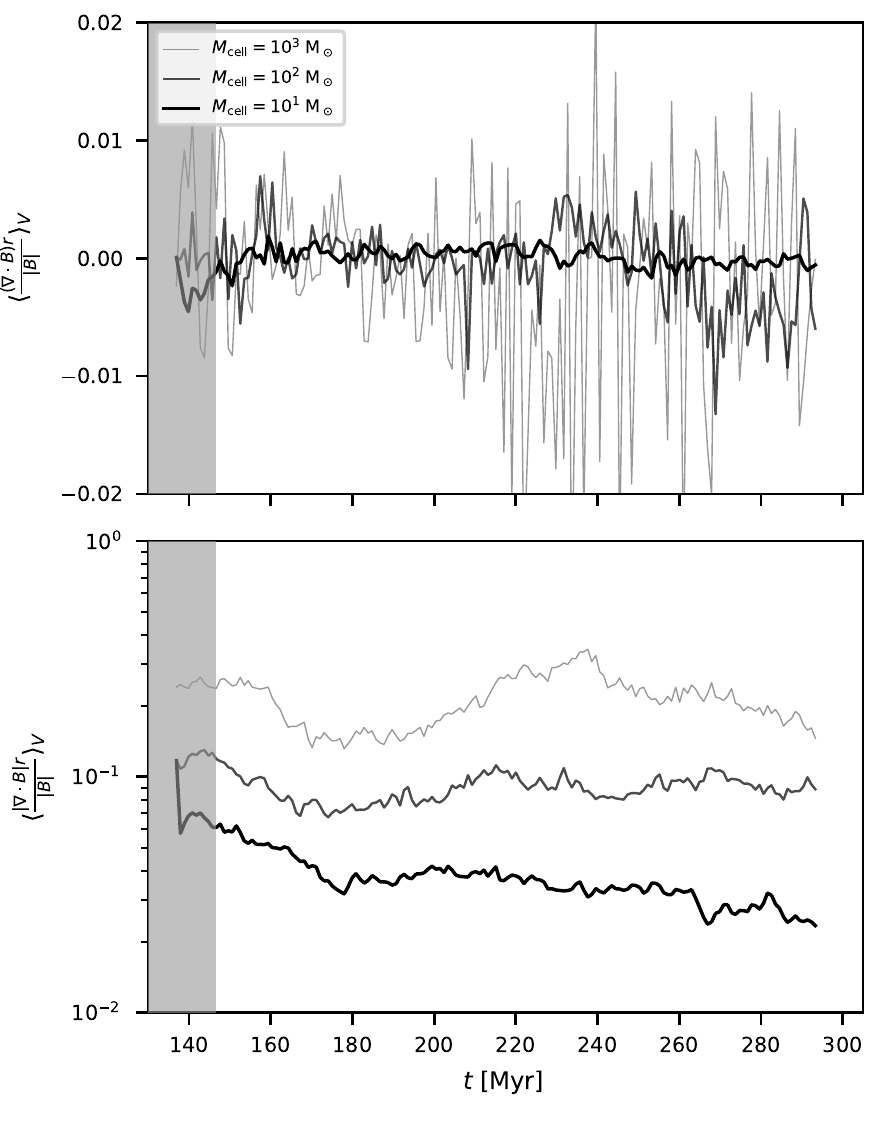}
        \caption{Relative divergence error of the magnetic field, volume-averaged over the central $500$~pc, as a function of time for the different simulations at different resolution. \emph{Top}: the divergence error is shown considering its sign as well. \emph{Bottom}: volume-average of the absolute value of the divergence error in $\log$ scale.} 
    \label{fig:DivBvst}
\end{figure}

In Fig.~\ref{fig:DivBvst}, we show the volume averaged divergence error in the central $500$~pc as a function of time. When averaged considering its sign (top panel), the error is always much less than 1\%, and is not diverging in time, but oscillates around the zero value. 

Even if the average is performed by taking the absolute value of the relative divergence error (bottom panel), it is contained to a few percent. 

In Fig.~\ref{fig:DivBvst} we also included the values computed for simulations at lower resolutions. As expected, the divergence error reduces with resolution, and even at the lowest resolution the divergence error is never significant. 

\end{appendix}

\end{document}